\documentclass[aps,prx,singlecolumn,superscriptaddress,floatfix]{revtex4-2}
\usepackage[T1]{fontenc}
\usepackage[latin1]{inputenc}
\usepackage{graphicx,amsfonts}
\usepackage{amsmath,mathrsfs}
\usepackage{xcolor}
\usepackage{cancel}
\newcommand{\be}{\begin{equation}}
\newcommand{\ee}{\end{equation}}
\newcommand{\bea}{\begin{eqnarray}}
\newcommand{\eea}{\end{eqnarray}}
\newcommand{\bml}{\begin{subequations}}
\newcommand{\eml}{\end{subequations}}

\newcommand{\G}{\Gamma}

\newcommand{\bbm}{\begin{bmatrix}}
\newcommand{\ebm}{\end{bmatrix}}

\newcommand{\V}{\mathcal{V}}
\newcommand{\W}{\mathcal{W}}

\usepackage{enumitem}

\begin{document}

\title{First-Order General-Relativistic Viscous Fluid Dynamics}

\label{today}

\author{F\'abio S.\ Bemfica}
\affiliation{Escola de Ci\^encias e Tecnologia, Universidade Federal do Rio Grande do Norte, 59072-970, Natal, RN, Brazil}
\email{fabio.bemfica@ect.ufrn.br}

\author{Marcelo M.\ Disconzi}
\affiliation{Department of Mathematics, Vanderbilt University, Nashville, TN, USA}
\email{marcelo.disconzi@vanderbilt.edu}

\author{Jorge Noronha}
\affiliation{Illinois Center for Advanced Studies of the Universe \& Department of Physics, 
University of Illinois at Urbana-Champaign, Urbana, IL 61801, USA}
\email{jn0508@illinois.edu}

\begin{abstract}
We present the first generalization of 
 Navier-Stokes theory to relativity
that satisfies all of the following properties: (a) the system coupled to Einstein's
equations is causal and strongly hyperbolic; (b) equilibrium states are stable;
(c) all leading dissipative contributions are present, i.e., shear viscosity, bulk viscosity, and thermal conductivity;
(d) non-zero baryon number is included; (e) entropy production is non-negative in the regime of validity of the theory;
(f) all of the above 
holds in the nonlinear regime without any simplifying 
symmetry assumptions. These properties are accomplished using a generalization of Eckart's theory containing
only the hydrodynamic variables, so that no new extended degrees of freedom are needed
as in M\"uller-Israel-Stewart theories. Property (b), in particular, follows from a more general result
that we also establish, namely, sufficient conditions that when added to stability in the fluid's rest frame
imply stability in any reference frame obtained via a Lorentz transformation. All our results are mathematically rigorously established. 
The framework presented here provides the starting point for systematic investigations of general-relativistic viscous phenomena in neutron star mergers.
\end{abstract}

\maketitle


\section{Introduction} 

Relativistic fluid dynamics has been successfully used as an effective description of long wavelength, long time phenomena in a multitude of different physical systems, ranging from cosmology \cite{WeinbergCosmology} to astrophysics \cite{RezzollaZanottiBookRelHydro} and also high-energy nuclear physics \cite{Romatschke:2017ejr}. In the latter, relativistic viscous fluid dynamics has played an essential role in the description of the dynamical evolution of the quark-gluon plasma formed in ultrarelativistic heavy-ion collisions \cite{Heinz:2013th} and also in the quantitative extraction of its transport properties (see, for instance,  \cite{Bernhard:2019bmu}). More recently, with the observation of binary neutron star mergers \cite{TheLIGOScientific:2017qsa,Monitor:2017mdv,GBM:2017lvd}, the modeling of the different dynamical stages experienced by the hot and dense matter formed in these collisions requires extending of our current understanding of viscous fluids towards the strong gravity regime where general relativistic effects are important (see, e.g., \cite{Duez_et_al_viscosity_2004,Shibata:2017jyf,Shibata:2017xht,Alford:2017rxf,Radice:2018ghv,Most:2021zvc}).

The ubiquitousness of fluid dynamics stems from the existence of general conservation laws (such as energy, momentum, and baryon number) and their consequences to systems where there is a large separation of scales, such that the macroscopic behavior of conserved quantities can be understood without precise knowledge of all the details that govern the system's underlying microscopic properties \cite{LandauLifshitzFluids}. Ideal fluid dynamics is the extreme situation where dissipative effects are neglected and the theory's basic properties in this limit are reasonably well understood,
both in a fixed background as well as when coupling to Einstein's equations 
is taken into account \cite{RezzollaZanottiBookRelHydro,DisconziRemarksEinsteinEuler,Choquet-BruhatFluidsExistence}. We remark that because all sources of dissipation
relevant for our discussion stem from bulk viscosity, shear viscosity, and heat conduction, 
and following standard practice in the field
\cite{Romatschke:2017ejr},
we will use the terms viscous fluid and dissipative fluid interchangeably. In particular,
other sources of dissipation, such as anomalous
dissipation \cite{
Eyink:2017zfz,Eyink:2017xtw}, will not be discussed.

When dissipative effects are taken into account, the behavior of fluids is far less understood (unless stated otherwise, fluids, hydrodynamics, etc. henceforth mean 
relativistic fluids, relativistic hydrodynamics, etc.), despite the importance of viscous dissipation in cutting-edge scientific experiments 
such as in studies of the quark-gluon plasma or their
expected relevance for neutron star mergers, as mentioned above. Historically, a stumbling block has been the difficulty
of modeling dissipative phenomena while preserving 
\emph{causality.} Causality is a central postulate in special and general relativity, stating that the speed which information can propagate in any system cannot be larger than the speed of light \cite{HawkingEllisBook}. This implies that a solution to the equations of motion at a given space-time point $x$ is completely determined by the spacetime region that is in the past of and causally connected to $x$ \cite{HawkingEllisBook,ChoquetBruhatGRBook,WaldBookGR1984}.
Of course,
this property must hold in relativity regardless of whether dissipation is present or not \cite{HawkingEllisBook}.
While causality is typically not an 
issue for most matter models under reasonable assumptions \cite{ChoquetBruhatGRBook}, including the case of ideal fluids
\cite{AnileBook}, ensuring causality of fluid theories
in the presence of dissipation turned out to be a major challenge  \cite{RezzollaZanottiBookRelHydro}. 

The challenges one encounters when modeling fluids with
dissipation, however, are not restricted to enforcing
causality. Another hallmark property of dissipative
fluids is \emph{stability.} By this we mean that perturbations of a system that is in thermodynamic equilibrium should decay in time. This expresses the
basic intuition that if dissipation is present, 
the system will dissipate energy and, consequently, 
small deviations from equilibrium will be damped, leading the dynamics to return to equilibrium within
some characteristic time scale. Naturally, in order
to implement this idea in a given formalism one needs
to specify what is meant by equilibrium and perturbations.
We will consider homogeneous (non-rotating) equilibrium states and our perturbations will be plane-wave
solutions to the equations of motion linearized
about such homogeneous states. Although this is 
not the most general definition of equilibrium
\cite{Hiscock_Lindblom_stability_1983},
it captures the most basic intuition about how
deviations from equilibrium should behave in a dissipative
theory and, consequently, in practice this has
been the definition most often used in the literature
\cite{Hiscock_Lindblom_instability_1985,Pu:2009fj}. Like causality, stability
is a property that is difficult to incorporate
in theories of relativistic fluids with dissipation.

Aside from causality and stability, a third
fundamental property required for a theory
of relativistic viscous fluids is that the equations
of motion be \emph{locally well-posed.} This means
that given initial conditions, there must exist one
and only one solution to the equations of motion
taking the prescribed initial conditions \cite{F:Sobolev} 
and defined for some time  \footnote{We say some time instead of all times because solutions can develop singularities and, thus, not be defined
for all times. The fact that nonlinear equations
of physical significance can indeed develop singularities, e.g., shock waves 
\cite{RezzollaZanottiBookRelHydro} in fluid dynamics or spacetime singularities in general
relativity \cite{HawkingEllisBook}, 
is well-known. The question of how to deal with such singularities, e.g., how to meaningfully extend the solutions past the singularity or replace
the singular theory by a more complete one
where singularities can be avoided, while of 
uttermost physical importance, is beyond
the scope of this work.}. Physically,
this means that the system has a well-defined
evolution determined by the initial conditions.
Like causality, local well-posedness is a property
required of any field theory
\cite{ChoquetBruhatGRBook,RezzollaZanottiBookRelHydro,HawkingEllisBook,WaldBookGR1984}, but we emphasize
it here since, also like causality, this is a property that is difficult to achieve in theories
of fluids with dissipation.

Needless to say, there is little 
use for a theory of fluids that is causal,
stable, and locally well-posed if it is not
able to make connections with real physical
phenomena. Thus, a theory of relativistic
viscous fluids must in addition be 
suitable for empirical studies. 
This means, at the least, that the theory
must agree with well-established physical
facts, but also that one needs
to be able to extract quantitative predictions from such a theory.

The interplay between theory and experiment
is, of course, at the heart of physics. 
In the context of relativistic fluid dynamics,
such interplay has been heavily guided
by complex numerical simulations  
\cite{Romatschke:2017ejr}. Moreover, 
it is clear that simulations will 
continue to be at the center of developments
in the field, particularly when it comes
to the investigations of viscous
effects in neutron star mergers. In this regard, while there is no one-size-fits-all
approach for implementing numerical
simulations of general relativistic systems
\cite{RezzollaZanottiBookRelHydro,baumgarte_shapiro_2010}, in the numerical general
relativity community one concept that has
been very important for the construction
of numerical algorithms is that
of \emph{strong hyperbolicity}
\cite{ReulaStrongHyperbolic}. This means
that the principal part of the equations of motion can be diagonalized; see 
Section \ref{sec:hyperbolicity} for details.
Although a discussion of the role of strong hyperbolicity
in general relativistic numerical simulations
is beyond the scope of this work (the reader 
can consult the above references for details),
we stress that strong hyperbolicity is a highly
desirable feature for numerical 
studies of general relativistic systems  (see 
also 
\cite{GuermondetalNumerical} for more discussion on potential caveats of numerical simulations).

\begin{quote}
In sum, a physically meaningful theory
of relativistic viscous fluids must be:
\begin{enumerate}[label=(\Roman*),series=properties]
    \item Causal.
    \label{I:Causality}
    \item Stable.
    \label{I:Stability}
    \item Locally well-posed.
    \label{I:LWP}
\end{enumerate}
In addition, it is highly desirable 
to have a theory that is
\begin{enumerate}[resume*=properties]
    \item Strong hyperbolic.
\label{I:Strong_hyperbolicity}
\end{enumerate}
\end{quote}

While property \ref{I:Stability} is, by definition, concerned with the equations linearized about equilibrium in Minkowski background, 
we emphasize that whenever referring to causality, local well-posedness, and strong hyperbolicity, i.e., properties
\ref{I:Causality}, \ref{I:LWP}, and \ref{I:Strong_hyperbolicity},
we are always talking about the equations of motion in the 
\emph{full nonlinear regime.} It is important to stress this point
because a substantial body of theoretical work in relativistic viscous fluids
is restricted to analyzing the equations linearized about equilibrium and, thus, the corresponding claims about causality etc. are restricted to this particular, linearized-about-equilibrium case
(see Section \ref{S:Overview}).
Furthermore, for applications in general relativity
(in particular the study of viscous effects in neutron star mergers), one is interested in the case where properties \ref{I:Causality}--\ref{I:Strong_hyperbolicity} hold with dynamical coupling to Eintein's equations (again, with exception of property \ref{I:Stability}).

At this point we should stress that when we say that a theory is causal, stable, etc., we do not mean it unconditionally, but rather under a specific set of assumptions. Obviously, one is interested in cases
where the assumptions are physically reasonable, even if 
they do not cover all cases of physical interest. For simplicity, however, in the remaining of this Introduction and in Section \ref{S:Background_discussion}, we will avoid discussion of specific hypotheses. Thus, when we say that a certain theory is causal, etc., we mean ``causal under a specific set of assumptions,'' and unless stated otherwise, it will be implicitly understood that the assumptions in question are of physical interest.
An exception to this will be made only later in 
Section \ref{S:Overview}, when we will summarize the  extent to which different theories of viscous fluids
satisfy one or more of the properties \ref{I:Causality}--\ref{I:Strong_hyperbolicity}, since in this case mentioning the assumptions under which such theories fulfil some of these requirements will be important for comparison among them and also with our results. Even in this case, however, we will refer to those assumptions only at a high level (e.g., we will say that a certain property holds for non-zero shear viscosity but without specifying the precise range of non-zero values which is in fact required for the result to hold).
We believe that this will suffice to give the reader a panoramic view of the state-of-affairs in the field. All the precise assumptions for the results that will be discussed can be found in the references we provide or, in the case of the results of this paper, in the Sections that follow the introduction.

The goal of this work is to provide the first
example of a theory of relativistic viscous fluids that simultaneously satisfies all the properties \ref{I:Causality}-\ref{I:Strong_hyperbolicity}.
All our results are mathematically rigorous, hold 
with or without dynamical coupling to Einstein's equations, are valid in the full nonlinear regime, and
do not make any symmetry or simplifying assumptions.  We establish these
results without the need for additional (extended) variables (see Section \ref{S:Overview} for details).

Section \ref{S:Background_discussion} provides
a more or less self-contained exposition of our results and how they fit within studies of relativistic fluids with viscosity. We hope that such an exposition will be helpful to readers interested in the subject here investigated but who are not necessarily specialists in all the topics covered by our methods  \footnote{We thank the Editor and one of the anonymous referees for suggesting that we provide such a self-contained assessment.}. In order to keep our account as simple as possible, we will carry out the discussion in Section \ref{S:Background_discussion} at a high-level,
writing few formulas and omitting several details, but we will provide full references for interested readers. 
More precisely, 
in Section \ref{S:Frames}, we discuss some important 
concepts underlying the investigation of
relativistic viscous fluids. None of the ideas
discussed in Section \ref{S:Frames} are new but
they play a key role in our constructions. Therefore, it is convenient to revisit such
ideas here. In Section \ref{S:Overview}, we review the state-of-affairs in the field regarding properties \ref{I:Causality}--\ref{I:Strong_hyperbolicity}. This review is not intended to be exhaustive; rather, our goal is to provide enough context for our results. 
Finally, in Section \ref{S:Summary},
we provide a summary and discussion of our results. 
Specialists might skip Section
\ref{S:Background_discussion}
without compromising understanding (although some specialists might still be interested in some aspects 
of the discussion in Section \ref{S:Summary}).

\medskip
\emph{Definitions}: The spacetime metric $g_{\mu\nu}$ has a mostly plus signature $(-+++)$. Greek indices run from 0 to 3, Latin indices from 1 to 3. The space-time covariant derivative is denoted as $\nabla_\mu$. We use natural units: $c = \hbar = k_B = 1$.

\subsection*{Organization of the paper}

This paper is organized as follows. 
In Section \ref{S:Background_discussion} we provide an overview of our results and the context surrounding them. In Section \ref{sec:theory},
we formulate a generalization of Navier-Stokes theory using the Bemfica-Disconzi-Noronha-Kovtun (BDNK) formalism \cite{Bemfica:2017wps,Bemfica:2019knx,Kovtun:2019hdm,Hoult:2020eho}. In Section \ref{sec:causality} we provide necessary and sufficient conditions that must be fulfilled by the parameters of the theory for causality to hold. In Section \ref{sec:hyperbolicity} we prove that the full nonlinear system of equations in general relativity is strongly hyperbolic, the solutions are unique, and the initial-value problem is well-posed in general relativity. A new theorem concerning the linear stability properties of relativistic fluids in flat spacetime is given in \ref{sec:linear_stability_theorem}. We employ this theorem in Section \ref{sec:stability} to obtain conditions that ensure that the new theory presented here is  stable. The rigorous mathematical proofs of Theorem I, Proposition I, Theorem II, and Theorem III are found in Appendix \ref{Theorem_I}, \ref{Proposition_I}, \ref{Theorem_II}, and \ref{Theorem_III}, respectively. Our conclusions and outlook can be found in \ref{sec:conclusions}. 

\section{Background and Discussion}\label{S:Background_discussion}

\subsection{Definition of out-of-equilibrium variables: hydrodynamic frames}
\label{S:Frames}

In the modern perspective,
relativistic fluid dynamics is understood as an effective theory for the 
evolution of
conserved densities, such as the
energy-momentum tensor $T^{\mu\nu}$. (We could include, in this introductory part, other conserved quantities
such as the baryon current $J^\mu$ and those
associated with higher moments. In fact, conservation of $J^\mu$ will be implicitly understood later in the discussion of Sections \ref{S:Frames}--\ref{S:Overview} and thereafter since we will often refer to the presence of a chemical potential. For simplicity, however, we
will often refer only to $T^{\mu\nu}$ in this part, since this 
will suffice for the aspects we want to highlight.) To say that $T^{\mu\nu}$ is conserved means
that 
\begin{equation*}
\nabla_\mu T^{\mu\nu} = 0,
\end{equation*}
which provides equations of motion 
governing the dynamics of the fluid.

The energy-momentum tensor $T^{\mu\nu}$ is understood
as the expectation value of the microscopic quantum
operator $\hat{T}^{\mu \nu}$, which is an
observable that can be defined for any 
non-equilibrium state. In equilibrium,
the state of the system can be parametrized
by the temperature $T_{eq}$, the flow 
velocity $u^\mu_{eq}$ (observe that this is the four-velocity of the fluid, although we will often refer to it simply as the velocity; the fluid velocity is always assumed to be normalized, see Section \ref{S:Overview}), and the chemical potential $\mu_{eq}$.
One of the assumptions that
forms the basis of a fluid dynamics description is that for states not very far from 
equilibrium, the physical observable 
$T^{\mu\nu} = \langle \hat{T}^{\mu\nu} \rangle$ can still be parametrized
in terms of a ``temperature'' $T$, a ``flow
velocity'' $u^\mu$, and a ``chemical potential''
$\mu$ that \emph{reduce to
$T_{eq}$, 
$u^\mu_{eq}$, and $\mu_{eq}$ in equilibrium}.
We write quotation marks to emphasize the fact that 
the
quantities $T$, $u^\mu$, and $\mu$
\emph{have no first-principles microscopic
definitions.} Therefore, 
while 
it is useful to interpret 
$T$, $u^\mu$, and $\mu$ as out-of-equilibrium
macroscopic temperature, velocity, and chemical potential, since they are close 
to $T_{eq}$, 
$u^\mu_{eq}$, and $\mu_{eq}$ and reduce
to the latter in equilibrium, we should
\emph{ultimately understand 
$T$, $u^\mu$, and $\mu$ as auxiliary variables
that are used to parametrize the 
physical observable $T^{\mu\nu}$
-- the latter enjoying a first-principles, microscopic definition even when the system is out of equilibrium.} 

It follows that
there exists an ambiguity in the definition
of the out-of-equilibrium quantities 
$T$, $u^\mu$, and $\mu$, since
there are different ways of
parametrizing $T^{\mu\nu}$ subject
to the constraint that one recovers the 
unambiguous parametrization in terms of 
$T_{eq}$, 
$u^\mu_{eq}$, and $\mu_{eq}$ in equilibrium. 
In other words, different \emph{out-of-equilibrium} choices of 
$T$, $u^\mu$, and $\mu$ to parametrize 
$T^{\mu\nu}$ are allowed as long as they agree in equilibrium. This is sometimes expressed by
saying that $T$, $u^\mu$, and $\mu$ correspond
to a ``fictitious'' temperature, flow velocity,
and chemical potential \cite{RezzollaZanottiBookRelHydro,MIS-6}.

A particular choice of parametrization of 
$T^{\mu\nu}$ in terms of $T$, $u^\mu$, and $\mu$
has been historically called a choice of a \emph{hydrodynamic frame,} or simply frame.
(It is a bit unfortunate the word ``frame'' has also
other meanings in relativity theory, e.g., reference frames related
by a Lorentz transformation, frames in a tetrad formalism, null-frames, or a local rest frame, etc. However, all these different meanings can be distinguished from the context.) \emph{A choice of frame is, therefore, 
a definition of what one means by temperature,
velocity, and chemical potential out of equilibrium.}
Consequently, \emph{a choice of frame is always involved
whenever we describe a fluid out of equilibrium
in terms of temperature, velocity, and chemical potential.} This is still the case even if further,
extended variables are introduced (see Section \ref{S:Overview} for
the notion of extended variables). The notion of hydrodynamic frame
and how it represents a choice of out-of-equilibrium variables
is discussed extensively in the literature. An incomplete list is given
by the references \cite{MIS-2, MIS-6,degroot,Cercignani,Rischke:1998fq,
Tsumura:2006hn,Tsumura:2007wu,Tsumura:2009vm,Bhattacharyya:2007vjd,Bhattacharya:2011tra,Tsumura:2011cj,Van:2011yn,Kovtun:2019hdm,Kovtun:2012rj,Tsumura:2012kp,Kikuchi:2015swa,Gavassino:2020ubn}. References \cite{Kovtun:2019hdm,Kovtun:2012rj}, 
in particular, contain a detailed discussion 
of the topic.

Observe that once $T^{\mu\nu}$ is cast in terms
of $T$, $u^\mu$, and $\mu$, the energy-momentum
conservation equations $\nabla_\mu T^{\mu\nu} = 0$ can be equivalently written as evolution equations for
those quantities. We also remark that one can choose other thermodynamic quantities, e.g., the energy density or the pressure, to parametrize $T^{\mu\nu}$, and we will in fact do so later on in the paper. Of course, not all thermodynamic scalars are independent; they are connected by the first-law of thermodynamics and a prescription of an equation of state \cite{RezzollaZanottiBookRelHydro}.
Obviously, the non-uniqueness in the definition
of the variables used to parametrize $T^{\mu\nu}$
out of equilibrium remains if we choose a parametrization in terms of other thermodynamic variables such as the energy density, etc.

In order to pass from this qualitative
argument about the ambiguity of 
$T$, $u^\mu$, and $\mu$ away from equilibrium
to a more precise assessment of such 
ambiguity, one needs to be more specific about how
one formalizes the idea that fluid dynamics arises
as a long time, long wavelength limit of an underlying microscopic
theory, i.e., as a description of
the macroscopic dynamics of the system
for small deviations from equilibrium. 
Such a formalization can be accomplished
in the framework of the so-called \emph{gradient 
expansion,} which was used a century ago by Chapman and Enskog in the derivation of fluid dynamics from the (non-relativistic) Boltzmann equation
and that has since then been adapted to the relativistic setting \cite{degroot}. We remark that the gradient expansion is not the
only way to formalize the idea that fluid dynamics is an effective description that emerges from a more fundamental microscopic behavior; see
Section \ref{S:Overview} for a discussion of ideas involving the
so-called moment expansion and holographic techniques.
Nevertheless, the gradient expansion, while not fundamental, is a very
convenient and powerful formalism based on effective field theory ideas that allows one to
track how different parametrizations of $T^{\mu\nu}$ lead
to different fluid descriptions.

The gradient expansion is based on the idea that one can write
\begin{align}
    T^{\mu\nu} = O(1) + O(\partial) +
    O(\partial^2) + \dots,
    \nonumber
\end{align}
where $O(\partial^n)$ denotes terms with 
$n$ derivatives of $T$, $u^\mu$, and $\mu$ (so,
e.g., $O(\partial^2)$ involves both terms of the
form $\partial^2 T$ and $\partial T \partial \mu$, etc.) and $O(1)$ corresponds to the terms that
reduce to $T^{\mu\nu}_{eq}$, the energy-momentum
tensor parametrized in terms of $T_{eq}$, $u^\mu_{eq}$, and $\mu_{eq}$. Schematically, this is an expansion in powers of the Knudsen number $\mathrm{Kn}\sim \ell_\mathrm{micro}\partial$, i.e., the ratio between the relevant microscopic scale $\ell_\mathrm{micro}$ and the inverse macroscopic scale $L$, associated with the derivative of the hydrodynamic fields. In this sense, the gradient expansion corresponds to the well-known Knudsen number expansion used in the description of kinetic systems \cite{Cercignani,degroot}. In particular, 
since the expansion truncated at $O(1)$ corresponds to ideal hydrodynamics, \emph{viscous contributions require considering at least $O(\partial)$ terms,} which is consistent with the basic intuition that dissipation is a phenomenon associated with deviations from equilibrium.

In order to construct a fluid theory out of the gradient
expansion, one truncates it at a certain order. This truncation necessarily defines a scale at which the effective description is supposed to be valid, with higher order
effects encoded by the terms neglected in the expansion
which are considered outside the limit of validity of
the truncated theory. Aside from the truncation order, one also 
needs to specify the \emph{constitutive relations,} i.e.,
the specific form of each term $O(\partial^n)$
in terms of $T,u^\mu, \mu$, up to the truncation order
(see Sections \ref{S:Overview} and \ref{sec:theory}
for examples). \emph{By specifying the truncation order and the constitutive relations,
one is in fact defining what is meant by $T$, $u^\mu$, and
$\mu$ out of equilibrium, i.e., one is making a choice of hydrodynamic frame.}

Different frame choices, therefore, correspond to different effective descriptions of the same truncated theory.
At this point, it seems almost unnecessary 
to talk about ``frames,'' and one might be
tempted to simply say that one has distinct theories of fluids. The key word
here, however, is \emph{effective.} Indeed,
when we consider two distinct constitutive
relations truncated at a given order, 
\begin{align*}
    T^{\mu\nu} = T^{\mu\nu}(T, u^\alpha, \mu)
    \, \text{ and } \,
    \tilde{T}^{\mu\nu} = \tilde{T}^{\mu\nu}(\tilde{T}, \tilde{u}^\alpha, \tilde{\mu}),
\end{align*}
one obviously have different fluid theories:
the equations of motion $\nabla_\mu T^{\mu\nu}=0$ and $\nabla_\mu \tilde{T}^{\mu\nu}=0$ are not the same.
Consequently (upon writing these conservation laws in terms of $T, u^\alpha, \mu$ and
$\tilde{T}, \tilde{u}^\alpha, \tilde{\mu}$, respectively), the quantities 
 $T, u^\alpha, \mu$ and
$\tilde{T}, \tilde{u}^\alpha, \tilde{\mu}$, satisfy different evolution equations and,
thus, cannot represent the same definition
of temperature, fluid velocity, and chemical potential. However, one needs to keep in mind
that the temperature, flow velocity, and chemical potential are not fundamental quantities, whereas the energy-momentum
tensor is (it does have a first-principle definition). Thus, 
$T^{\mu\nu}(T, u^\alpha, \mu)$ and 
$\tilde{T}^{\mu\nu}(\tilde{T}, \tilde{u}^\alpha, \tilde{\mu})$ differ because
they represent \emph{distinct} coarse-grained 
or low-energy limits of the actual, microscopically
uniquely defined, energy-momentum tensor. Therefore,
the language of frames signals the key fact that one is always considering one possible effective description
among many.

Summarizing, \emph{there exists an intrinsic ambiguity in
how one parametrizes $T^{\mu\nu}$ in terms
of out-of-equilibrium temperature $T$, velocity $u^\mu$, and chemical potential $\mu$.} 
Such ambiguity
simply expresses the fact that \emph{these quantities do not have first-principle microscopic definitions
away from equilibrium.} What is not ambiguous
away from equilibrium is the definition of $T^{\mu\nu}$.
One resolves this ambiguity by choosing a definition
of $T, u^\mu, \mu$. Such a choice is known as a frame
choice. Different parametrizations of $T^{\mu\nu}$, therefore, correspond to different frame choices. Not all frame choices, however, are equally useful.
In our work, we explore suitable 
definitions of temperature, flow velocity, and
chemical potential to construct effective theories describing fluids that lead to sensible theories
in terms of satisfying properties \ref{I:Causality}-\ref{I:Strong_hyperbolicity}.

At this point, the attentive reader will probably have noticed that much of the above discussion
does not depend on relativistic principles. In other words, the fact that there is no first-principles definition of out-of-equilibrium quantities such as temperature, flow velocity, and chemical potential, applies to non-relativistic theories as well. 
In the non-relativistic setting, however, there exists a highly successful theory of dissipative (Newtonian)
fluids, namely, the Navier-Stokes-Fourier theory. In light of its success, it is fair to say that for all practical purposes, one can take the definitions
of out-of-equilibrium quantities in the Navier-Stokes-Fourier theory as the correct ones in a non-relativistic context. Had an equivalently successful theory of relativistic viscous fluids been available (where success would in particular incorporate properties \ref{I:Causality}--\ref{I:Strong_hyperbolicity}),
we could similarly take the definitions of out-of-equilibrium quantities in such a theory as the correct ones for all practical purposes. Nevertheless, as we will explain in the next section, there is not, at the moment, a theory of relativistic viscous fluids that can claim such a level of success. Hence, exploring how different frame choices can lead to different fluid descriptions becomes a topic of uttermost interest (see Section \ref{S:Summary}).

\subsection{A brief overview of viscous theories\label{S:Overview}}

The first proposal for a relativistic viscous
fluid theory was done by Eckart \cite{EckartViscous} in 1940, with a closely related
formulation by  Landau and Lifshitz \cite{LandauLifshitzFluids} in the '50s. In these
works, the authors postulated a form for the 
energy-momentum tensor (and also of the baryon current $J^\mu$, but, as in the previous Section, here we simplify the discussion by focusing on $T^{\mu\nu}$ only) based on ideas from thermodynamics
and following a covariant generalization of the non-relativistic Navier-Stokes-Fourier theory.
For example, in Eckart's theory, one has
\begin{align*}
    T^{\mu\nu}_{Eckart} =
    \varepsilon u^\mu u^\nu + (P
    - \zeta \nabla_\lambda u^\lambda )\Delta^{\mu\nu} 
    + q^\mu u^\nu + q^\nu u^\mu - 2\eta \sigma^{\mu\nu},
\end{align*}
where, $\varepsilon$, $T$, and $u^\mu$ are the (out-of-equilibrium) energy density  \footnote{Note that here the energy density of the system is defined to match that of an equilibrium state. This energy density is then related to the temperature by the (relativistic version of the) first law of thermodynamics.}, temperature, and velocity of the fluid, with the latter normalized  \footnote{The fluid velocity is normalized because in relativity observers are characterized by their world-line; different parametrizations of the same world-line can produce different tangent vectors. We fix this ambiguity by normalizing the tangent vectors to unity. The normalization $u^\mu u_\mu = -1$ is assumed in every fluid theory that we discuss.} by $u^\mu u_\mu = -1$, 
$\Delta^{\mu\nu} = g^{\mu\nu} + u^\mu u^\nu$ is the projection onto the space orthogonal to $u^\mu$,
$P$ is the equilibrium pressure (see below) given by an equation of state (the choice of which depends on the nature of the fluid, for example, for a conformal fluid one has $P=\frac{1}{3} \varepsilon$), $\zeta$ is the coefficient of bulk viscosity, $\eta$ is the coefficient of shear viscosity, 
$q_\mu = -\kappa T ( \Delta_\mu^\nu \nabla_\nu \ln T + u^\nu \nabla_\nu u_\mu )$
represents energy diffusion, with $\kappa$
being the coefficient of heat conduction, and
$\sigma^{\mu\nu} = \Delta^{\mu\nu\alpha\beta}\nabla_\alpha u_\beta$ is the shear tensor, with $\Delta^{\mu\nu}_{\alpha\beta} = \frac{1}{2} \left(\Delta^\mu_\alpha \Delta^\nu_\beta + \Delta^\mu_\beta \Delta^\nu_\alpha-\frac{2}{3}\Delta^{\mu\nu}\Delta_{\alpha\beta}\right)$ (so $\Delta^{\mu\nu}_{\alpha\beta}$ 
projects a two-tensor on the space  of two-tensors traceless and orthogonal to $u^\mu$).
In the absence of viscous effects,
when $\zeta = \eta = \kappa = 0$,
one recovers the energy-momentum tensor
of an ideal fluid.

According to the standard physical interpretation of the energy-momentum tensor of a fluid, the fluid's total pressure is given by $\frac{1}{3} \Delta_{\mu\nu} T^{\mu\nu}$. It is convenient
to write the total pressure as a sum of an ``equilibrium'' part, which is assumed to be given by an equation of state whose functional form follows that assigned to the fluid in the limit when viscous effects are absent, and a ``non-equilibrium'' part that contains explicitly the viscous contributions. In the case of $T^{\mu\nu}_{Eckart}$, the latter is given by $-\zeta \nabla_\mu u^\mu$. This term clearly illustrates the fact that only terms of first order in Knudsen number where kept in this case because $\zeta/P$ gives the relevant microscopic length scale associated with particle-number changing processes, while  $\nabla_\mu u^\mu$ accounts for the inverse length scale associated with the gradient of the hydrodynamic fields.

As said, Eckart and Landau and Lifshitz were
seeking a covariant version of the non-relativistic Navier-Stokes equation
compatible with thermodynamic principles, most 
notably, the second law of thermodynamics, i.e., their choice of $T^{\mu\nu}$ ensured
that entropy production (for a suitable definition of out-of-equilibrium entropy) is non-negative.
From a modern perspective, however, these theories
are better understood as effective theories that
arise from a gradient expansion truncated at first
order and with a specific choice of hydrodynamic
frame, i.e., a specific choice of constitutive
relation that parametrizes the energy-momentum tensor
in terms of out-of-equilibrium variables. In fact, it is possible to show that 
the Eckart and Landau and Lifshitz theories
can be obtained from kinetic theory
as an expansion in gradients truncated at first order
\cite{degroot}. Constraints on the coefficients that appear in such truncated series are found by imposing
the second law of thermodynamics.
In accordance with the notion of hydrodynamic frames,
the specific choices that lead to Eckart's and Landau
and Lifshitz's theories are known in the literature
as the Eckart and Landau and Lifshitz frames
\cite{RezzollaZanottiBookRelHydro}.
One can immediately see that other frame choices are
possible for an energy-momentum tensor truncated at
first order upon noticing that $T^{\mu\nu}_{Eckart}$ does
not contain all possible terms that are linear
in derivatives of $T$, $u^\mu$, and $\mu$ -- terms that
are allowed in a truncation at first order.
Theories arising from a gradient expansion truncated at first order are known as \emph{first-order theories.} 
The Eckart and Landau theories are, thus, examples of first-order theories.

The Eckart and
Landau and Lifshitz theories are very
intuitive and natural at first sight. They correspond
to immediate covariant generalizations of the non-relativistic Navier-Stokes-Fourier theory (in fact, they recover it in the non-relativistic limit), satisfy the second law of thermodynamics, preserve many features present in the ideal case (e.g., the energy density is recovered from the energy-momentum tensor by double contraction
with the velocity), are relatively simple, and, as already said, can be derived from kinetic theory.
Yet, they are remarkably at odds with fundamental physical principles in that they are known to violate causality and are unstable \cite{Hiscock_Lindblom_instability_1985,PichonViscous}.
Consequently,
the Eckart and Landau and Lifshitz theories cannot be taken
as viable theories of relativistic viscous fluids.
In fact, a large class of first-order theories,
of which Eckart's and Landau and Lifshitz's are particular cases, are known to be acausal and unstable \cite{Hiscock_Lindblom_instability_1985}.
One naturally wonders what are the root causes of the failures of these theories, especially when at first sight they look very intuitive. We return to this point
in Section \ref{S:Summary}.

A different approach for the construction of relativistic viscous fluid theories was taken by Israel and Stewart in a series of works
\cite{MIS-2, MIS-3, MIS-4, MIS-5, MIS-6}, adapting ideas
developed by M\"uller in the non-relativistic setting
\cite{MIS-1}. The resulting theory is referred to
Israel-Stewart or M\"uller-Israel-Stewart (MIS) theory, or sometimes simply Israel-Stewart theory. In the MIS theory,  the energy-momentum takes the form
\begin{align*}
    T^{\mu\nu}_{MIS} =
    \varepsilon u^\mu u^\nu + (P
    +\Pi )\Delta^{\mu\nu} 
    + \mathcal{Q}^\mu u^\nu + \mathcal{Q}^\nu u^\mu + \pi^{\mu\nu}.
\end{align*}
The quantities $\Pi$, $\pi^{\mu\nu}$, and $\mathcal{Q}^\mu$ represent
the bulk viscosity, shear viscosity, and energy diffusion of the fluid, and are referred as viscous fluxes. We see that $T^{\mu\nu}_{Eckart}$ corresponds to the choices where the bulk scalar
$\Pi = -\zeta \nabla_\mu u^\mu$, the shear-stress tensor is given by
$\pi^{\mu\nu}=-2\eta \sigma^{\mu\nu}$, and the energy diffusion reads
$\mathcal{Q}^\mu = q^\mu \equiv   -\kappa T ( \Delta_\mu^\nu \nabla_\nu \ln T + u^\nu \nabla_\nu u_\mu )$. In the MIS theory, however,
the viscous fluxes are taken to be 
\emph{new variables on the same par}
as the ``ordinary'' variables $T$, 
$u^\mu$, etc. (see below). Because $\Pi, \pi^{\mu\nu}, \mathcal{Q}^\mu$ add to the number of variables, hence extending
the state space, they are known as \emph{extended
(thermodynamic) variables} and theories that 
investigate extended variables are referred
to as \emph{extended (thermodynamic) theories}
\cite{JouetallBook, MuellerRuggeriBook}.
An important point to make (already alluded to earlier) is that \emph{one cannot dispense with a choice of hydrodynamic frame even in extended theories,} since one still needs to make a definition of out-of-equilibrium temperature, flow velocity, and chemical potential.

At this point, it is convenient to make the following definition. The variables $T$, $u^\mu$, $\mu$ and those derived from them via
the first law of thermodynamics and a choice of equation of state are known as \emph{hydrodynamic variables or fields.} In other words, the hydrodynamic variables are the ``ordinary'' fields already present in the case of an ideal fluid (although, the physical interpretation of these variables is not precisely the same as in the ideal fluid case; as discussed, the meaning of, e.g., temperature is different in or out of equilibrium). In this language, we can say that the Eckart and Landau and Lifshitz theories involve only the hydrodynamic variables, whereas the MIS theory involves both hydrodynamic and extended fields. In addition, the gradient expansion is always an expansion in the hydrodynamic variables  \footnote{More precisely, while in principle it makes sense to consider gradient expansions involving extended variables, all instances of the gradient expansion we consider here are expansions involving only the hydrodynamic fields. Thus, for our purposes we can say that a gradient expansion is always an expansion in the hydrodynamic variables.}.

Because the MIS formalism introduces new variables in addition to the hydrodynamic fields, it also requires new equations of motion besides the standard conservation laws such as $\nabla_\mu T^{\mu\nu}_{MIS}=0$. The desired equations are postulated to be relaxation-type
equations whose precise form is chosen so that entropy production is non-negative -- where the entropy current is also extended from its usual form used in ideal fluids to include the extended variables $\Pi, \pi^{\mu\nu}, \mathcal{Q}^\mu$. For example, $\Pi$ satisfies 
\begin{align*}
    \tau_\Pi u^\mu \nabla_\mu \Pi + \Pi 
    = - \zeta \nabla_\mu u^\mu 
    -\frac{1}{2} \zeta T \Pi \nabla_\mu
    \left(\frac{\tau_\Pi}{\zeta T} u^\mu \right),
\end{align*}
where $\tau_\Pi$ is a relaxation time.
See, e.g., \cite{RezzollaZanottiBookRelHydro},
for the full set of equations satisfied by
$\Pi, \pi^{\mu\nu}, \mathcal{Q}^\mu$, the form of the entropy current including these fields, and the
derivation of the equations of motion from the second law of thermodynamics  \footnote{We remark that a simplified version of the MIS equations where certain terms are neglected are known as the Maxwell-Cattaneo equations, see \cite{MaartensLecturesDissipative}.}.

The MIS theory enjoys the following good properties: the equations of motion are stable, thus satisfying property \ref{I:Stability}, and their linearization about equilibrium states is causal, thus satisfying
property \ref{I:Causality} 
\cite{Hiscock_Lindblom_stability_1983,Olson:1989ey}. Also, it can, in certain limits, be derived from kinetic theory \cite{MIS-4, MIS-5, RezzollaZanottiBookRelHydro}.

We next discuss three other theories of great interest
that employ extended variables:  DNMR, resumed BRSSS (rBRSSS), and a-hydro theories. The Denicol-Niemi-Molnar-Rischke (DNMR) theory is  an effective theory derived from kinetic theory via
an expansion in moments \cite{Denicol:2012cn}. The moment expansion goes back to Grad in his work 
on non-relativistic fluids \cite{Grad-1948, Grad-1949}. Applying this formalism to the relativistic Boltzmann equations, together with a new power-counting scheme involving Knudsen and inverse Reynolds number expansions, DNMR arrived at a set of equations for the hydrodynamic fields
and a set of extended variables $\Pi$, $\pi^{\mu\nu}$, and $\mathcal{Q}^\mu$ that represent 
the bulk viscosity, shear viscosity, and energy diffusion, similarly to the MIS equations. Also similar to the MIS equation is the fact that the equations satisfied by the viscous fluxes in the DNMR theory are relaxation-type equations. Despite their similarities, it is important to stress that the MIS and DNMR equations are not the same.

The DNMR theory enjoys many good properties. It is stable and its linearization about equilibrium states  \footnote{Although the DNMR and MIS equations are different, they coincide when linearized about equilibrium \cite{Brito:2020nou,Gavassino:2021kjm}.} is causal \cite{Pu:2009fj}. When only bulk viscosity is present, the DNMR theory is causal, locally well-posed, and strongly hyperbolic; these properties hold with and without dynamical coupling to Einstein's equations
\cite{BemficaDisconziNoronha_IS_bulk}. When all viscous fluxes are present,  but chemical potential is absent, the DNMR equations
have recently been shown to be causal (again, with or without coupling to Einstein's equations) 
\cite{Bemfica:2020xym} (see \cite{Denicol:2008ha, Pu:2009fj,Floerchinger:2017cii} for related results under symmetry assumptions). Hence, property \ref{I:Stability} holds in general for the DNMR equations; properties
\ref{I:Causality}, \ref{I:LWP}, and \ref{I:Strong_hyperbolicity} hold if shear viscosity and heat conduction are absent (with or without dynamical coupling to Einstein's equations); and property \ref{I:Causality} holds with all viscous fluxes present but in the absence of chemical potential \cite{F:Quasi-analytic} (with or without dynamical coupling to Einstein's equations).
Most importantly, the DNMR theory has been very successful in phenomenological studies of the quark-gluon plasma, particularly in numerical simulations of its dynamical behavior, e.g. \cite{Bernhard:2019bmu,Ryu:2015vwa}. 

We now move to discuss the resumed 
Baier-Romatschke-Son-Starinets-Stephanov
(rBRSSS) theory \cite{Baier:2007ix}. 
In order to do so, we need to start with the (plain, not resumed) BRSSS theory  \cite{Baier:2007ix}.
This is an effective theory obtained from the gradient expansion truncated at second order. As such, it involves only the hydrodynamic fields and the equations of motion were chosen in \cite{Baier:2007ix} to be defined in the Landau frame. This effective theory-based approach was originally developed for conformal fluids in \cite{Baier:2007ix} and the same equations of motion for a conformal system were concurrently derived in \cite{Bhattacharyya:2007vjd} through the fluid/gravity correspondence, a powerful technique introduced in that work which was motivated by the holographic duality of string theory \cite{Maldacena:1997re}. In order to address the issues with causality and stability, \cite{Baier:2007ix} proposed a MIS-like theory with transport coefficients that ensure its agreement with the gradient expansion at second order. In the context of \cite{Baier:2007ix}, this approach provides a resummation of higher order terms and the latter explains the differences found, for instance, between rBRSSS and DNMR. 
However, at the linearized level, this resummed BRSSS theory shares the same properties of DNMR. Furthermore, the techniques used in \cite{Bemfica:2020xym} can be adapted to establish causality for this theory in the nonlinear regime. The local well-posedness and hyperbolicity aspects of rBRSSS have not yet been established.

Because the MIS, DNMR, and rBRSSS theories share many properties, in particular the use of extended variables that satisfy similar relaxation-type equations, and their linearizations about equilibrium agree, they are sometimes collectively referred to 
as Israel-Stewart or M\"uller-Israel-Stewart theor\emph{ies,} 
Israel-Stewart-like or M\"uller-Israel-Stewart-like theories, or yet
generalized Israel-Stewart or M\"uller-Israel-Stewart theories.  They are sometimes also collectively referred to as second-order theories. While there is no harm
in grouping these theories together in this fashion, especially if one is concerned only with their general qualitative behavior, it is important to note that when it comes to specific features, including properties \ref{I:Causality}--\ref{I:Strong_hyperbolicity}, the exact form of the equations matters and, therefore, the differences among these theories become important.

The fourth extended theory we would like to briefly discuss is the anisotropic hydrodynamics theory (a-hydro)
 \cite{Florkowski:2010cf,Martinez:2010sc,Bazow:2013ifa,Almaalol:2018gjh,Alqahtani:2017mhy}. The latter is, in principle, more general than most approaches as it investigates the problem of small deviations around a given anisotropic non-equilibrium state. Formally, this approach involves a resummation in both Knudsen and inverse Reynolds numbers, which may be interpreted as a generalization of DNMR's power-counting ideas \cite{Strickland:2017kux}. The equations of motion, which are in practice derived using kinetic theory, can be approximated to give rise to a MIS-like theory. As such, causality and stability in the linearized regime follow from previous results. Nothing is known about causality in the nonlinear regime of this theory. The local well-posedness and hyperbolicity aspects of a-hydro have not yet been  established.

The above summary highlights how the use of extended variables has led to many successes in the study of relativistic viscous fluids. These accomplishments seem even more impressive when they are contrasted with the fact already mentioned that first-order theories (which do not employ extended variables) had been largely ruled out for decades due to instabilities and lack of causality \cite{Hiscock_Lindblom_instability_1985, PichonViscous}. Such successes nonetheless, it is important to keep in mind several actual or potential limitations of the extended theories discussed above, as we now discuss.

First of all, observe that none of the theories MIS, DNMR, rBRSSS, or a-hydro is known to satisfy all the properties \ref{I:Causality}-\ref{I:Strong_hyperbolicity}. To the extent that they satisfy some of these properties, this happens under \emph{restrictive} assumptions. Indeed, in the case of the quark-gluon plasma it is abundantly clear that one needs to 
consider situations when all viscous fluxes
are present and the chemical potential is non-zero \cite{Denicol:2018wdp} (and it is likely that this is also true in 
neutron star mergers \cite{Alford:2017rxf,Most:2021zvc}), in which case none of these theories  is known to be causal and locally well-posed. 
Moreover, while numerical simulations of the dynamics of the quark-gluon plasma based on the DNMR equations have been carried out for a long time \cite{Muronga:2001zk,Muronga:2003ta,Romatschke:2007mq}, only recently, with the aforementioned causality
results \cite{BemficaDisconziNoronha_IS_bulk, Bemfica:2020xym},
one can determine regions in the parameter and state spaces for which
causality holds or fails. When such constraints are taken into consideration, it is found that state-of-the-art numerical simulations
of the quark-gluon plasma violate causality \cite{Chiu:2021muk,Plumberg:2021bme}, especially at early times  \cite{Plumberg:2021bme}. Although
further research is required to find out 
the implications of such causality violations to our current understanding of those properties of the quark-gluon plasma that have been extracted from numerical simulations, such results should serve
as a definite cautionary tale about running numerical simulations of relativistic viscous fluids whose causality properties are poorly understood. Furthermore, if causality violations can be a real issue in numerical simulations of the quark-gluon plasma, which are carried out in flat spacetime, the situation is even more precarious in simulations of general relativistic viscous fluids, such as in neutron star mergers. While some simulations have been implemented in this setting \cite{Shibata:2017xht}, they rely on a formulation
for which the key properties \ref{I:Causality}, \ref{I:LWP}, and \ref{I:Strong_hyperbolicity}
are not known to hold.

Another potential limitation of the extended theories discussed above is that they do not seem appropriate for describing 
shock-waves \cite{HiscockShocksMIS,GerochLindblomCausal,Olson:1991pf}. 
This is a potentially important limitation given 
the preponderance of shock-waves in fluid dynamics,
which is aggravated by the recent discovery that
solutions to 
    MIS-like equations can become singular in finite time \cite{Disconzi:2020ijk}. Additionally, MIS-like and a-hydro theories are only expected to describe the transient regime of dilute gases as their derivation is most naturally understood within kinetic theory \cite{MIS-6,Denicol:2012cn}. Therefore, their use in other types of systems, such as in strongly coupled relativistic fluids, is a priori not justified. In fact, it is known that MIS-like equations do not generally describe the complex transient regime of holographic strongly coupled gauge theories \cite{Denicol:2011fa,Heller:2014wfa,Florkowski:2017olj} (see \cite{Grozdanov:2018fic} for the case of higher-derivative corrections). In this aspect, we anticipate that the causal and stable first-order theory developed here does not describe this transient regime either, despite satisfying properties \ref{I:Causality} through  \ref{I:Strong_hyperbolicity}. However, this is not an issue per se given that the description of such a far-from-equilibrium state is certainly beyond the regime of applicability of first-order hydrodynamics.  

Finally, MIS-like theories lack the degree of universality expected to hold in hydrodynamics as the equations of motion themselves change depending on the derivation. For instance, the equations of motion in \cite{Baier:2007ix} have different terms than in \cite{Denicol:2012cn}, which is explained by the different power-counting scheme employed in those works. This situation should be contrasted 
with theories derived from the gradient expansion: although, of course, a plethora of different effective theories can be derived in the gradient expansion formalism, these different theories can always be viewed as particular  
cases, obtained via different frame choices, of the most general expansion truncated at a certain order. In fact, 
an approach of this type is employed in this manuscript, see Section \ref{S:Summary}.

Summarizing, despite its undeniable success in advancing our understanding of relativistic viscous fluids in general, and of the quark-gluon plasma in particular, MIS-like and a-hydro theories still face many challenges, especially when it comes to settings where general relativity is involved. Thus, it is extremely important to also consider alternative theories of relativistic
viscous fluids. This is especially the case when pursuing the study of viscous effects in
neutron star mergers  \cite{Alford:2017rxf,Alford:2019qtm,Alford:2020pld,Most:2021zvc} and, as already mentioned, it is far from clear that the MIS-like and a-hydro approach are the correct approaches for this setting.

In view of the above, it is not surprising that researchers have explored other theories of relativistic viscous fluids than those discussed so far. A natural place to start such an investigation is the gradient expansion, and the simplest  possibility that includes viscous effects is that of first-order theories, i.e., effective fluid descriptions arising as a truncation of the gradient expansion at first order. On the other hand, since, as said, large classes of first-order theories
are acausal and unstable, one might naturally wonder whether such an approach would be doomed to fail. In order to answer this, it is important to understand the assumptions involved. While it is true that the acausality and instability results
\cite{Hiscock_Lindblom_instability_1985, Olson:1989ey} cover large classes of first-order theories, 
\emph{these results apply only to theories that satisfy}
\begin{align}
    u_\mu u_\nu T^{\mu\nu} = \varepsilon,
    \tag{$\star$}
    \label{E:Energy_density_LRF}
\end{align}
i.e., \emph{only to frame choices that preserve the relation
\eqref{E:Energy_density_LRF}.} In other words, the latter means that an observer moving with the fluid always sees the energy density as if it were in equilibrium, even for states where entropy is produced. Therefore, the construction of stable and causal first-order theories remains a distinct possibility as long as one avoids constitutive relations that imply
\eqref{E:Energy_density_LRF}. First-order theories for which \eqref{E:Energy_density_LRF} holds are often collectively referred to as the (relativistic) Navier-Stokes (NS) theory \cite{Denicol:2012cn}, although there is no universal agreement on the terminology \cite{Hoult:2020eho}. 

The physical meaning of \eqref{E:Energy_density_LRF}, as well as of not satisfying it, will be discussed in Section \ref{S:Summary}. We also remark that the assumptions in \cite{Hiscock_Lindblom_instability_1985, Olson:1989ey} imply other special relations than
\eqref{E:Energy_density_LRF}. But here, for simplicity, we focus only on \eqref{E:Energy_density_LRF}, since our goal is not to have a detailed discussion of the assumptions involved in those works but rather 
to illustrate how their conclusions apply only for a particular class of theories that employ very specific frame choices and, therefore, say
nothing about first-order theories that employ other hydrodynamic frames.
In other words, here the reader can take \eqref{E:Energy_density_LRF} as a placeholder for the class of frames that are assumed in the instability and acausality results \cite{Hiscock_Lindblom_instability_1985, Olson:1989ey}. Such a class of frames if far from exhaustive. Consequently, the results in 
\cite{Hiscock_Lindblom_instability_1985, Olson:1989ey} simply do not apply is different constitutive relations are used.

This motivated researchers to construct stable and causal first-order theories of viscous fluids. Important attempts in this direction go back to the first decade of this century \cite{Van:2007pw,Tsumura:2007wu,Van:2011yn, ChoquetBruhatGRBook}. The first more formal indication that causal and stable first-order theories could be constructed if the frame choice \eqref{E:Energy_density_LRF} is avoided is given in the works \cite{TempleViscous, TempleViscous2, TempleViscous3}. These works
were also the first ones to carry out a systematic study of viscous shocks in relativistic theories, a topic that in fact seems to be one of the main goals in these references.

The first construction of a stable and causal first-order theory of viscous fluids
was carried out by the authors in \cite{Bemfica:2017wps} for the case 
of conformal fluids (see also \cite{DisconziExistenceCausalityConformal} for some of the mathematical details of \cite{Bemfica:2017wps}). These results hold
with or without dynamical coupling to Einstein's equations.
Although \cite{Bemfica:2017wps} was restricted to conformal fluids, \emph{it provided an unequivocal proof that first-order stable and causal theories are possible, provided that one avoids the frame choice \eqref{E:Energy_density_LRF}.} Soon thereafter, causal and stable first-order theories were obtained by Kovtun \cite{Kovtun:2019hdm} and by the authors 
\cite{Bemfica:2019knx} for the case of non-conformal fluids
without a chemical potential  \footnote{The results
of \cite{Bemfica:2019knx} also allow for dynamical coupling to Einstein's equations.} -- although stability was obtained only
with help of a numerical investigation, so it might be more precise to say that stability was only strongly suggested and not established.
The resulting first-order theory became known in the literature as the BDNK theory
\cite{Hoult:2020eho}. Its local well-posedness
and strong hyperbolicity
was established in \cite{DisconziBemficaGraber, DisconziBemficaRodriguezShaoSobolevConformal}.
The stability and causality of the BDNK theory
in the presence of a chemical potential was obtained in \cite{Hoult:2020eho} (again, stability in this case was inferred only numerically). We also mention the closely related results 
\cite{Freistuhler-2020, Taghinavaz:2020axp}.
Of course, all these results are obtained using frame choices different than \eqref{E:Energy_density_LRF}.
Perhaps not surprisingly, after these results, the community took a renewed interest in first-order theories. See, e.g., the works
\cite{Das:2020gtq, Grozdanov:2018fic,Romenski:2019qzs,Gavassino:2020ubn,Dore:2020jye,Andersson:2019ezz,Poovuttikul:2019ckt,Das:2020fnr,Shokri:2020cxa,Erschfeld:2020blf,Celora:2020pzs} and references therein. 
We remark that choices of frames other than 
\eqref{E:Energy_density_LRF} have been studied before 
BDNK in \cite{Stewart:1972hg,Van:2007pw,Tsumura:2007wu,TempleViscous}, but, as said, the first construction
of a stable and causal first-order theory was done in \cite{Bemfica:2017wps} in the case of a conformal fluid. We will return to the BDNK theory in Section \ref{S:Summary}. In what follows, we will continue with our brief review of viscous theories.

Another first-order theory of interest is the Lichnerowicz theory \cite{Lichnerowicz_GR_book}, introduced in the '50s
but not investigated in detail until recently (see references that follow). The Lichnerowicz theory has been shown to be causal in the (very special) case of irrotational fluids \cite{DisconziViscousFluidsNonlinearity} by the second author of this paper (see also \cite{DisconziCzubakNonzero}). While irrotationality is too strong of a constraint to be useful for most physical applications, the work \cite{DisconziViscousFluidsNonlinearity} is of interest because it initiated the techniques that have since then been employed to study the causality of the BDNK theory, including the techniques employed in this work. We should also mention that the Lichnerowicz theory has found some interesting applications in the study of dissipative cosmological models  \cite{Disconzi_Kephart_Scherrer_2015,DisconziKephartScherrerNew,Acquaviva:2018rqi,Montani:2016hmf}.

Another formalism of importance in the study of viscous theories is that of divergence-type (DT) theories \cite{GerochLindblomDivergenceType}. In this approach, all the conserved quantities describing the dynamics of the fluid are obtained from a single generating function $\chi$ which 
is a function of a dynamical set of variables
$\zeta_A = (\zeta, \zeta_\mu, \zeta_{\mu\nu})$ (with $\zeta_{\mu\nu}$ trace-free and symmetric) representing the degrees of freedom of the fluid.
For example, in the DT approach the energy-momentum tensor is obtained as
\begin{align*}
    T_{DT}^{\mu\nu} &= \frac{\partial \chi}{\partial \zeta_\mu \partial \zeta_\nu}.
\end{align*}
DT theories provide a far-reaching subject with many important contributions to the physics
of fluids, kinetic theory, and out-of-equilibrium phenomena. Here, we limit ourselves to discuss 
DT theories with respect to properties \ref{I:Causality}--\ref{I:Strong_hyperbolicity}.
See \cite{GerochLindblomDivergenceType, GerochLindblomCausal, Geroch-RelativisticDissipative, RezzollaZanottiBookRelHydro, MuellerRuggeriBook, Kreiss_et_al, Nagy_et_all-Hyperbolic_parabolic_limit} for further discussion of DT theories and 
\cite{RamosCalzettaDT1, RamosCalzettaDT2, Lehner:2017yes} for applications of DT theories to the quark-gluon plasma.

All information of DT theories is contained in the generating function $\chi$. Unfortunately, there is no prescription on how to construct $\chi$, even more on how to construct a generating function that leads to a theory satisfying \ref{I:Causality}--\ref{I:Strong_hyperbolicity}. In fact, we think it would be more accurate to consider the DT approach as a general formalism instead of a precisely defined theory or set of theories.
That is because radically different theories, such as Eckart's and certain types of extended theories, can be cast in divergence-type by the choice of a suitable generating function \cite{GerochLindblomDivergenceType}.

Properties \ref{I:Causality}--\ref{I:Strong_hyperbolicity} have been investigated in the context of DT theories in \cite{GerochLindblomDivergenceType}. The authors constructed a DT theory that satisfies \ref{I:Causality}--\ref{I:Strong_hyperbolicity}
for states in equilibrium, i.e., when
$\zeta_A = \left. \zeta_A \right|_{eq}$. Next, they argued that, by continuity, these properties will also hold for $\zeta_A$ sufficiently close to 
$\left. \zeta_A\right|_{eq}$. However, 
no estimate is obtained for how close 
to $\left. \zeta_A\right|_{eq}$ the state $\zeta_A$ needs to be. Thus, given \emph{any} non-equilibrium state $\zeta_A$, this continuity result \emph{does not provide any information on whether this specific system satisfies the desired properties \ref{I:Causality}--\ref{I:Strong_hyperbolicity}.}
In particular,
without a quantitative estimate on how small 
$\zeta_A - \left. \zeta_A\right|_{eq}$ needs to be,
one does not know whether the states $\zeta_A$ for which properties \ref{I:Causality}--\ref{I:Strong_hyperbolicity}  hold include states of physical interest.
It could in principle happen that this continuity argument only guarantees the desired properties in a neighborhood of $\left. \zeta_A\right|_{eq}$ that is  orders of magnitude smaller than the size of any deviation from equilibrium that one typically considers in viscous fluid dynamics.

Another way of saying this is that the results in \cite{GerochLindblomDivergenceType} are purely qualitative, not providing a quantitative assessment of their applicability to physical systems. This should be contrasted with the precise quantitative results we establish here (see Sections \ref{sec:causality}--\ref{sec:linear_stability_theorem}) and in the predecessor works \cite{Bemfica:2020xym,Bemfica:2017wps,Bemfica:2019knx}, which are obtained by employing substantially more refined techniques than a general continuity argument. 
In \cite{Lehner:2017yes,RamosCalzettaDT1, GerochLindblomCausal,Kreiss_et_al, Nagy_et_all-Hyperbolic_parabolic_limit}, further results have been obtained, but they are all of the same qualitative nature as above, relying on precisely the same continuity argument. Thus, 
we believe that a fair assessment of DT theories 
is that they can in principle 
accommodate properties \ref{I:Causality}--\ref{I:Strong_hyperbolicity},
but precise conditions ensuring that such properties hold -- in particular conditions
that allow application to concrete physical problems -- are yet unknown.

We finally briefly mention
recent formulations of viscous fluids \cite{Andersson:2013jga,Gavassino:2020kwo} inspired by Carter's formalism and the variational principle \cite{Romenski:2019qzs}. Such formulations address some of the properties \ref{I:Causality}--\ref{I:Strong_hyperbolicity} but do not establish them in completeness. 

Although the review here provided is not exhaustive, we believe that it suffices to get across the following main point, namely, \emph{despite intense work on the subject and many different proposals made in the last 80 years,
one still does not have a theory of relativistic viscous fluids that incorporates all relevant viscous fluxes and chemical potential while satisfying all the properties \ref{I:Causality}--\ref{I:Strong_hyperbolicity}.}
Constructing such a theory is the goal of the present paper.

\subsection{Summary and discussion of our results\label{S:Summary}}
In this paper we consider the BDNK theory
with chemical potential and all relevant viscous fluxes, namely, bulk viscosity, shear viscosity, and heat conduction, and show that it satisfies all the properties \ref{I:Causality}--\ref{I:Strong_hyperbolicity}, i.e., causality, stability, local well-posedness, and strong hyperbolicity. Our results hold 
in the full nonlinear regime for the fluid equations in a fixed background or dynamically coupled to Einstein's equations. We work in $3+1$ dimensions and do not make any symmetry or simplifying assumptions. As explained in the previous section, this is the first time that a theory of relativistic viscous fluids with all these properties is constructed. In addition, all our results are mathematically rigorous and 
we provide a set of precise inequalities among scalar quantities (e.g., shear and bulk viscosity) that determine the regions in parameter and state space for which properties \ref{I:Causality}--\ref{I:Strong_hyperbolicity} hold. Such inequalities are useful for numerical simulations as they allow us to check, at each time step, whether conditions for causality and stability are fulfilled.

The key conceptual ingredient that allows us to establish our 
results is the realization that the causality and stability 
properties of a theory are intrinsically tied to its 
hydrodynamic frame. This happens because different choices affect the properties of the corresponding PDEs that describe the evolution of the fluid. In particular, we avoid the frame choice 
\eqref{E:Energy_density_LRF}, which in first-order theories 
leads to acausality and instability. The frame choice \eqref{E:Energy_density_LRF} has a natural intuitive appeal, namely, it states that the energy density measured by an observer moving with the fluid (i.e., in the fluid's local rest frame), $u_\mu u_\nu T^{\mu\nu}$, can be parametrized by a single scalar that can be identified with the energy density of the fluid in equilibrium (notice that \eqref{E:Energy_density_LRF} holds for an ideal fluid). It is not surprising, therefore, that Eckart and Landau and Lifshitz adopted frames satisfying \eqref{E:Energy_density_LRF}. On the other hand, such a simplicity in the definition of the hydrodynamic fields out of equilibrium, while desirable, is by no means a fundamental property. 
\emph{The key idea underlying the BDNK theory is that one should let the fundamental principle of causality (and also of stability and local well-posedness) dictate which frame choices (i.e., parametrizations of $T^{\mu\nu}$) are allowed,} rather than choose a frame based on non-fundamental principles and only then investigate properties such as causality.
In passing, we note that the MIS-like theories discussed in this section also adopt \eqref{E:Energy_density_LRF}, although, as just said, other frame choices can be made. Different frames have been recently investigated in the context of extended theories in \cite{Noronha:2021syv, Rocha:2021lze}.

The idea of exploring different frame choices to construct a first-order theory that satisfies properties \ref{I:Causality}--\ref{I:Strong_hyperbolicity} is not entirely new to this work. It was, in fact, the key idea employed in the earlier versions of the BDNK theory that have been showed to satisfy those properties in some particular cases (see Section \ref{S:Overview}). We will next explain what the new aspects of this work are, but in order to do so, we need to first review some other key ideas employed in the earlier constructions of the BDNK theory.

Since we do not want to make premature frame choices, our first step is to consider the most general frame, i.e., we write down the most general expression for $T^{\mu\nu}$ (and also $J^\mu$ in the case of the present work since we here consider non-zero chemical potential) compatible with the gradient expansion truncated at first 
order; see \eqref{generalTmunu} and 
\eqref{generaldefKovtun} for the precise expression. 
By considering the most general constitutive relations compatible 
with the symmetries of the problem as our starting point, we are in fact applying the basic tenets behind the construction of effective theories \cite{Polchinski:1992ed,Manohar:1996cq,Bedaque:2002mn,Kovtun:2012rj} to formulate hydrodynamics as a classical effective theory that describes the near equilibrium, long time/long wavelength behavior of many-body systems in terms of the same variables $\{T,\mu,u^\nu\}$ already present in equilibrium. For completeness, we remind the reader that an effective theory is constructed to capture the most general dynamics among low-energy degrees of freedom that is consistent with the assumed symmetries. When this procedure is done using an action principle, the action must include all possible fields consistent with the underlying symmetries up to a given operator dimension and the coefficients of this expansion  can then be computed from the underlying microscopic theory. These coefficients are ultimately constrained by general physical principles such as unitarity, CPT invariance, and vacuum stability. Analogously, in an effective theory formulation of relativistic viscous hydrodynamics, the equations of motion must take into account all the possible terms in the constitutive relations up to a given order in derivatives that describe deviations from equilibrium. The coefficients that appear in this expansion can then be computed from the underlying microscopic theory (using, for instance, linear response theory \cite{Kovtun:2012rj}), being ultimately constrained by general physical principles such as causality in the case of relativistic fluids \cite{HawkingEllisBook} and also by the fact that the equilibrium state must be stable, i.e. small disturbances from  equilibrium in an interacting (unitary) many-body system should decrease with time \cite{forster1995hydrodynamic}.  

Observe that by considering the most general energy-momentum tensor at first order, we are allowing viscous corrections to the equilibrium energy density, i.e., one has 
\begin{align*}
    u_\mu u_\nu T^{\mu\nu} = \varepsilon + \partial (T,\mu).
\end{align*}
(See \eqref{generaldefKovtun} for the precise expression.) Even though this is in sharp
contrast with \eqref{E:Energy_density_LRF}, in hindsight it seems the natural thing to do. After all,
it is standard to do precisely the same with the pressure, i.e., to split $\frac{1}{3} \Delta_{\mu\nu} T^{\mu\nu}$ into an ``equilibrium'' part and a ``viscous part'' (see Section \ref{S:Overview})  \footnote{Note that in this paper we have $u_\mu u_\nu T^{\mu\nu}=\varepsilon+\mathcal{A}$, where $\mathcal{A}$ is given in \eqref{10c}. It is worth mentioning that this choice gives $\mathcal{A}=\mathcal{O}(\partial^2)$ on-shell, as can be seeing from Eqs.\ \eqref{EOMcurrent}, which means that the on-shell out of equilibrium energy $u_\mu u_\nu T^{\mu\nu}=\varepsilon+\mathcal{O}(\partial^2)$ is a positive defined quantity up to the order of validity of the theory. This is in accordance with the weak energy condition.}.
There is no reason not to follow a similar recipe for the energy density seen by a co-moving observer.

We next investigate how causality constrains the constitutive relations. 
The idea that one should let causality determine which frames are allowed in a theory, while conceptually powerful, does not tell us how to in practice find the appropriate frames. Causality
of a theory can be determined by computing its characteristics \cite{Courant_and_Hilbert_book_2}.
Roughly, the characteristics are hypersurfaces in spacetime that correspond to the propagation modes of a theory. For example, in the case of Einstein's equations, the characteristics are simply the light-cones $g_{\mu\nu} v^\mu v^\nu = 0$. While in principle we can always compute the characteristics of a system of PDEs, in practice a brute-force calculation of the characteristics seems unattainable for a nonlinear system
of PDEs as complex as the BDNK system. In order to be able to compute the characteristics, \emph{we take a cue from the system's underlying geometric properties.} Inspired
by structures found in the case of ideal fluids by the second author and Speck in \cite{DisconziSpeckRelEulerNull}, which need to be recovered
in the ideal limit, we look for acoustical-metric-like structures. In addition, knowing what the characteristics of the system should be in some particular limit (e.g., in the conformal case that had already been treated), is also helpful to guide the calculations. In the case treated here, in particular, 
we already know what needs to be recovered in the limit of zero chemical potential. Finally, physical intuition also tells us what kinds of modes of propagation should be present in the system. In a nutshell, 
by relying on geometrical and physical intuition and an understanding of the causal properties of the theory in some particular limits, we can have a good educated guess for what the characteristics should look like. This allows us to look for a specific factorization of the characteristic determinant that points in that direction. This is the reason why, in our calculations, we group certain terms in certain ways, leading to expressions that can be managed in the end. Naturally, a brute-force approach would not be able to anticipate how one should group and factor terms in a way that would allow an explicit determination of the  characteristics.

The next step is to carry out a diagonalization of the principal part of the equations of motion in order to establish strong hyperbolicity. We are able to do so because we have a precise understanding of the system's characteristics. Even so, in order to carry out the diagonalization, we need to write the system as a system of first-order PDEs (notice that $\nabla_\mu T^{\mu\nu} = 0$ is a system of second-order PDEs because
$T^{\mu\nu}$ involves up to first derivatives of the hydrodynamic fields). In doing so, there is the risk of introducing spurious characteristics. For example, in the standard linear wave equation the characteristics are the light-cones. However, when one writes it as a first-order system in the standard way, the resulting system has a spurious characteristic (it corresponds, in the language of eigenvalues that can be applied to first-order systems, to a zero eigenvalue). While the presence of spurious characteristics per se is not an obstacle to diagonalization, the more of them there are, the more likely there will be obstacles to the diagonalization. Thus, we seek to \emph{choose as variables for our first-order system quantities that have direct physical or geometrical meaning,} so that the roots of the resulting characteristic polynomial resemble as closely as possible the ones of the original system. Of course, this does not guarantee diagonalizability. We still need to carry out some work mostly technical in nature to assure that the system is diagonalizable. But mutilating the equations upon rewriting them as first order by introducing new, fake features, is likely to only make the technical work harder or even insurmountable.

With diagonalization at hand, we can proceed to establish local well-posedness. The basic idea is that once the system is diagonalized, one can rely on techniques of diagonal systems of PDEs. There is a catch, though. The diagonalization of the system is at the level of the so-called principal symbol (i.e., it is a purely algebraic procedure that does not deal directly with differential operators). In order to apply it to the actual system of PDEs, one needs to introduce pseudo-differential operators, and the quasilinear nature of the equations causes
further complications as we need to deal with pseudo-differential operators with limited smoothness. While
there are results available in the literature for such situations (e.g., \cite{TaylorPDE3}), we have not found
a result that could be directly applied to our case. Thus, the first and second authors developed 
(with Rodriguez, Shao, and Graber) the necessary tools in \cite{DisconziBemficaGraber, DisconziBemficaRodriguezShaoSobolevConformal} with applications to the BDNK equations with zero chemical potential in mind. From these techniques and the diagonalization, local well-posedness
follows.

Finally, let us address stability. For this, one needs to find the roots of the polynomial determining the Fourier modes of the perturbations. More precisely, only the sign of the roots is relevant. Since the corresponding polynomial is of high order, there is little hope of determining its roots exactly and even the analysis of the 
sign of the roots is very challenging. Moreover, differently than what happens to the causality analysis, geometrical intuition is not of much help here because the Fourier modes are not covariant quantities. Because of these difficulties, in previous works the stability of the BDNK equations was not determined rigorously, being
obtained numerically or only in the homogeneous Lorentz boosted frame \cite{Hoult:2020eho,Bemfica:2019knx}. Due to a new result demonstrated in this paper, this limitation
is eliminated, as we shall discuss  below.

We are now ready to discuss specific novelties of the present work. While we continue to employ the ideas described above and in fact improve on them, especially with respect to some of the technical aspects that are more challenging for the complete system here considered, we want to highlight what are the truly new aspects introduced in this work. First, we are able to completely and rigorously determine the stability of the system. For this, we rely on a 
\emph{new stability theorem,} which roughly says that stability in the fluid's local rest frame (which can in general be determined because in this case the polynomial for the modes simplifies considerably) implies stability in any Lorentz boosted frame provided that the system is causal and strong hyperbolic; see Section \ref{sec:linear_stability_theorem} for the precise assumptions and statement of the theorem.
The theorem thus establishes a close relationship between causality and stability. While connections
between causality and stability have been discussed before, see \cite{Hiscock_Lindblom_stability_1983,Pu:2009fj} and references therein, these results focused on specific theories, thus making unclear whether they were due to the specific form
of the equations of motion or if they were examples of a yet undiscovered connection between causality and stability as general physical principles. \emph{Our theorem, in contrast, is a general theorem that can be applied to many different systems,} showing that the relationship between causality and stability runs deeper and is not a feature of specific systems. In fact, we obtain stability of the BDNK system by showing that it satisfies the assumptions of the general theorem. 

Interestingly, after the first version of this manuscript became available, a related theorem was proven in 
\cite{Gavassino:2021owo}, albeit using entirely different methods. The results in \cite{Gavassino:2021owo} 
also provide further physical intuition on the relationship between causality and stability, showing that
lack of causality allows that dissipation in one Lorentz frame be viewed as
``anti-dissipation'' (i.e., dissipation running ``backwards in time'') in another Lorentz frame. We also remark the related work \cite{Gavassino:2021kjm}. Combined, our paper and the works
\cite{Gavassino:2021owo, Gavassino:2021kjm} provide a comprehensive picture of the relationship between
causality and stability, an idea that was hinted several times in the literature before (see above references) but that had eluded the community until now.
 
We now discuss strong hyperbolicity. While strong hyperbolicity has been obtained for 
the BDNK theory before in the absence of a chemical potential \cite{Bemfica:2019knx,DisconziBemficaRodriguezShaoSobolevConformal, DisconziBemficaGraber}, the introduction
of a chemical potential causes new severe difficulties and the approach used in the case without chemical potential does not seem to work. Indeed, in \cite{Bemfica:2019knx,DisconziBemficaRodriguezShaoSobolevConformal, DisconziBemficaGraber}, the choice of variables to write the system as first-order was based primarily on their physical interpretation. For example, the viscous correction to the equilibrium energy density was one of the variables chosen.
As just said, a similar approach does not work here. While it is often a good idea to consider variables with a physical meaning, the first-order reduction we seek to establish itself
does not need to carry much physical meaning, so an approach employing easily identifiable physical variables might not bear any fruit. The first-order system does carry, however, some intrinsic \emph{geometric} properties, such as 
natural decompositions in the directions parallel and perpendicular to $u^\mu$ or the fact that the characteristics of the original system are preserved by the reduction to first-order. Thus, a choice
of geometric variables seems more appropriate. That is what we have done, considering new variables 
that involve several tensorial decompositions of the original variables. This has the extra advantage
that several tensorial and geometric properties of the fields can be used to carry out the difficult calculations needed to diagonalize the system. Yet another advantage is that while the previous physical choice of variables was specific to the form of the BDNK equations, the geometric approach is much more general and, thus, can be adapted to other theories in that similar tensorial decompositions hold for several fluids equations.
Therefore, a second novel aspect of this work is a \emph{new framework to investigate strong hyperbolicity in relativistic fluids.}
We remark that once the diagonalization is carried out, we can rely on the techniques developed in
\cite{DisconziBemficaGraber, DisconziBemficaRodriguezShaoSobolevConformal} to establish local well-posedness. Thus, while local well-posedness is probably the most technical and mathematical aspect of our results, we were able to rely more on previous techniques than any other of the results we obtain here.

In addition, it should by no means be overlooked that, although the proof of causality provided here follows
similar ideas as in our earlier work \cite{Bemfica:2019knx}, the fact that we are now considering the full set of equations makes the analysis much more difficult. Thus, a third novelty of our work is a \emph{substantial improvement of the techniques previously employed to analyze causality.}
From our causality analysis, it follows that the characteristics of the BDNK are the flow lines, sound waves,
the so-called second sound, corresponding to the propagation of temperature perturbations \cite{Hiscock_Lindblom_stability_1983}, and  shear waves (plus heat diffusion). In addition, when coupling to Einstein's equations
is considered, we find another set of characteristics corresponding to gravitational waves.

Finally, as already stressed many times, the main end product of this paper is itself a major novelty,
namely, the first construction of a viscous theory containing all relevant fields and satisfying \ref{I:Causality}--\ref{I:Strong_hyperbolicity}. We accomplish so by building and expanding on several previous ideas and also by introducing a series of novel ones, as described above.

Having discussed the new aspects of our work, we move to discuss how they combine with other aspects of the BDNK theory to provide a promising theoretical tool for the study of general relativistic viscous phenomena.
We begin by pointing out that the BDNK theory has been shown to be derivable from kinetic theory
and holographic arguments \cite{Bemfica:2017wps, Bemfica:2019knx, Hoult:2021gnb}. While derivation from kinetic theory
by itself is not guarantee that a theory is physically meaningful since the coarse-grain procedure
might introduce non-physical features -- indeed, recall that the Eckart and Landau-Lifshitz theories are derivable from kinetic theory --, it is reassuring to establish this connection with a microscopic theory. As shown in \cite{Hoult:2021gnb}, the derivation of BDNK theory from holography can be done in the context of the fluid/gravity correspondence \cite{Bhattacharyya:2007vjd} by carefully taking into account the presence of zero modes of the corresponding differential operators in the holographic bulk.

Next, we should point out that, contrary to MIS-like theories, the BDNK theory is capable of handling shocks.
By this, we mean that Rankine-Hugoniot-type conditions can in principle be obtained for the BDNK theory simply due to the fact that the BDNK equations are written as the conservation laws $\nabla_\mu T^{\mu\nu} = 0$ and $\nabla_\mu J^\mu = 0$. Aside from this simple observation, viscous shocks have been recently studied for the BDNK theory in the case of a conformal fluid using numerical methods in 
\cite{Pandya:2021ief}, while mathematically rigorously properties were established in \cite{Freistuhler-2021}.

At this point, we need to explain the role of shocks in the BDNK theory. Since the BDNK theory is an effective
theory truncated at first order in the gradient expansion, it is expected to be valid when gradients are not very large, which is precisely the opposite of shocks. In order to explain what we mean by a description of shocks in the BDNK formalism, let us consider for a moment
an ideal fluid. 
In this case, one also is assuming that gradients
are small. Alternatively, one may also see this as the limit where microscospic length scales are much smaller than the length scales associated with the gradients. However, shocks are known to develop
in solutions of ideal hydrodynamics, and the 
study of shocks is indeed an important topic
within the community. To what extent such shocks are accurate depictions of the state of the physical system is a legitimate question.
Nevertheless, once we have decided
to study shocks in the context of ideal hydrodynamics,
the formalism allows us to do so in that the equations
of motion of ideal fluids can accommodate weak solutions
(a.k.a distributional solutions) using the Rankine-Hugoniot conditions \cite{AnileBook}.
The same situation happens with BDNK: the formalism in principle allows for the study of shocks.
Whether or not
such solutions are physical, or accurate in the sense that the results would change significantly if the formalism was extended to second order, is an important question that
is beyond the scope of our paper. However, the point we are making is that we can, in principle, study shock solutions in the BDNK theory.

In other words, while the derivation of BDNK theory 
rests on the assumption
of small gradients, one might try to apply it to situations where in principle gradients are not small 
(like shocks), just like it was done before in the context of ideal fluids. Although this seems inconsistent,
it is precisely what it is done when one employs the
equations of ideal fluids to the study of shocks.
Moreover, it is also the case that MIS-like theories are often applied to situations where gradients are
not so small, e.g., \cite{Schenke:2012wb,Niemi:2014wta,Shen:2014vra,Noronha-Hostler:2015coa,Heller:2015dha,Strickland:2017kux, Plumberg:2021bme}. It is an intriguing, almost philosophical, question why one can sometimes still
obtain meaningful results in such cases, even though shocks are formally beyond the regime of validity of any known approach to viscous fluids -- an important question, however, that is beyond our scope here.

We now discuss another aspect of importance in viscous theories, which is entropy production. Naturally, one needs
the second law of thermodynamics to be satisfied, i.e., entropy production for physically realizable states of the system must be non-negative. Before addressing this point in the BDNK theory, however, some important points need to be highlighted. Strictly speaking, there is no universally understood expression for the entropy of a given system out of equilibrium, aside from the one given
by the Boltzmann equation. Thus, while it is useful to define an out-of-equilibrium entropy (which must, of course, reduce to the definition of equilibrium entropy in the absence of dissipation) we need to keep in mind that such a definition is not fundamental or even unique.
Moreover, the requirement that entropy production be non-negative on-shell unconditionally, i.e., to all orders in gradients, is certainly too stringent. In fact, since a fluid description is an effective description, it has a certain limit of applicability. Therefore, one should require that entropy production be non-negative only within the regime of validity of the theory (which is constructed within a certain approximation scheme). This point was stressed in \cite{TempleViscous} and discussed in detail in \cite{Kovtun:2019hdm}. In fact, enforcing non-negative entropy production even in the presence of any size of gradients was part of the Eckart and Landau and Lifshitz theories, but the resulting theory is unstable and acausal, as seen, showing that this requirement by itself is not guaranteed to lead to sensible theories in the context of the gradient expansion. Non-negative entropy production to all gradients is also a guiding principle in the construction of the MIS theory, but so far properties \ref{I:Causality}, \ref{I:LWP}, and \ref{I:Strong_hyperbolicity} remain open for it. On the other hand, the DNMR equations, that are stable, causal (in the absence of chemical potential), and are extensively used in numerical simulations of the quark-gluon plasma, do not have entropy production non-negative to all orders in Knudsen and inverse Reynolds numbers, but they should have non-negative entropy production within the limit of validity of the theory \cite{Denicol:2012cn}. The same is true for the BDNK theory, as pointed out in \cite{Kovtun:2019hdm} and shown in Section \ref{S:Entropy}. A thorough discussion of the role of entropy in viscous theories can be found  in \cite{Gavassino:2020ubn}.

We finally comment on the ability of the BDNK theory to describe realistic physical systems. 
In order to go beyond theoretical aspects and make connection with experiments, one needs to carry out realistic numerical simulations of the BDNK equations. Not surprisingly, given how recent the theory is,
such investigations are at an initial stage, but the results so far have been encouraging. In \cite{Pandya:2021ief}, the authors carry out numerical simulations of the BDNK theory in $1+1$ dimensions in the case of a conformal fluid and compare the results with simulations of MIS (rBRSSS) equations in the same setting. They found that for small 
values of the coefficient of shear viscosity, BDNK and MIS provide essentially the same evolution, but their dynamics differ for larger viscosity values. Given that small viscosity is one of the main regimes of interest of both theories (higher order corrections might become relevant in both theories if viscosity is not small), this shows that at least in this test case the BDNK theory reproduces the well-studied and considerably successful behavior of MIS theory. In addition, the BDNK theory also reproduces well-known
behavior considering Bjorken \cite{Bjorken:1982qr} and Gubser \cite{Gubser:2010ze,Gubser:2010ui,Marrochio:2013wla} flows, including the presence of a hydrodynamic attractor \cite{Bemfica:2017wps}. Further numerical studies of BDNK theory can be found in \cite{Pandya:2022pif,Bantilan:2022ech}.

We also stress the obvious point that 
being a causal, stable, and locally well-posed theory are themselves 
fundamental properties that need to be satisfied as a pre-requisite for describing actual physical phenomena.
Thus, while on the one hand a theory possessing these properties is only of formal interest if it is not connected to experiments, on the other hand a theory that has some phenomenological success but violates, say, causality, cannot be taken as an accurate description of real relativistic physical phenomena.
In this regard, we once more remark that, in view of the results presented in this paper, the BDNK theory is currently the only theory that satisfies the fundamental requirements \ref{I:Causality}--\ref{I:LWP} and the additional property \ref{I:Strong_hyperbolicity} when all viscous contributions and chemical potential are incorporated, including in the case when dynamical coupling to Einstein's equations is considered.


\section{Generalized Navier-Stokes Theory}
\label{sec:theory}

We consider a general-relativistic fluid described by an energy-momentum tensor $T^{\mu\nu}$ and a timelike conserved current $J^\mu$ associated with a global $U(1)$ charge that we take to represent baryon number. In our approach, the equations of relativistic fluid dynamics are given by the conservation laws 
\be
\label{conservEOM}
\nabla_\mu J^\mu = 0 \qquad \textrm{and} \qquad \nabla_\mu T^{\mu\nu} = 0,
\ee
which are dynamically coupled to Einstein's field equations
\be
\label{EinsteinEOM}
\qquad R_{\mu\nu}-\frac{R}{2}g_{\mu\nu}=8\pi G\,T_{\mu\nu}.
\ee
For the sake of completeness, we begin by recalling the case of a fluid in local equilibrium \cite{RezzollaZanottiBookRelHydro}. In this limit, one uses the following expressions in the conservation laws
\be
T^{\mu\nu} = \varepsilon u^\mu u^\nu + P\Delta^{\mu\nu}\qquad \textrm{and} \qquad J^\mu = n u^\mu,
\label{defineTmunuJmuideal}
\ee 
where $\varepsilon$ is the equilibrium energy density, $n$ is the equilibrium baryon density, $P = P (\varepsilon,n)$ is the thermodynamical pressure defined by the equation of state, and $u^\mu$ is a normalized timelike vector (i.e., $u_\mu u^\mu = -1$) called the flow velocity, and $\Delta_{\mu\nu}=g_{\mu\nu}+u_\mu u_\nu$ is a projector onto the space orthogonal to $u^\mu$. The thermodynamical quantities in equilibrium are connected via the first law of thermodynamics $\varepsilon+P = Ts +\mu\, n$, where $T$ is the temperature, $s$ is the equilibrium entropy density, and $\mu$ is the chemical potential associated with the conserved baryon charge.  We note that $u^\mu \nabla_\mu \varepsilon=0$ and $u^\mu \nabla_\mu n = 0$ in global equilibrium. These are much stronger constraints on the dynamical variables than in the case of local equilibrium where, e.g. only the combination $u^\mu \nabla_\mu \varepsilon + (\varepsilon+P)\nabla_\mu u^\mu$ vanishes. In local equilibrium, both $u_\mu T^{\mu\nu}$ and  $J^\nu$ are proportional to $u^\nu$ and, thus, the flow velocity may be defined using either quantity \cite{RezzollaZanottiBookRelHydro}.   

The system of equations \eqref{conservEOM} and \eqref{EinsteinEOM} for an ideal fluid [defined by \eqref{defineTmunuJmuideal}] is causal in the full nonlinear regime. Furthermore, given suitably defined initial data for the dynamical variables, solutions for the  nonlinear problem exist and are unique. The latter properties establish that the equations of motion of ideal relativistic fluid dynamics are locally well-posed in general relativity \cite{Choquet-BruhatFluidsExistence, DisconziRemarksEinsteinEuler}.

Let us now consider the effects of dissipation. Without any loss of generality, one may decompose the current and the energy-momentum tensor in terms of an arbitrary future-directed unit timelike vector $u^\mu$ as follows \cite{Kovtun:2012rj}
\be
J^\mu = \mathcal{N} u^\mu + \mathcal{J}^\mu
\label{generalJmu}
\ee
\be
T^{\mu\nu}= \mathcal{E} u^\mu u^\nu+ \mathcal{P}\Delta^{\mu\nu}+u^\mu\mathcal{Q}^\nu+u^\nu\mathcal{Q}^\mu+\mathcal{T}^{\mu\nu}
\label{generalTmunu}
\ee
where $ \mathcal{N} = -u_\mu J^\mu$, $\mathcal{E} = u_\mu u_\nu T^{\mu\nu}$, and $\mathcal{P} = \Delta_{\mu\nu}T^{\mu\nu}/3$ are Lorentz scalars while the vectors $ \mathcal{J}^\nu = \Delta^\nu_\mu J^\mu$, $\mathcal{Q}^\nu  = - u_\mu T^{\mu \lambda} \Delta_\lambda^\nu$, and the traceless symmetric tensor $\mathcal{T}^{\mu\nu} = \Delta^{\mu\nu}_{\alpha\beta}T^{\alpha\beta}$, with $\Delta^{\mu\nu}_{\alpha\beta} = \frac{1}{2} \left(\Delta^\mu_\alpha \Delta^\nu_\beta + \Delta^\mu_\beta \Delta^\nu_\alpha-\frac{2}{3}\Delta^{\mu\nu}\Delta_{\alpha\beta}\right)$, are all transverse to $u_\nu$. 
Observe that this decomposition is purely algebraic and simply expresses the fact that a vector
and a symmetric two-tensor can be decomposed relatively to a future-directed unit timelike
vector. The physical content of the theory is prescribed by relating the several components
in this decomposition to physical observables, which will then evolve  \footnote{
General constraints on the variables may be imposed by considering, for instance, energy conditions \cite{HawkingEllisBook,WaldBookGR1984}. In fact, we note that the dominant energy condition \cite{HawkingEllisBook} imposes that for all future-directed timelike vectors $z^\mu$ the vector
$Y^\nu  = - z_\mu T^{\mu\nu}$ must also be a future-directed timelike or null vector \cite{WaldBookGR1984}.}  according to
\eqref{generalJmu} and \eqref{generalTmunu}.

The general decomposition in Eqs.\ \eqref{generalJmu} and \eqref{generalTmunu} expresses  $\{J^\mu,T^{\mu\nu}\}$ in terms of 17 variables $\{\mathcal{E},\mathcal{N},\mathcal{P},u^\mu,\mathcal{J}^\mu,\mathcal{Q}^\mu, \mathcal{T}^{\mu\nu}\}$ and the conservation laws in Eq.\ \eqref{conservEOM} give 5 equations of motion for these variables. Therefore, additional assumptions must be made to properly define the evolution of the fluid. As mentioned before, 
the NS theory, including 
the standard approach in Refs.\ \cite{LandauLifshitzFluids,EckartViscous}, assumes that $\mathcal{E}=\varepsilon$ and $\mathcal{N} = n$. The same assumption is usually made in the MIS theory \cite{MIS-6}, though different prescriptions can be easily defined in the context of kinetic theory \cite{Tsumura:2011cj,Denicol:2012cn,Monnai:2018rgs}. A further constraint is usually imposed on the transverse vectors, i.e., either $\mathcal{J}^\mu = 0$ or $\mathcal{Q}^\mu=0$ throughout the evolution. For instance, the former gives $J^\mu = n u^\mu$ and $T^{\mu\nu} = \varepsilon u^\mu u^\nu+ (P+\Pi)\Delta^{\mu\nu}+u^\mu\mathcal{Q}^\nu+u^\nu\mathcal{Q}^\mu+\mathcal{T}^{\mu\nu}$, where $\Pi$ is the bulk viscous pressure (in equilibrium, $\Pi = 0$, $\mathcal{Q}^\nu=0$, and $\mathcal{T}^{\mu\nu}=0$). In this case, in an extended variable approach such as MIS \cite{MIS-6}, $\Pi$, $\mathcal{Q}^\nu$, and $\mathcal{T}^{\mu\nu}$ obey additional equations of motion that must be specified and solved together with the conservation laws, whereas
in the NS approach these quantities are expressed in terms of $u^\mu$, $\varepsilon$,
and its derivatives.

In this paper we investigate the problem of viscous fluids in general relativity using the BDNK formulation of relativistic fluid dynamics.
See Sections \ref{S:Overview} and \ref{S:Summary} for a detailed discussion of the origins of the BDNK theory and the conceptual framework that it entails. As explained in those sections, the starting point in the formulation of the BDNK theory is the most general expression for the energy-momentum tensor and the baryon current at first order.

In practice, the most general expressions for the \emph{constitutive relations} that define the quantities in \eqref{generalJmu} and \eqref{generalTmunu}, truncated to first order in derivatives, are (following the notation in \cite{Kovtun:2019hdm})
\bml
\label{generaldefKovtun}
\bea
\mathcal{E}&=& \varepsilon + \varepsilon_1 \frac{u^\alpha\nabla_\alpha T}{T}+\varepsilon_2 \nabla_\alpha u^\alpha+\varepsilon_3 u^\alpha\nabla_\alpha(\mu/T),\\
\mathcal{P}&=&P+\pi_1 \frac{u^\alpha\nabla_\alpha T}{T}+\pi_2 \nabla_\alpha u^\alpha+\pi_3 u^\alpha\nabla_\alpha(\mu/T),\\
\mathcal{N}&=& n+\nu_{1} \frac{u^\alpha\nabla_\alpha T}{T}+\nu_2 \nabla_\alpha u^\alpha+\nu_3 u^\alpha\nabla_\alpha(\mu/T),\\
\mathcal{Q}^\mu&=&\theta_{1}\frac{\Delta^{\mu\nu}\nabla_\nu T}{T}+\theta_{2} u^\alpha\nabla_\alpha u^\mu+\theta_{3} \Delta^{\mu\nu}\nabla_\nu(\mu/T),\\
\mathcal{J}^\mu&=&\gamma_1\frac{\Delta^{\mu\nu}\nabla_\nu T}{T}+\gamma_2 u^\alpha\nabla_\alpha u^\mu+\gamma_3 \Delta^{\mu\nu}\nabla_\nu(\mu/T)\\
\mathcal{T}^{\mu\nu}&=&  -2 \eta \sigma^{\mu\nu},
\eea
\eml
where $\sigma^{\mu\nu} = \Delta^{\mu\nu\alpha\beta}\nabla_\alpha u_\beta$ is the shear tensor. The transport parameters $\{\varepsilon_i,\pi_i,\theta_i,\nu_i,\gamma_i\}$ and the shear viscosity $\eta$ are functions of $T$ and $\mu$. Thermodynamic consistency of the equilibrium state (i.e., that $\varepsilon$, $P$, and $n$ have the standard interpretations of equilibrium quantities connected via well-known thermodynamic relations) imposes that $\gamma_1 = \gamma_2$ and $\theta_1 = \theta_2$ \cite{Kovtun:2019hdm}. The final equations of motion for $\{T,\mu,u^\alpha\}$, which are of second-order in derivatives, are found by substituting the expressions above in the conservation laws. 
In the language of Section \ref{S:Frames}, 
expressions \eqref{generaldefKovtun} for 
\eqref{generalJmu} and \eqref{generalTmunu} correspond to the most general choice of a hydrodynamic frame for a first-order theory. As stressed in \cite{Kovtun:2019hdm}, it is of course impossible to not choose a hydrodynamic frame since the latter actually defines the meaning of the variables $\{T,\mu,u^\mu\}$ out of equilibrium (see Section \ref{S:Frames} for details).

In fact, in the regime of validity of the first-order theory, one may shift $\{T,\mu,u^\mu\}$ by adding terms that are of first-order in derivatives, shifting also the transport parameters $\{\varepsilon_i,\pi_i,\theta_i,\nu_i,\gamma_i\}$, without formally changing the physical content of $T^{\mu\nu}$ and $J^\mu$ \cite{Kovtun:2019hdm}. However, there are combinations of the transport parameters that remain invariant under these field redefinitions. In fact, the shear viscosity $\eta$ and the combination of coefficients that give the bulk viscosity $\zeta$ and charge conductivity $\sigma$ are invariant under first-order field redefinitions, as explained in \cite{Kovtun:2019hdm}. Additional constraints among the transport parameters appear when the underlying theory displays conformal invariance, as discussed in detail in Ref.\ \cite{Bemfica:2017wps} at $\mu=0$, and at finite chemical potential in \cite{Kovtun:2019hdm,Hoult:2020eho} (see also \cite{Taghinavaz:2020axp}). 

Hoult and Kovtun \cite{Hoult:2020eho} investigated \eqref{generaldefKovtun} at nonzero chemical potential using a class of hydrodynamic frames where $\varepsilon_3 = \pi_3 = \theta_3=0$. This corresponds to the case where there are non-equilibrium corrections to both the conserved current and the heat flux. This choice is useful when considering relativistic fluids where the net baryon density is not very large, as in high-energy heavy-ion collisions. Conditions for causality were derived and limiting cases were studied that strongly indicated that this choice of hydrodynamic frame is stable against small disturbances around equilibrium. Further studies are needed to better understand the nonlinear features of its solutions (well-posedness) and also the stability properties of this class of hydrodynamic frames at nonzero baryon density in a wider class of equilibrium states.

In this paper we consider another class of  hydrodynamic frames that we believe can be more naturally implemented in simulations of the baryon rich matter formed in neutron star mergers or in low energy heavy-ion collisions. Our choice for the hydrodynamic frame is closer to Eckart's as we define the flow velocity using the baryon current, i.e., $J^\mu = n u^\mu$ holds throughout the evolution ($\gamma_i=\nu_i=0$). Clearly, this limits the domain of applicability of the theory to problems where there are many more baryons than anti-baryons so the net baryon charge is large. 

In this case, it is more convenient to use $\varepsilon$ and $n$ as dynamical variables instead of $T$ and $\mu/T$ because the most general expressions for the Lorentz scalar contributions to the constitutive relations involve only linear combinations of $u^\mu \nabla_\mu \varepsilon$ and $ \nabla_\mu u^\mu$, given that current conservation implies that the replacement $u^\lambda \nabla_\lambda n = -n \nabla_\lambda u^\lambda$ is valid. For simplicity, we choose to parametrize the out of equilibrium corrections to the scalars as follows  (we note that $\theta_1=\theta_2$ and $\gamma_1 = \gamma_2$ and
in practice, 8 out of the 14 parameters in \eqref{generaldefKovtun} can be set using first-order field redefinitions \cite{Kovtun:2019hdm}, so one is then left with $\eta$, $\zeta$, $\sigma$, and three other parameters)
\bml
\label{newscalars}
\bea
\mathcal{E}&=& \varepsilon + \tau_\varepsilon\left[ u^\lambda \nabla_\lambda \varepsilon +  (\varepsilon+P) \nabla_\lambda u^\lambda\right]\\
\mathcal{P}&=&P-\zeta \nabla_\lambda u^\lambda+\tau_P\left[u^\lambda \nabla_\lambda \varepsilon +  (\varepsilon+P) \nabla_\lambda u^\lambda\right] ,
\eea
\eml
where $\tau_\varepsilon$ and $\tau_P$ have dimensions of a relaxation time and $\zeta$ is the bulk viscosity transport coefficient. When evaluated on the solutions of the equations of motion, one can see that these quantities assume their standard form as in Eckart's theory up to second order in derivatives because $\mathcal{E} \sim \varepsilon + \mathcal{O}(\partial^2)$ and $\mathcal{P} = P - \zeta \nabla_\mu u^\mu+ \mathcal{O}(\partial^2)$ on shell  (we follow traditional terminology where a given quantity is said to be on shell when it is evaluated using the solutions to the equations of motion). 

In fact, we remind the reader that in Eckart's theory \cite{EckartViscous} the energy-momentum tensor is given by $T_{\mu\nu} = \varepsilon u_\mu u_\nu + \left(P-\zeta \nabla_\lambda u^\lambda\right)\Delta_{\mu\nu} -2\eta \sigma_{\mu\nu} + u_\mu \mathcal{Q}_\nu + u_\nu \mathcal{Q}_\mu$, with heat flux $\mathcal{Q}_\mu = -\kappa T \left(u^\lambda \nabla_\lambda u_\mu + \Delta_\mu^\lambda \nabla_\lambda T/T\right)$ where $\kappa = (\varepsilon+P)^2 \sigma/(n^2 T)$ is the thermal conductivity coefficient. However, as remarked in \cite{Kovtun:2019hdm}, in the domain of validity of the first-order theory one may rewrite the Eckart expression for the heat flux as $\mathcal{Q}_\nu = \sigma T \frac{(\varepsilon+P)}{n}\Delta^{\lambda}_\nu\nabla_\lambda (\mu/T)$ plus second-order terms. This is done by noticing that $(\varepsilon+P)u^\lambda \nabla_\lambda u^\mu + \Delta^{\mu\lambda}\nabla_\lambda P  = 0 + \mathcal{O}(\partial^2)$ on shell, which implies that one may write, using the standard thermodynamic relation $\frac{dP}{\varepsilon+P} = \frac{dT}{T} + \frac{n T}{\varepsilon+P}d\left(\frac{\mu}{T}\right)$, 
\be
u^\lambda \nabla_\lambda u^\alpha + \frac{\Delta^{\alpha\lambda}\nabla_\lambda T}{T} =- \frac{nT}{\varepsilon+P}\Delta^{\alpha\lambda}\nabla_\lambda (\mu/T)+\mathcal{O}(\partial^2).
\ee
Therefore, one can always choose the coefficients such that the heat flux $\mathcal{Q}^\mu$ has the same physical content of Eckart's theory plus terms that are of second order on shell. We use this to write this quantity as
\be
\mathcal{Q}_\nu = \sigma T \frac{(\varepsilon+P)}{n}\Delta^{\lambda}_\nu\nabla_\lambda (\mu/T) + \tau_Q \left[(\varepsilon+P)u^\lambda \nabla_\lambda u_\nu + \Delta_\nu^\lambda \nabla_\lambda P\right],
\ee
where $\tau_Q$ has dimensions of a relaxation time.

In this work, we make the following choice for the constitutive relations that give the energy-momentum tensor and the baryon current:
\bml
\label{finaltheory}
\bea
J^\mu &=& n u^\mu\\
T^{\mu\nu} &=& \left(\varepsilon+\mathcal{A}\right)u^\mu u^\nu + (P+\Pi)\Delta^{\mu\nu} - 2\eta \sigma^{\mu\nu} + u^\mu \mathcal{Q}^\nu + u^\nu \mathcal{Q}^\mu\\
\mathcal{A} &=& \tau_\varepsilon\left[ u^\lambda \nabla_\lambda \varepsilon +  (\varepsilon+P) \nabla_\lambda u^\lambda\right]\label{10c}\\
\Pi&=&-\zeta \nabla_\lambda u^\lambda+\tau_P\left[u^\lambda \nabla_\lambda \varepsilon +  (\varepsilon+P) \nabla_\lambda u^\lambda\right] \\
\mathcal{Q}^\nu &=& \tau_Q(\varepsilon+P)u^\lambda \nabla_\lambda u^\nu + \beta_\varepsilon \Delta^{\nu\lambda}\nabla_\lambda \varepsilon + \beta_n \Delta^{\nu\lambda}\nabla_\lambda n
\eea
\eml
where
\bml
\label{definebetas}
\bea
\beta_\varepsilon &=& \tau_Q \left(\frac{\partial P}{\partial \varepsilon}\right)_n + \frac{\sigma T (\varepsilon+P)}{n} \left(\frac{\partial  (\mu/T) }{\partial \varepsilon}\right)_n\\
\beta_n &=& \tau_Q \left(\frac{\partial P}{\partial n}\right)_\varepsilon + \frac{\sigma T (\varepsilon+P)}{n} \left(\frac{\partial  (\mu/T) }{\partial n}\right)_\varepsilon,
\eea
\eml
and $\tau_\varepsilon$, $\tau_P$, and $\tau_Q$ quantify the magnitude of second order corrections to the out of equilibrium contributions to the energy-momentum tensor given by the energy density correction $\mathcal{A}$, the bulk viscous pressure $\Pi$, and the heat flux $\mathcal{Q}^\mu$. 
In other words, \eqref{finaltheory}--\eqref{definebetas} correspond to the frame we consider in this work, thus they provide a definition of what we mean by the non-equilibrium hydrodynamic fields.

The reason for considering the constitutive relations \eqref{finaltheory}--\eqref{definebetas} is that they lead to a theory satisfying properties \ref{I:Causality}--\ref{I:Strong_hyperbolicity}, as it will be shown below. We refer the reader to Section \ref{S:Summary} for a discussion of the ideas and techniques that led to the particular choice \eqref{finaltheory}--\eqref{definebetas}.

The equations of motion for the fluid variables are obtained from the conservation laws and they can be written explicitly as
\bml
\label{EOMcurrent}
\bea
\label{EOMbaryon}
&& u^\lambda \nabla_\lambda n + n \nabla_\lambda u^\lambda = 0, \\ 
\label{EOMenergy}
&&u^\lambda \nabla_\lambda \varepsilon + (\varepsilon+P)\nabla_\lambda u^\lambda =- u^\lambda \nabla_\lambda \mathcal{A} -  (\mathcal{A}+\Pi)\nabla_\lambda u^\lambda - \nabla_\mu \mathcal{Q}^\mu -  \mathcal{Q}^\mu u^\lambda \nabla_\lambda u_\mu +2\eta \sigma_{\mu\nu} \sigma^{\mu\nu} ,\\ \label{EOMvector}
&& \left(\varepsilon+P\right)u^\nu \nabla_\nu u^\beta + \Delta^{\beta\lambda}\nabla_\lambda P = -\left(\mathcal{A}+\Pi\right)u^\nu \nabla_\nu u^\beta - \Delta^{\beta\lambda}\nabla_\lambda \Pi +\Delta^\beta_\lambda \nabla_\mu (2\eta\sigma^{\mu\lambda})\nonumber \\  &-& u^\lambda\nabla_\lambda \mathcal{Q}^\beta - \frac{4}{3}\nabla_\lambda u^\lambda \mathcal{Q}^\beta - \mathcal{Q}_\mu \sigma^{\mu\beta} -  \mathcal{Q}_\mu \omega^{\mu\beta}, 
\eea
\eml
where $\omega_{\mu\nu} = \dfrac{1}{2}\left(\Delta_\mu^\lambda \nabla_\lambda u_\nu - \Delta_\nu^\lambda \nabla_\lambda u_\mu\right)$ is the kinematic vorticity tensor \cite{RezzollaZanottiBookRelHydro}. The equations above show that, on shell, $\mathcal{A} \sim 0 + \mathcal{O}(\partial^2)$, $\Pi \sim -\zeta \nabla_\mu u^\mu + \mathcal{O}(\partial^2)$, and $\mathcal{Q}_\nu = \sigma T \frac{(\varepsilon+P)}{n}\Delta^{\lambda}_\nu\nabla_\lambda (\mu/T) + \mathcal{O}(\partial^2)$. Eqs.\ \eqref{finaltheory}, \eqref{definebetas}, and \eqref{EOMcurrent} define a causal and stable generalization of Eckart's theory that is fully compatible with general relativity, as we shall prove in the next sections. We remark that when one neglects the effects of a conserved current altogether, the theory reduces to the case studied in Refs.\ \cite{Kovtun:2019hdm,Bemfica:2019knx}. For additional discussion about the case without a chemical potential, including far from equilibrium behavior and also the presence of analytical solutions, see Refs.\ \cite{Das:2020fnr,Shokri:2020cxa,Das:2020gtq}.

\subsection{Entropy Production\label{S:Entropy}}

It is instructive to investigate how the second law of thermodynamics is obeyed in this general first-order approach. This was discussed in detail by Kovtun in \cite{Kovtun:2019hdm} and, more recently, by other authors in Ref.\ \cite{Gavassino:2020ubn}. 

The standard covariant definition of the entropy current based on the first law of thermodynamics $T\,\mathbb{S}^\mu  = P u^\mu - u_\nu T^{\nu\mu} - \mu J^\mu$ \cite{MIS-6}, together with \eqref{finaltheory}, can be used to show that the entropy density measured by a co-moving observer is given by
\be
- u_\mu \mathbb{S}^\mu  = s + \frac{\mathcal{A}}{T}.
\ee
Note that in our system one finds that $\mathcal{A} = 0 + \mathcal{O}(\partial^2)$ on shell. Furthermore, using Eqs.\ \eqref{finaltheory}  and \eqref{EOMcurrent} one finds that the divergence of the entropy current is given by
\bea
\nabla_\mu \mathbb{S}^\mu &=& \frac{2\eta \sigma_{\mu\nu}\sigma^{\mu\nu}}{T} -\frac{\Pi}{T}\nabla_\mu u^\mu+ \frac{n}{\varepsilon+P}\mathcal{Q}^\nu \Delta_{\nu}^\lambda \nabla_\lambda (\mu/T) - \frac{\mathcal{Q}^\nu}{T}\left[u^\lambda \nabla_\lambda u_\nu + \frac{\Delta^{\lambda}_\nu\nabla_\lambda P}{\varepsilon+P}\right] - \frac{\mathcal{A}}{T}\,\frac{u^\lambda \nabla_\lambda T}
{T}.
\label{entropyproduction1}
\eea
It is crucial to note \cite{Kovtun:2019hdm} that in a first-order approach $\nabla_\mu \mathbb{S}^\mu$ can only be correctly determined up to second order in derivatives (recall  that in this argument terms such as $\nabla_\mu \nabla_\nu \phi$ and $(\nabla_\mu \phi)(\nabla_\nu \phi)$, for any field $\phi$, count as second order terms; see Section \ref{S:Frames}). This means that not all the terms in \eqref{entropyproduction1} actually contribute to this expression at second order. For instance, when evaluating \eqref{entropyproduction1} \emph{on shell} one must keep in mind that the last two terms in \eqref{entropyproduction1} are already at least of third order and must, thus, be dropped. A similar argument can be used to show that the term $\Pi\nabla_\mu u^\mu = -\zeta (\nabla_\mu u^\mu)^2 + \mathcal{O}(\partial^3)$. Therefore, one can see that 
\be
\nabla_\mu \mathbb{S}^\mu= \frac{2\eta \sigma_{\mu\nu}\sigma^{\mu\nu}}{T} +\frac{\zeta(\nabla_\mu u^\mu)^2}{T} + \sigma T \left[\Delta_{\nu}^\lambda \nabla_\lambda (\mu/T)\right]\left[\Delta^{\nu\alpha} \nabla_\alpha (\mu/T)\right]+\mathcal{O}(\partial^3),
\ee
which is non-negative when $\eta, \zeta, \sigma \geq 0$. Hence, there are no violations of the second law of thermodynamics in the domain of validity of the first-order theory - higher order derivative terms $\mathcal{O}(\partial^3)$  in the entropy production can only be understood by considering terms of higher order in derivatives in the constitutive relations in $T^{\mu\nu}$ and $J^\mu$, which is beyond the scope of the first-order approach.    

 
\section{Causality}
\label{sec:causality}


In order to determine the conditions under which causality holds in this theory,
we need to understand the system's characteristics. Our system is a mixed first-second order
system of PDEs. While the principal part and characteristics of systems of this form can be
investigated using Leray's theory \cite{ChoquetBruhatGRBook,DisconziFollowupBemficaNoronha,Leray_book_hyperbolic}, here it is simpler
to transform our equations into a system where all equations are of second-order. We thus
apply $u^\mu \nabla_\mu$ on \eqref{EOMbaryon}.
In this case, the conservation laws \eqref{conservEOM} coupled to Einstein's equations \eqref{EinsteinEOM}
written in harmonic gauge, $g^{\mu\nu}\Gamma_{\mu\nu}^\alpha=0$, read
\bml
\label{EOM}
\bea
&& u^\beta u^\alpha\partial^2_{\alpha \beta} n+
n\delta^\alpha_\nu u^\beta \partial^2_{\alpha\beta} u^\nu+\tilde{\mathcal{B}}_1(n,u,g)\partial^2 g=
\mathcal{B}_1(\partial n, \partial u, \partial g),\\
&&(\tau_\varepsilon u^\alpha u^\beta+\beta_\varepsilon\Delta^{\alpha\beta})\partial^2_{\alpha\beta}\varepsilon+\beta_n\Delta^{\alpha\beta}\partial^2_{\alpha\beta}n+\rho(\tau_\varepsilon+\tau_Q)u^{(\alpha}\delta^{\beta)}_\nu\partial^2_{\alpha\beta}u^\nu+\tilde{\mathcal{B}}_2(\varepsilon,n,u,g)\partial^2 g=\mathcal{B}_2(\partial \varepsilon,\partial n,\partial u,\partial g), \\
&&(\beta_\varepsilon+\tau_P) u^{(\alpha}\Delta^{\beta)\mu}\partial^2_{\alpha\beta}\varepsilon+\beta_n u^{(\alpha}\Delta^{\beta)\mu}\partial^2_{\alpha\beta}n
+\mathcal{C}^{\mu\alpha\beta}_\nu\partial^2_{\alpha\beta} u^\nu+\tilde{\mathcal{B}}_3^\mu(\varepsilon,n,u,g)\partial^2 g=\mathcal{B}_3^\mu(\partial\varepsilon,\partial n,\partial u,\partial g),\\
&&g^{\alpha\beta}\partial^2_{\alpha\beta}g^{\mu\nu}=\mathcal{B}_4^{\mu\nu}(\partial\varepsilon,\partial n,\partial u,\partial g),
\eea
\eml
where $\partial^2_{\alpha\beta} = \partial_\alpha\partial_\beta$ (using standard partial derivatives), $\rho = (\varepsilon+P)$, and $A_{(\alpha}B_{\beta)}=(A_\alpha A_\beta+A_\beta B_\alpha)/2$.  The remaining notation is as follows. We use $\partial^\ell \phi$ to indicate that a term depends
on at most $\ell$ derivatives of $\phi$. A term of the form 
$\mathcal{B}(\partial^{\ell_1} \phi_1, \dots, \partial^{\ell_k} \phi_k) \partial^\ell  \phi_i$, $i \in 
\{1,\dots,k\}$, indicates an expression that is linear in $\partial^\ell \phi_i$ with coefficients
depending on at most $\ell_1$ derivatives of $\phi_1$,..., $\ell_k$ derivatives of $\phi_k$.
For example, the term  $(u^\mu \partial_\mu \varepsilon + \partial_\mu u^\mu )
g^{\alpha\beta} \partial^2_{\alpha\beta} g_{\gamma\delta}$ would be written
as $\mathcal{B}(\partial \varepsilon,\partial u, g) \partial^2 g$
(a term of this form is not present in our system, we write it
here only for illustration).
The terms $\tilde{\mathcal{B}}$ above are top-order in derivatives of $g$ 
and thus belong to the principal part, although, as we will see, their explicit form is not needed
for our argument, whereas the $\mathcal{B}$ terms are lower order and do not contribute to the principal part.
We have also defined  
\be
\mathcal{C}^{\mu\alpha\beta}_\nu=\left(\tau_P \rho -\zeta -\frac{\eta}{3}\right ) \Delta^{\mu(\alpha}\delta^{\beta)}_\nu+(\rho\tau_Q u^\alpha u^\beta-\eta\Delta^{\alpha\beta})
\delta^\mu_\nu.
\ee
We notice that by taking $u^\mu \nabla_\mu$ of \eqref{EOMbaryon} we are not introducing
new characteristics in the system. This can be viewed from the characteristic determinant computed  below
which contains an overall factor of $u^\mu \xi_\mu$ to a power greater than one. Theorem I below establishes necessary and sufficient conditions for causality to hold in our system of equations.
We show that the assumptions of Theorem I are not empty in section \ref{S:Non_empty}. Throughout this paper, we use the following definition for the speed of sound $c_s$:
\be
\label{soundspeed}
c_s^2=\left (\frac{\partial P}{\partial \varepsilon}\right )_{\bar{s}}=\left (\frac{\partial P}{\partial \varepsilon}\right )_n+\frac{n}{\rho}\left (\frac{\partial P}{\partial n}\right )_\varepsilon,
\ee
where $\bar{s}$ is the equilibrium entropy per particle. Also, we define 
\be
\kappa_s=\frac{\rho^2T}{n}\left [\frac{\partial(\mu/T)}{\partial\varepsilon}\right ]_{\bar{s}}=\frac{\rho^2 T}{n}\left [\frac{\partial(\mu/T)}{\partial\varepsilon}\right ]_n+T\rho\left [\frac{\partial(\mu/T)}{\partial n}\right ]_\varepsilon.
\ee

\medskip

\noindent \textbf{Theorem I.}
{\it Let $(\varepsilon,n,u^\mu, g_{\alpha\beta})$ be a solution to 
\eqref{EinsteinEOM} and \eqref{EOMcurrent}, with $u^\mu u_\mu = -1$,
defined in a globally hyperbolic spacetime $(M,g_{\alpha\beta})$.
Assume that:

\medskip
(A1) $\rho=\varepsilon+P,\tau_\varepsilon,\tau_Q,\tau_P>0$ and $\eta,\zeta,\sigma\ge0$.
\medskip

\noindent Then, causality holds for $(\varepsilon,n,u^\mu, g_{\alpha\beta})$ if, and only if,
the following conditions are satisfied:
\bml
\label{C_conditions}
\bea
&&\rho\tau_Q>\eta,\label{C_condition_a}\\
&&\left[ \tau_\varepsilon \left(\rho  c_s^2 \tau _Q+\zeta +\frac{4 \eta}{3} +\sigma \kappa _s\right)+\rho  \tau _P \tau _Q\right]^2\ge 
4\rho\tau_\varepsilon\tau_Q\left[ \tau_P \left(\rho  c_s^2 \tau_Q+\sigma  \kappa_s\right)-\beta_\varepsilon \left ( \zeta +\frac{4 \eta}{3}\right  )\right ]\ge0,\label{C_condition_b}\\
&&2\rho\tau_\varepsilon\tau_Q > \tau_\varepsilon \left(\rho  c_s^2 \tau _Q+\zeta +\frac{4 \eta}{3} +\sigma \kappa _s\right)+\rho  \tau _P \tau _Q\ge0,\label{C_condition_d}\\
&&\rho\tau_\varepsilon\tau_Q+ \sigma  \kappa_s\tau_P> \tau_\varepsilon \left(\rho  c_s^2 \tau _Q+\zeta +\frac{4 \eta}{3} +\sigma \kappa _s\right)+\rho  \tau _P \tau _Q(1-c_s^2)+\beta_\varepsilon \left ( \zeta +\frac{4 \eta}{3}\right  ).\label{C_condition_e}
\eea
\eml
The same result holds true for equations  \eqref{EOMcurrent} if the metric is not dynamical.
} 
 
\medskip
\noindent \emph{Proof.} The proof can be reduced to a computation of the characteristics of 
\eqref{EOM} \cite{Leray_book_hyperbolic}. Technical details are found in Appendix\ \ref{Theorem_I}.

\section{Strong hyperbolicity and local well-posedness}
\label{sec:hyperbolicity}

In this section we investigate the initial-value problem for equations
\eqref{EinsteinEOM} and \eqref{EOMcurrent}.
The goal is to show that the system is causal and locally well-posed under 
very general conditions. First, we briefly discuss the 
initial data required to solve the system of equations. Then, we  
re-write our system as a first-order system. We show that this first-order system is 
diagonalizable in the sense of Proposition I. This means, in particular, that the 
system is \emph{strong hyperbolic} according to the usual definition of the term, as in, e.g., \cite{AnileBook,RezzollaZanottiBookRelHydro}. The importance of having strongly hyperbolic
equations is due to its implications for the initial-value problem. As already mentioned, one 
is generally interested in evolution equations that are locally well-posed  \footnote{See  \cite{F:Sobolev} for technical remarks on function spaces and local well-posedness.}. For equations with constant coefficients, local well-posedness 
is equivalent to strong hyperbolicity \cite{KasaharaYamagutiStrongHyperbolicity}. For non-constant coefficients and
nonlinear systems, such an equivalence does not hold \cite{KajitaniStrongHyperbolicity,MetivierStrongHyperbolicity,StrangNecessaryConditionsCauchy}. However, 
there remains a close connection between strong hyperbolicity and local well-posedness. For most reasonable systems, once diagonalizability is available, one can
use known techniques to derive energy estimates which, in turn, can be used to 
prove local well-posedness, see Section \ref{S:Summary} for more discussion on the techniques involved. This is precisely the case for our system of equations.
Even though our equations consist of a system of second order PDEs, we can use
the diagonalized system of first-order equations to derive energy estimates.
Once these estimates are available, we use a standard approximation argument as in
\cite{Choquet-BruhatFluidsExistence,Lichnerowicz_MHD_book} to obtain 
local well-posedness (see Theorem II).

\subsection{Initial data}
Equations \eqref{EOMcurrent} are second order in $\varepsilon$, $n$, and 
$u^\mu$. Thus, initial data 
along a non-characteristic hypersurface consist of the values of 
$\varepsilon$, $n$, $u^\mu$ and their first-order time derivatives. Clearly, the initial
$u^\mu$ has to satisfy $u^\mu u_\mu = -1$. Also, it is important to note that Eq.\ \eqref{EOMbaryon} is first-order and, thus, the initial-data cannot be arbitrary but must satisfy a compatibility condition
ensuring that \eqref{EOMbaryon} holds at $t=0$. Therefore, one can use \eqref{EOMbaryon} to write the time derivative of $n$ in terms of the time derivative of $u^\mu$ (this feature would also appear in Navier-Stokes theory in the Eckart hydrodynamic frame). 

A natural choice to determine the initial conditions for the matter sector is to set an initial state that is within the regime of validity of the first-order theory and closely reproduces Eckart's theory. First, one can directly extract $n$ and $u^\mu$  from $J^\mu$ at the initial spacelike hypersurface. Then, one sets the non-equilibrium correction to the energy density $\mathcal{A}$ in \eqref{finaltheory} to zero in the initial state, so then the initial value for $\varepsilon$ equals $T^{\mu\nu}u_\mu u_\nu$ and the first-order time derivative of $\varepsilon$ is defined in terms of the first-order time derivative of the flow velocity (plus spatial derivatives that are known in the initial state). Clearly, $\mathcal{A}$ will be different than zero later during the actual evolution, and its value can be used to check if the simulations remain within the regime of validity of the first-order approach (i.e., $|\mathcal{A}|/\varepsilon$ must remain less than unity). Finally, the time derivative of the flow velocity can be set by imposing that the second-order on shell term $(\varepsilon+P)u^\lambda \nabla_\lambda u^\nu + \Delta^{\nu\lambda}\nabla_\lambda P$ vanishes. Hence, one can obtain the time derivative of the flow velocity and all the other required initial data in the regime of validity of the first-order approach, emulating Eckart's theory as much as possible. 

We recall that the initial-data for the gravitational sector has to further satisfy the well-known 
Einstein constraint equations. We briefly make some comments on this in Section \ref{sec:conclusions}.

\subsection{Diagonalization and Eigenvectors}
\label{S:diagonalization}
In this section we write equations \eqref{EinsteinEOM} and \eqref{EOMcurrent} as a first-order
system, as discussed above. For this, we begin defining the 
variables $V=u^\alpha\partial_\alpha \varepsilon$,
$\V^\mu=\Delta^{\mu\alpha}\partial_\alpha \varepsilon$,
$W=u^\alpha\partial_\alpha n$,
$\W^\mu=\Delta^{\mu\alpha}\partial_\alpha n$,
$S^\mu=u^\alpha\nabla_\alpha u^\mu$,
$\mathcal{S}^\nu_\lambda=\Delta^{\alpha}_\lambda\nabla_\alpha u^\nu$,
$F_{\mu\nu}=u^\alpha\partial_\alpha g_{\mu\nu}$, and
$\mathcal{F}^\lambda_{\mu\nu}=\Delta^{\lambda\alpha}\partial_\alpha g_{\mu\nu}$.
Then, the equations of motion can be cast as
\bml
\label{Sobolev_4}
\bea
&&\tau_\varepsilon u^\alpha\partial_\alpha V+\tau_Q\rho\partial_\nu S^\nu+\tau_\varepsilon\rho u^\alpha\partial_\alpha S^\nu_\nu+\beta_\varepsilon\partial_\nu \V^\nu+\beta_n\partial_\nu \W^\nu=r_1,\label{Sobolev_4a}\\
&&\tau_P \Delta^{\mu\alpha}\partial_\alpha V
+\tau_Q\rho u^\alpha\partial_\alpha S^\mu+\beta_\varepsilon u^\alpha\partial_\alpha \V^\mu+\beta_n u^\alpha\partial_\alpha \W^\mu
+\eta \Pi^{\mu\lambda\alpha}_\nu \partial_\alpha S^\nu_\lambda=r_2^\mu,\label{Sobolev_4b}\\
&& u^\alpha \partial_\alpha \V^\mu-\Delta^{\mu\alpha}\partial_\alpha V=r_3^\mu,\label{Sobolev_4c}\\
&& u^\alpha \partial_\alpha \W^\mu+n\Delta^{\mu\alpha}\partial_\alpha\mathcal{S}^\nu_\nu =r_4^\mu,\label{Sobolev_4d}\\
&& u^\alpha\partial_\alpha\mathcal{S}^\nu_\lambda-\Delta^\alpha_\lambda\partial_\alpha S^\nu-\mathcal{X}^{\nu A\alpha}_\lambda \partial_\alpha F_A
-\mathcal{Y}^{\nu A\alpha}_{\lambda\delta}\partial_\alpha\mathcal{F}^\delta_{A}=r_{5\lambda}^\nu,\label{Sobolev_4e}\\
&&u^\alpha\partial_\alpha F_{A}-\Delta^\alpha_\delta\mathcal{F}^\delta_A=r_{6A},\label{Sobolev_4g}\\
&&u^\alpha\partial_\alpha \mathcal{F}^\delta_{A}-\Delta^{\delta\alpha}\partial_\alpha F_{A}=r_{7A}^\delta,\label{Sobolev_4f}\\
&&u^\alpha\partial_\alpha \varepsilon=r_8,\label{Sobolev_4h}\\
&&u^\alpha\partial_\alpha n=r_9,\label{Sobolev_4i}\\
&& u^\alpha \partial_\alpha u^\mu=r_{10}^\mu,\label{Sobolev_4j}\\
&& u^\alpha\partial_\alpha g_A= r_{11A},\label{Sobolev_4k}
\eea
\eml
where the $r'$s are functions of the fields $\varepsilon,u^\nu,\cdots,\mathcal{F}_{\mu\nu}^ \lambda$ but not its derivatives and $A=\sigma\beta$ for $\sigma\ge \beta$, i.e., $A$ takes the 10 independent values $00,01,02,03,11,12,13,22,23,33$ with repeated index $A$ summing from 00 to 33, 
\bml
\label{Sobolev_5}
\bea
\Pi^{\mu\lambda\alpha}_\nu&=&-\eta(\Delta^{\mu\lambda}\delta^\alpha_\nu+\Delta^{\alpha\lambda}\delta^\mu_\nu)+\left (\rho\tau_P-\zeta+\frac{2\eta}{3}\right )\Delta^{\mu\alpha}\delta^\lambda_\nu,\\
\mathcal{X}^{\nu A(=\sigma\beta)\alpha}_\lambda&=&\frac{1}{2}\left [g^{\nu(\sigma}\Delta^{\beta)}_\lambda u^\alpha-u^{(\sigma}\Delta^{\beta)}_\lambda g^{\nu\alpha}
-u^{(\sigma}\Delta^{\beta)\nu}\Delta^\alpha_\lambda\right ](2-\delta_A),\\
\mathcal{Y}^{\nu A(=\sigma\beta)\alpha}_{\lambda\delta}&=&\frac{1}{2}u^{(\sigma}u^{\beta)}\Delta^\alpha_{\lambda}\delta^\nu_\delta(2-\delta_A).
\eea
\eml
By $\delta_A$ we mean the Kr\"onecker delta in the sense that when $A=\sigma\beta$ then $\delta_A=\delta_{\sigma\delta}$, which equals one when $\sigma=\beta$ and zero otherwise. Also, the terms $r$ may be functions of the 95 variables. The equations in \eqref{Sobolev_4} were obtained as follows: Eqs.\ \eqref{Sobolev_4a} and \eqref{Sobolev_4b} come from the conservation law $\nabla_\nu T^{\mu\nu}=0$ when projected into the directions parallel and perpendicular to $u^\nu$, respectively. Eqs.\ \eqref{Sobolev_4c}, \eqref{Sobolev_4d}, \eqref{Sobolev_4e}, and \eqref{Sobolev_4f} correspond, respectively, to the identities $\nabla_\alpha\nabla_\beta\varepsilon-\nabla_\beta\nabla_\alpha\varepsilon=0$, $\nabla_\alpha\nabla_\beta n-\nabla_\beta\nabla_\alpha n=0$, $\nabla_\alpha\nabla_\beta u^\nu-\nabla_\beta\nabla_\alpha u^\nu=R^\nu_{\alpha\beta\sigma}u^\sigma=(\partial_\alpha\Gamma^\nu_{\beta\sigma}-\partial_\beta\Gamma^\nu_{\alpha\sigma})u^\sigma+$ terms of order zero in derivatives, and $\partial_\alpha\partial_\beta g_{\mu\nu}-\partial_\beta\partial_\alpha g_{\mu\nu}=0$, all contracted with $u^\alpha\Delta^{\beta}_\lambda$. Eq.\ \eqref{Sobolev_4g} is the Einstein equation in the harmonic gauge, i.e., $g^{\alpha\beta}\partial_\alpha\partial_\beta g_{\mu\nu}=$ terms of lower order in derivatives, while \eqref{Sobolev_4h}--\eqref{Sobolev_4k} are the definitions of $V$, $W$ (also using the identity
$u^\alpha\nabla_\alpha n+n\nabla_\alpha u^\alpha=W+n \mathcal{S}^\alpha_\alpha=0$ to eliminate $W$ thoroughly), $S^\mu$, and $F_A$, respectively. We may now define the $95\times 1$ column vectors $\Psi$ and $\mathfrak{B}$ as
\begin{align}
\Psi=\bbm \psi_m\\\psi_g\\\psi_d\ebm
\label{E:vector_U}
\end{align}
and $\mathfrak{B}=(r_1,\cdots, r_{11A})^T$, where $\psi_m=(V,S^\nu,\V^\nu,\W^\nu,\mathcal{S}^\nu_0,\mathcal{S}^\nu_1,\mathcal{S}^\nu_2,\mathcal{S}^\nu_3)^T\in\mathbb{R}^{29}$, $\psi_g=(F_A,\mathcal{F}_A^0,\mathcal{F}_A^1,\mathcal{F}_A^2,\mathcal{F}_A^3)^T\in\mathbb{R}^{50}$, and $\psi_d=(\varepsilon,n,u^\nu,g_A)^T\in\mathbb{R}^{16}$, to write the quasi-linear first order system \eqref{Sobolev_4} in matrix form as
\be
\label{Sobolev_6}
\mathfrak{A}^\alpha\partial_\alpha \Psi=\mathfrak{B},
\ee  
where, here, $\mathfrak{A}^\alpha=\mathbb{A}^\alpha\oplus u^\alpha I_{16}$ ($\oplus$ being the direct sum). The matrix $\mathbb{A}^\alpha$ is split in the following way 
\be
\label{Sobolev_7}
\mathbb{A}^\alpha=\bbm \mathbb{A}^\alpha_m & -L^\alpha\\ 0_{50\times 29} & \mathbb{A}^\alpha_g \ebm,
\ee
where
\be
\label{Sobolev_8}
\mathbb{A}^\alpha_m=\bbm
\tau_\varepsilon u^\alpha & \rho\tau_Q\delta^\alpha_\nu & \beta_\varepsilon\delta^\alpha_\nu & \beta_n \delta^\alpha_\nu & \rho\tau_\varepsilon u^\alpha \delta_\nu^0 & \rho\tau_\varepsilon u^\alpha \delta_\nu^1 & \rho\tau_\varepsilon u^\alpha \delta_\nu^2 & \rho\tau_\varepsilon u^\alpha \delta_\nu^3\\
\tau_P \Delta^{\mu\alpha} & \rho\tau_Q u^\alpha\delta^\mu_\nu & \beta_\varepsilon u^\alpha \delta^\mu_\nu & \beta_n u^\alpha \delta^\mu_\nu & \Pi^{\mu0\alpha}_\nu & \Pi^{\mu1\alpha}_\nu & \Pi^{\mu2\alpha}_\nu & \Pi^{\mu3\alpha}_\nu\\
-\Delta^{\mu\alpha} & 0_{4\times 4} & u^\alpha \delta^\mu_\nu & 0_{4\times 4} & 0_{4\times 4} & 0_{4\times 4} & 0_{4\times 4} & 0_{4\times 4}\\
0_{4\times 1} & 0_{4\times 4} & 0_{4\times 4} & u^\alpha \delta^\mu_\nu & n\Delta^{\mu\alpha}\delta_\nu^0 & n\Delta^{\mu\alpha}\delta_\nu^1 & n\Delta^{\mu\alpha}\delta_\nu^2 & n\Delta^{\mu\alpha}\delta_\nu^3\\
0_{4\times 1} & -\Delta^{\alpha}_0\delta^\mu_{\nu} & 0_{4\times 4} & 0_{4\times 4} & u^\alpha \delta^\mu_\nu & 0_{4\times 4} & 0_{4\times 4} & 0_{4\times 4}\\
0_{4\times 1} & -\Delta^{\alpha}_1\delta^\mu_\nu & 0_{4\times 4} & 0_{4\times 4} & 0_{4\times 4} & u^\alpha \delta^\mu_\nu & 0_{4\times 4} & 0_{4\times 4}\\
0_{4\times 1} & -\Delta^{\alpha}_2\delta^\mu_\nu & 0_{4\times 4} & 0_{4\times 4} & 0_{4\times 4} & 0_{4\times 4} & u^\alpha \delta^\mu_\nu & 0_{4\times 4}\\
0_{4\times 1} & -\Delta^{\alpha}_3\delta^\mu_\nu & 0_{4\times 4} & 0_{4\times 4} & 0_{4\times 4} & 0_{4\times 4} & 0_{4\times 4} & u^\alpha \delta^\mu_\nu\\
\ebm,
\ee
while
\be
\label{Sobolev_9}
\mathbb{A}^\alpha_g=\bbm 
u^\alpha I_{10} & -\Delta^\alpha_0I_{10} & -\Delta^\alpha_1I_{10} & -\Delta^\alpha_2I_{10} & -\Delta^\alpha_3I_{10}\\
-\Delta^{0\alpha}I_{10} & u^\alpha I_{10} & 0_{10\times 10} & 0_{10\times 10} & 0_{10\times 10}\\
-\Delta^{1\alpha}I_{10} & 0_{10\times 10} & u^\alpha I_{10} & 0_{10\times 10} & 0_{10\times 10}\\
-\Delta^{2\alpha}I_{10} & 0_{10\times 10} & 0_{10\times 10} & u^\alpha I_{10} & 0_{10\times 10}\\
-\Delta^{3\alpha}I_{10} & 0_{10\times 10} & 0_{10\times 10} & 0_{10\times 10}  & u^\alpha I_{10}
\ebm
\ee
and
\be
\label{Sobolev_10}
L^\alpha=\bbm
0_{1\times 10} & 0_{1\times 10} & 0_{1\times 10} & 0_{1\times 10} & 0_{1\times 10} \\
0_{4\times 10} & 0_{4\times 10} & 0_{4\times 10} & 0_{4\times 10} & 0_{4\times 10} \\
0_{4\times 10} & 0_{4\times 10} & 0_{4\times 10} & 0_{4\times 10} & 0_{4\times 10} \\
0_{4\times 10} & 0_{4\times 10} & 0_{4\times 10} & 0_{4\times 10} & 0_{4\times 10} \\
\mathcal{X}^{\mu A\alpha}_0 & \mathcal{Y}^{\mu A\alpha}_{0 0} & \mathcal{Y}^{\mu A\alpha}_{01} & \mathcal{Y}^{\mu A\alpha}_{02} & \mathcal{Y}^{\mu A\alpha}_{03}\\
\mathcal{X}^{\mu A\alpha}_1 & \mathcal{Y}^{\mu A\alpha}_{10} & \mathcal{Y}^{\mu A\alpha}_{1 1} & \mathcal{Y}^{\mu A\alpha}_{12} & \mathcal{Y}^{\mu A\alpha}_{13}\\
\mathcal{X}^{\mu A\alpha}_2 & \mathcal{Y}^{\mu A\alpha}_{20} & \mathcal{Y}^{\mu A\alpha}_{2 1} & \mathcal{Y}^{\mu A\alpha}_{22} & \mathcal{Y}^{\mu A\alpha}_{23}\\
\mathcal{X}^{\mu A\alpha}_3 & \mathcal{Y}^{\mu A\alpha}_{30} & \mathcal{Y}^{\mu A\alpha}_{3 1} & \mathcal{Y}^{\mu A\alpha}_{32} & \mathcal{Y}^{\mu A\alpha}_{33}\\
\ebm .
\ee

We are now ready to establish that, when written as a first-order system as above, 
the equations of motion are strongly hyperbolic. In section \ref{S:Non_empty}, we show that 
the assumptions of Proposition I are not empty.

\medskip

\noindent \textbf{Proposition I.} 
{\it
\noindent Consider the system \eqref{Sobolev_4}.
Assume that (A1) with $\eta>0$ holds and that \eqref{C_conditions} in Theorem I holds in strict form, i.e., with $>$
instead of $\geq$. Let $\xi$ be a timelike co-vector. Then:

\medskip

(i) $\det(\mathfrak{A}^\alpha \xi_\alpha) \neq 0$;

\medskip

(ii) For any spacelike vector $\zeta$, the eigenvalue problem 
$(\zeta_\alpha + \Lambda \xi_\alpha)\mathfrak{A}^\alpha  R = 0$
has only real eigenvalues $\Lambda$ and a complete set of right eigenvectors $R$.
}

\medskip

\noindent \emph{Proof.} The proof of this proposition is very lengthy and we refer the interested reader to check all the details and the proof presented in  Appendix \ref{Proposition_I}.

\bigskip

\subsection{Local well-posedness}
\label{localwellposedness_section}
In this section we establish the local existence and uniqueness of solutions to the nonlinear equations of motion in \eqref{EinsteinEOM} and \eqref{EOMcurrent}. 

We begin by noticing that  \eqref{EOMcurrent} used the normalization $u^\mu u_\mu=-1$ to project the divergence
of $T_{\mu\nu}$ and $J^\mu$ onto the directions parallel and orthogonal to $u^\mu$. In order to
show that the condition $u^\mu u_\mu=-1$ is propagated by the flow, it is more convenient 
to work directly with \eqref{conservEOM} and \eqref{EinsteinEOM}. In order to complete the system,
we differentiate $u^\mu u_\mu=-1$ twice in the $u^\mu$ direction,
\begin{align}
u^\beta\nabla_\beta \left[u^\alpha \nabla_\alpha (u^ \alpha u_\alpha)\right]=0.
\label{E:u_unit_differentiated_twice}
\end{align}
We also differentiate $\nabla_\mu J^\mu=0$ once, as in section \ref{sec:causality},
\begin{align}
u^\mu \nabla_\mu \left(\nabla_\nu J^\nu\right)= 0.
\label{E:div_J_differentiated}
\end{align}
Observe that \eqref{E:u_unit_differentiated_twice} and \eqref{E:div_J_differentiated} imply
that  $u^\mu u_\mu=-1$ and $\nabla_\mu J^\mu=0$ hold at later times if these
hold at the initial time.

The main result of this section can be found below.

\medskip
\noindent \textbf{Theorem II.}
{\it 
Let $(\Sigma, \mathring{g}_{\alpha\beta}, \widehat{\kappa}_{\alpha\beta}, \mathring{\varepsilon},
\widehat{\varepsilon}, \mathring{n}, \widehat{n}, \mathring{u}^\alpha, \widehat{u}^\alpha)$
be an initial-data set for the system comprised of Einstein's equations \eqref{conservEOM}
and $\nabla_\mu J^\mu = 0$, where $T_{\alpha\beta}$ and $J^\mu$ are given in
\eqref{finaltheory}. Assume that $\mathring{u}^\mu \mathring{u}_\mu = -1$,
$\mathring{n} > 0$  \footnote{If we work with uniformly local
Sobolev spaces \cite{KatoQuasiLinear} instead of the ordinary Sobolev spaces, we can in fact assume $\mathring{n} \geq C > 0$, where $C$ is a constant. The same is the case if $\Sigma$ is compact. We remark that in view of the finite-speed propagation property assured by the causality of the equations, different asymptotic conditions can be considered for the initial data without significantly changing the conclusions of Theorem II.}, and that $\nabla_\mu J^\mu=0$ holds for the initial data.
Assume (A1) with $\eta>0$ and suppose that \eqref{C_conditions} of Theorem I hold in strict form 
and that the transport coefficients are analytic functions of their arguments.
Finally, assume that 
$\mathring{g}_{\alpha\beta}, \mathring{\varepsilon},
 \mathring{n},  \mathring{u}^\alpha \in H^N(\Sigma)$
and that $\widehat{\kappa}_{\alpha\beta},\widehat{\varepsilon},\widehat{n},\widehat{u}^\alpha
\in H^{N-1}(\Sigma)$, $N\geq 5$, where $H^N$ is the Sobolev space. Then, there exists a globally
hyperbolic development of the initial data. This globally hyperbolic development is unique if taken to 
be the maximum globally hyperbolic development of the initial data.
}

\medskip

\noindent \emph{Proof.} The proof is found in Appendix \ref{Theorem_II}.

\section{A new theorem about linear stability}
\label{sec:linear_stability_theorem}

Any ordinary fluid  must be stable against small deviations from the thermodynamic equilibrium state \cite{LandauLifshitzFluids}. (We only consider systems such that the equilibrium state is unique and has a finite correlation length. Therefore, in principle, our discussion does not apply to systems where the correlation length in equilibrium can become arbitrarily large, such as at a critical point.). We recall that in equilibrium $\beta_\mu  = u_\mu/T$ must be a Killing vector, i.e. $\nabla_\mu \beta_\nu + \nabla_\nu \beta_\mu = 0$, and also $\nabla_\alpha(\mu/T)=0$ \cite{MIS-6,Becattini:2012tc,Becattini:2016stj}. In Minkowski spacetime, non-rotating equilibrium corresponds to a class of states  with constant $T$ and $\mu$ and background flow velocity $u^\mu = \gamma(1,\mathbf{v})$ defined by a constant sub-luminal 3-velocity $\mathbf{v}$, where $\gamma = 1/\sqrt{1-\mathbf{v}^2}$. 
 (In this paper we neglect the constant thermal vorticity term, see \cite{Becattini:2012tc} for a nice discussion of its physical content and consequences.)
In the local rest frame (LRF) $\mathbf{v}=0$ and the background flow is simply $u^\mu = (1,0,0,0)$. In a stable theory, small disturbances from the general equilibrium state $T \to T + \delta T(t,\mathbf{x})$, $\mu \to \mu + \delta \mu(t,\mathbf{x})$, and $u^\mu \to u^\mu + \delta u^\mu(t,\mathbf{x})$ (with $u_\mu \delta u^\mu = 0$) lead to small variations in the energy-momentum tensor and current, $\delta T^{\mu\nu}(t,\mathbf{x})$ and $\delta J^\mu(t,\mathbf{x})$, which decay with time. 

The standard theories from Eckart and Landau-Lifshitz are unstable, as shown by Hiscock and Lindblom many years ago \cite{Hiscock_Lindblom_instability_1985}. This instability appears because such theories possess exponentially growing, hence \emph{unstable}, non-hydrodynamic modes, which spoil linear stability around equilibrium even at vanishing wave number.
   (The frequency of a hydrodynamic mode, such as a sound wave, vanishes in a spatially uniform state. On the other hand, a non-hydrodynamic mode correspond to a collective excitation that possesses nonzero frequency even at zero wavenumber.)
For Landau-Lifshitz theory at zero chemical potential, this instability is only observed when considering a general equilibrium state with nonzero $\mathbf{v}$ \cite{Hiscock_Lindblom_instability_1985,Denicol:2008ha,Pu:2009fj}, while in the case of Eckart the instability already appears even when $\mathbf{v}=0$. The lack of causality in these approaches implies that it is not sufficient to investigate only the static $\mathbf{v}=0$ case in order to determine the stability properties of a general equilibrium state where $\mathbf{v}\neq 0$, even though such states are in principle connected via a simple Lorentz transformation. 

The necessity to investigate the stability properties of general equilibrium states where $\mathbf{v}\neq 0$ makes linear stability analyses of viscous hydrodynamic theories very complicated. Already in the local rest frame, finding whether the linear modes of the system are stable requires determining the sign of the imaginary part of the roots of a high order polynomial,  which becomes a daunting task when $\mathbf{v}\neq 0$ (see \cite{Hoult:2020eho} and \cite{Brito:2020nou} for recent examples of how complicated a $\mathbf{v}\neq 0$ analysis can become in BDNK and MIS theory, respectively).   

We prove below a new theorem that gives sufficient conditions for causal fluid dynamic equations to be linearly stable against disturbances of a general non-rotating equilibrium state with arbitrary background velocity. In this case, proving stability for the local rest frame implies stability in any other frame  (note that the word frame here is used in the standard context of special relativity, i.e.,
to refer to an inertial observer, and has nothing to do with the concept of a hydrodynamic frame discussed in previous sections, which concerned the definition of hydrodynamic variables out of equilibrium) connected to the local rest frame via a Lorentz transformation. This general feature is expected to hold in any interacting relativistic system, i.e., no issues should appear if one simply observes a given system in another inertial frame. We then use this theorem in Section \ref{sec:stability} to find conditions under which the hydrodynamic theory presented here is stable. We remark that our results can be used to establish stability at nonzero $\mathbf{v}\neq 0$ in other theories as well, e.g. MIS, as long as the conditions discussed below are fulfilled.    

\subsection{Transforming a second order system of linear differential equations into a first order one}
\label{firstorder}

We begin by showing how one may convert a system of linear second order PDE's into a first order one, as this is needed for the theory discussed in this paper. 
Let the system of linearized second order PDE's be given by
\be
\label{s_1}
\sum_b \mathfrak{M}(\partial)^a_b\delta\psi^b(X)=\mathfrak{N}(\partial\delta\psi)^a,
\ee
where $a$ and $b$ runs from 1 to $n$,  $\mathfrak{M}(\partial)_a^ b$ are differential linear operators of order 2, $\mathfrak{N}(\partial\Psi)$ are linear terms containing derivatives of the perturbed fields $\delta \Psi$ up to order 1, and $\delta\psi^1(X),\cdots,\delta\psi^n(X)$ are the perturbed fields (for instance, $\delta \varepsilon$, $\delta n$, and etc). We suppose that \eqref{s_1} arises from the conservation laws $-u_\alpha\partial_\beta\delta T^{\alpha\beta}=0$, $\Delta^\mu_\alpha\partial_\beta\delta T^{\alpha\beta}=0$, and $\partial_\alpha\delta J^\alpha=-u_\beta u^\alpha\partial_\alpha\delta J^\beta+\Delta^{\alpha\beta}\partial_\alpha \delta J_\beta =0$, where the first two come from $\partial_\alpha\delta T^{\alpha\beta}=0$, while the last equation appears only when $J^\mu$ is included. In this manner, the derivatives in the EOM's in \eqref{s_1} shall always appear as combinations of $u^\alpha\partial_\alpha$ and $\Delta^{\alpha\beta}\partial_\beta$ only. Thus, if the system in \eqref{s_1} has one or more second order equations, it can be rewritten as a first order system in the $N\equiv 5n$ new variables $\delta\bar{\psi}^a(X)=u^\alpha\partial_\alpha\psi^a(X)$ and $\delta\tilde{\psi}^a_\mu(X)=\Delta^\nu_\mu \partial_\nu\psi^a(X)$. These definitions automatically lead \eqref{s_1} to $n$ first order linear equations. One then needs to supplement those with the $4n$ dynamical equations that are missing. By means of the identity $\partial_\alpha\partial_\beta \psi^a(X)-\partial_\beta\partial_\alpha \psi^a(X)=0$, one may find the extra $4n$ dynamical equations $u^\alpha\partial_\alpha \delta\tilde{\psi}^a_\mu(X)-\Delta_\mu^\alpha\partial_\alpha \delta\bar{\psi}^a(X)=0$, giving the needed $5n$ first order dynamical equations, as required. In matrix form it becomes
\be
\label{s_2}
\mathbb{A}^\alpha\partial_\alpha \delta \Psi(X)+\mathbb{B}\delta\Psi(X)=0,
\ee 
where $\mathbb{A}^\alpha$ and $\mathbb{B}$ are $N\times N$ constant real matrices and $\delta\Psi(X)$ is a $N\times 1$ column vector with entries $\delta\bar{\psi}^1,\delta\tilde{\psi}^1_\nu,\cdots,\delta\bar{\psi}^n,\delta\tilde{\psi}^n_\nu$. This ends the procedure. However, if one of the equations in \eqref{s_1} is already of first order but contains variables that have second order derivative in other equations, then one can eliminate this equation by using it as a constraint to eliminate one of the variables. For example, consider the case of the ideal current $J^\mu=nu^\mu$. In this case, the conservation equation $\partial_\alpha J^\alpha=0$ becomes $u^\alpha\partial_\alpha \delta n(X)+n\partial_\alpha \delta u^\alpha(X)=0$. If $T^{\mu\nu}$ has shear or bulk contributions, for example, then the other equations must have second order derivatives of $\delta u^\mu$. Thus, one must write $\partial_\alpha\delta J^\alpha=0$ as $\delta\bar{\psi}+n \delta\tilde{\psi}^\mu_\mu=0$, where $\delta\tilde{\psi}^\mu_\nu=\Delta^\alpha_\nu\partial_\alpha u^\mu$ and $\delta\bar{\psi}=u^\alpha\partial_\alpha n$. This is a zeroth order equation in the new variables and, therefore, is just a constraint. One may use this constraint in order to eliminate the variable $\delta\bar{\psi}$ in the other dynamical equations. Then, in this case one ends up with $5n-1$ dynamical equations for the $5n-1$ fields. 

Finally, we remark that other approaches to viscous relativistic fluids, such as MIS, are already written in the format \eqref{s_2} in the linearized regime so the  procedure to reduce the order of the equations of motion described above is not needed and one can move directly to the part below. 

\subsection{New linear stability theorem}
\label{linearstability}
To study linear stability, let us expand the perturbed fields in the Fourier modes $K^\mu=(i\Gamma,k^i)$ by substituting $\delta\Psi(X)\to \exp(iK_\mu X^\mu)\delta\Psi(K)= \exp(\Gamma t+ik_ix^i)\delta\Psi(K)$ in \eqref{s_2}. The result is 
\be
\label{s_3-2}
iK_\mu \mathbb{A}^\mu\delta\Psi(K)+\mathbb{B}\delta\Psi(K)=0.
\ee 
Since $K^\mu$ appears, as aforementioned, as combinations of $-u^\alpha K_\alpha=\gamma(i\Gamma-k_i v^i)$ and $\Delta^{\mu\nu}K_\mu K_\nu=(u^\mu K_\mu)^2+\Gamma^2+k^2$, where $k^2=k_ik^i$, then the direction of $k^i$ is not relevant once one keeps $v^i$ arbitrary. Thus, we may write $K^\mu=-n^\mu n_\nu K^\nu+\zeta^\mu\zeta_\nu K^\nu$, where $n_\mu$ is timelike and $\zeta_\mu$ is spacelike, with $n^\mu n_\mu=-1$, $n_\mu\zeta^\mu=0$, and $\zeta_\mu\zeta^\mu=1$, [for example, it is common to choose $K^\mu=(K^0,k,0,0)$ so that $n_\mu$ and $\zeta_\nu$ are $(-1,0,0,0)$ and $(0,1,0,0)$, respectively]. In this case we define $\Omega=n_\alpha K^\alpha$ and $\kappa=\zeta_\alpha K^\alpha$ such that $K^\mu=-\Omega n^\mu+\kappa \zeta^\mu$ \cite{Brito:2020nou}. Then, \eqref{s_3-2} can be written as
\be
\label{s_3}
i\Omega(-n_\alpha \mathbb{A}^\alpha)\delta\Psi(K)=-i \kappa \zeta_\alpha \mathbb{A}^\alpha\delta\Psi(K)-\mathbb{B}\delta\Psi(K).
\ee  
The general form of the co-vectors $n$ and $\zeta$ is $n_\alpha=\gamma_n(-1,c^i)$ for any $c^i$ such that $0\le c^ic_i<1$ and where $\gamma_n=1/\sqrt{1-c^ic_i}\ge 1$, and $\zeta_\alpha=\gamma_\zeta(-\hat{d}^jc_j,\hat{d}^i)\ge1$, where $\hat{d}^i\hat{d}_i=1$ for an arbitrary unitary $\hat{d}^i$ and  $\gamma_\zeta=1/\sqrt{1-(\hat{d}^ic_i)^2}\ge1$. From the Cauchy-Schwarz inequality $(\hat{d}^ic_i)^2\le |c^i|^2$ (here $|c^i|=\sqrt{c^ic_i}$), then one obtains that
\be
\label{s_3-1}
\gamma_n\ge\gamma_\zeta.
\ee
Stability demands that the perturbed modes $\Gamma=\Gamma(k^i)$ are such that $\Gamma_R\le 0$. Now, consider the eigenvalue problem 
\be
\label{s_2-1}
(\Lambda n_\alpha +\zeta_\alpha) \mathbb{A}^\alpha \mathfrak{r}=0, 
\ee
where here $\Lambda$ is the eigenvalue associated with the right eigenvector $\mathfrak{r}$.

\medskip

\noindent \textbf{Proposition II.} 
{\it If \eqref{s_2} is causal, then the eigenvalues $\Lambda$ are real and lie in the range $[-1,1]$. Furthermore, $\det(n_\alpha \mathbb{A}^\alpha)\ne 0$.
}

\medskip

\noindent \emph{Proof.} Causality demands that the roots of $Q(\xi)=\det(\xi_\alpha \mathbb{A}^\alpha)=0$ are such that (i) $\xi_0=\xi_0(\xi_i)\in\mathbb{R}$ and that (ii) the curves $\xi_0$ lie outside or over the light-cone. In other words, $\xi^\alpha\xi_\alpha\ge 0$. If one writes $\xi_\alpha=\Lambda n_\alpha+\zeta_\alpha$, where $n$ and $\zeta$ are real, then condition (i) means that $\Lambda$ is real. On the other hand, since $n$ and $\zeta$ are orthonormal, then condition (ii) means that $\xi_\alpha\xi^\alpha=-\Lambda^2+1\ge 0$, which demands that $\Lambda^2\le 1$, i.e., $\Lambda\in[-1,1]$. Now, since $Q(\xi)=0$ if and only if $\xi$ is spacelike or lightlike, this means that $\det(n_\alpha \mathbb{A}^\alpha)\ne 0$.

\hfill $\Box$

\bigskip

\noindent \textbf{Theorem III.} 
{\it
Let \eqref{s_2-1} have a set of $N$ linearly independent real eigenvectors $\{\mathfrak{r}_1,\cdots,\mathfrak{r}_N\}$. If \eqref{s_2} is causal and stable in the local rest frame $\mathcal{O}$, then it is also stable in any other Lorentz frame $\mathcal{O}'$ connected to $\mathcal{O}$ by a Lorentz transformation.
}

\medskip

\noindent \emph{Proof.} The details of the proof are found in Appendix \ref{Theorem_III}. However, we summarize some steps here. Note that causality enables us to invert the matrix $(-n_\alpha \mathbb{A}^\alpha)$. Then, it is possible to rewrite \eqref{s_3} as
\bea
i\Omega \delta\Psi(K)^\dagger (R^T)^{-1}R^{-1}\delta\Psi(K)&=&-i \kappa \delta\Psi(K)^\dagger (R^T)^{-1}R^{-1} (-n_\alpha \mathbb{A}^\alpha)^{-1} (\zeta_\alpha \mathbb{A}^\alpha)\delta\Psi(K)\nonumber\\
&-&\delta\Psi(K)^\dagger (R^T)^{-1}R^{-1} (-n_\alpha \mathbb{A}^\alpha)^{-1}\mathbb{B}\delta\Psi(K),
\eea  
where ${}^\dagger$ stand for the matrix transpose and complex conjugate operations altogether, ${}^T$ stands for matrix transpose operation, while $R$ is the square matrix that diagonalizes $(-n_\alpha \mathbb{A}^\alpha)^{-1} (\zeta_\alpha \mathbb{A}^\alpha)$, since \eqref{s_2-1} has a complete set of real eigenvectors in $\mathbb{R}^ n$ with only real eigenvalues. Then, we can expand $\delta\Psi(K)$ in terms of these eigenvectors. In the proof, it is shown that $\delta\Psi(K)^\dagger (R^T)^{-1}R^{-1}\delta\Psi(K)$ and $\delta\Psi(K)^\dagger (R^T)^{-1}R^{-1} (-n_\alpha \mathbb{A}^\alpha)^{-1} (\zeta_\alpha \mathbb{A}^\alpha)\delta\Psi(K)$ are real   for any Lorentz frame. After some work, we demonstrate that, under the theorem's statements, stability reduces to the condition that the term $\delta\Psi(K)^\dagger (R^T)^{-1}R^{-1} (-n_\alpha \mathbb{A}^\alpha)^{-1}\mathbb{B}\delta\Psi(K)$ must be greater or equal to zero. Since this is proven to be a scalar under Lorentz boosts, it can be computed in any frame. Thus, this implies that if the theory is stable in the LRF and obeys the other conditions of the theorem, it is stable in any other Lorentz frame. 

We note that this result implies that the original system of linearized second order PDE's in \eqref{s_1} is stable under the stated assumptions. 

\subsubsection{Applying the stability theorem to a toy model\label{S:Toy_model_example}}

To illustrate the application of the stability theorem, consider the simple model described by the fields $\phi$ and $\psi^\mu$ that obey the first order dynamical linear equations
\bml
\label{toy1}
\bea
&&u^\alpha\partial_\alpha\phi-\alpha \Delta^{\alpha}_\nu\partial_\alpha \psi^\nu+\lambda \phi=0,\\
&&u^\alpha\partial_\alpha \psi^\mu-\beta \Delta^{\mu\alpha}\partial_\alpha \phi=0.
\eea
\eml
We consider the case where $u^\mu$ is constant [$u^\mu=\gamma(1,v^i)$ with $\gamma=1/\sqrt{1-v^2}$ and $v^2=v^iv_i < 1$] as done in the stability theorem of the last subsection. If we write \eqref{toy1} in matrix form as 
\be
\label{toy3}
\mathbb{A}^\alpha\partial_\alpha \Psi(X)+\mathbb{B}\Psi(X)=0,
\ee
where $\Psi(X)=(\phi,\psi^\nu)$ is a $5\times 1$ column vector, $\mathbb{B}=\bbm \lambda & 0_{1\times 4}\\0_{4\times 1} & 0_{4\times 4}\ebm$ and
\be
\label{toy4}
\mathbb{A}^\alpha=\bbm
u^\alpha & -\alpha \Delta^{\alpha}_\nu \\
-\beta \Delta^{\mu\alpha} & u^\alpha \delta^\mu_\nu  
\ebm
\ee
are $5\times 5$ matrices, the propagation modes $\omega=\omega(k^i)$ are obtained by means of the Fourier transform $\Psi(X)\to e^{iK_\mu X^\mu}\tilde{\Psi}(K)$, where $K^\mu=(\omega, k^i)$, and are the roots of $\det[iK_\alpha \mathbb{A}^\alpha+\mathbb{B}]=0$. Let us write $\omega=i\Gamma$. Then, stability requires that $\Re(\Gamma)\le0$. In the local rest frame (LRF), these equations are $\Gamma=0$ and $\Gamma^2+\lambda \Gamma+\alpha\beta k^2=0$, where $k^2=k_ik^i$.  Then, stability in the LRF implies the conditions 
\bml
\label{toy_LRF}
\bea
&&\alpha\beta\ge0,\\
&&\lambda\ge 0. 
\eea
\eml
As for the boosted frame obtained by the Lorentz transform $\Gamma\to\gamma(\Gamma+iv^ik_i)$ and $k^2\to \Gamma^2+k^2-\gamma^2(\Gamma+i v^ik_i)^2$, the first root is $\Gamma=-iv^ik_i$, which is stable, while the remaining two roots demand (after a long but straightforward computation) 
\bml
\label{toy_Boosted}
\bea
&&\lambda\ge 0,\\
&&0\le \alpha\beta\le 1.
\eea
\eml

To verify stability via the stability theorem proven in this paper, we must verify conditions where \eqref{toy1} is causal and if the matrix $\Phi_\alpha \mathbb{A}^\alpha$ (with $\Phi_\alpha=\Lambda n_\alpha+\zeta_\alpha$, $n$ and $\zeta$ are the unitary timelike and spacelike covectors defined in the text) has a complete set of eigenvectors in $\mathbb{R}^5$. Proposition I guarantees that if \eqref{toy1} is causal, then $\Lambda\in\mathbb{R}$. In order to study causality, we compute the characteristics $\xi_\alpha$ of the system, which reduces to the roots of $\det(\mathbb{A}^\alpha\xi_\alpha)=(u^\alpha\xi_\alpha)^3[(u^\beta\xi_\beta)^2 -\alpha\beta \Delta^{\mu\nu}\xi_\mu \xi_\nu]=0$. Causal roots must be real and obey $\xi_\mu \xi^\mu\ge 0$, which gives the conditions $0\le \alpha\beta\le1$. These conditions, together with stability in the LRF, coincides with the conditions obtained by means of the above direct calculation. However, if we did not know, a priori, the conditions for stability in any frame (which is the case when considering higher order polynomials for the modes), we would still have to obtain the eigenvectors of 
\be 
\Phi_\alpha \mathbb{A}^\alpha=\bbm
u^\alpha \Phi_\alpha & -\alpha \Delta^\alpha_\nu \Phi_\alpha \\
-\beta \Delta^{\mu\alpha}\Phi_\alpha & u^\alpha\Phi_\alpha \delta^\mu_\nu  
\ebm. 
\ee
We can do it firstly by obtaining the eigenvalues $\Lambda$, which may be easily obtained by changing $\xi_\alpha\to \Phi_\alpha$ in the computation of the characteristics. With that result one obtains the eigenvalue $\Lambda^{(1)}$ that is the root of $u^\alpha\Phi^{(1)}_\alpha=0$ with multiplicity 3 and the eigenvalue $\Lambda^{(2)}_\pm$, which give the 2 roots of $(u^\beta\Phi^{(2)}_{\pm\beta})^2 -\alpha\beta \Delta^{\mu\nu}\Phi^{(2)}_{\pm\mu} \Phi^{(2)}_{\pm\nu}=0$. The corresponding eigenvectors are:
\begin{itemize}
\item For $u^\alpha\Phi^{(1)}_\alpha=0$, the system $\Phi^{(1)}_\alpha \mathbb{A}^\alpha r^{(1)}_a=0$ has as eigenvectors the 3 linearly independent vectors given by 
\be
r^{(1)}_a=\bbm 0\\ w_a^\nu\ebm,
\ee
where $\{w_a^\nu\}_{a=1}^3$ is a set of 3 linearly independent vectors orthogonal to the vector $\Delta^{\mu\alpha}\Phi^{(1)}_\alpha$.

\item For $(u^\beta\Phi^{(2)}_{\pm\beta})^2 -\alpha\beta \Delta^{\mu\nu}\Phi^{(2)}_{\pm\mu}\Phi^{(2)}_{\pm\nu}=0$ we assume $\alpha\beta\ne 0$ and obtain the 2 eigenvectors  
\be
\label{toy5}
r^{(2)}_\pm=\bbm u^\alpha \Phi^{(2)}_{\pm\alpha}\\ \beta \Delta^{\nu\alpha}\Phi^{(2)}_{\pm \alpha}\ebm.
\ee
\end{itemize}
(Note that in the special case $\alpha\beta=0$ the root $u^\alpha\Psi_\alpha=0$ is the only root with multiplicity 5. We end up with two distinct situations: first, if $\alpha\ne0$ or $\beta\ne0$ with $\alpha\beta=0$, then one obtains 4 LI eigenvectors as can be seen from \eqref{toy3} and \eqref{toy4}. On the other hand, if $\alpha=\beta=0$, then the system is already diagonal and the theorem applies directly.)
Thus, \eqref{toy5} completes the remaining 2 linearly eigenvectors since $\Lambda_\pm$ are distinct eigenvectors, giving the 5 LI eigenvectors. Then, the stability theorem states that the system is stable if $\lambda\ge0$ and $0<\alpha\beta\le 1$ or if $\lambda\ge0$ and $\alpha=\beta=0$. Note that there is a slight difference from the condition obtained from the direct calculation. To wit, it does not include the case $\alpha\beta=0$ with $\alpha$ or $\beta$ different from zero. The conclusion is that stability in any frame does not necessarily imply strong hyperbolicity. However, strong hyperbolicity+causality+stability in the LRF implies stability in any boosted frame. In other words, stability may occur outside the conditions imposed by the theorem. 

\subsubsection{Applying the stability theorem to the MIS system\label{S:MIS_stability_example}}

As another example of the usefulness of Theorem III, let us briefly comment how it can be used to recover the stability conditions of the MIS equations \cite{Hiscock_Lindblom_stability_1983} in the presence of bulk viscosity. More precisely, we take the MIS-like equations studied in \cite{BemficaDisconziNoronha_IS_bulk} where only bulk viscous effects have been considered. In that case, it was proven that there exist conditions such that the system of PDEs is non-linearly causal and symmetric hyperbolic, hence the principal part of the equations is diagonalizable. The linear version of such equations 
forms a system that is also symmetric 
hyperbolic and the conditions for stability needed for the application of Theorem III
can be shown to agree with those 
found in
 \cite{Hiscock_Lindblom_stability_1983} for the case where only bulk viscosity is present.

\vspace{1cm}

\section{Conditions for linear stability}
\label{sec:stability}

We now apply the theorem proved in the last section to determine conditions that ensure the stability of the hydrodynamic theory proposed in this paper.  Let us first define
\be
D\equiv \rho c_s^2 (\tau_\varepsilon+\tau_Q)+\zeta +\frac{4\eta}{3}+\sigma\kappa_\varepsilon
\ee 
and 
\be
E\equiv\sigma\left[p^\prime_\varepsilon \kappa _s-c_s^2 \kappa_\varepsilon\right]=\sigma T\rho\left [\left (\frac{\partial P}{\partial\varepsilon}\right )_n\left (\frac{\partial (\mu/T)}{\partial n}\right )_\varepsilon-\left (\frac{\partial P}{\partial n}\right )_\varepsilon\left (\frac{\partial (\mu/T)}{\partial \varepsilon}\right )_n\right ],
\ee
where $\kappa_s=(T\rho^2/n)\left [\partial(\mu/T)/\partial\varepsilon\right ]_{\bar{s}}=\kappa_\varepsilon+\kappa_n$, $\kappa_\varepsilon=(T\rho^2/n)\left [\partial(\mu/T)/\partial\varepsilon\right ]_n$, $\kappa_n=(T\rho)\left [\partial(\mu/T)/\partial n\right ]_\varepsilon$, and $p^\prime_\varepsilon=(\partial P/\partial\varepsilon)_n$. Standard thermodynamic identities imply that $p^\prime_\varepsilon \kappa_s-c_s^2 \kappa_\varepsilon>0$, then $E\ge 0$ from (A1). By assuming the Cowling approximation \cite{CowlingApproximation} with $g_{\mu\nu} =\eta_{\mu\nu}=diag(-1,1,1,1)$ and $\delta g_{\mu\nu} = 0$, we find that: \\
\emph{The system described by \eqref{EOMcurrent} is linearly stable if it is causal within the strict form of the inequalities in \eqref{C_conditions} together with the additional restriction $\eta>0$ in (A1) and
\bml
\label{stability}
\bea
&&(\tau_\varepsilon+\tau_Q)|B|\ge \tau_\varepsilon\tau_Q D\ge\rho c_s^2\tau_\varepsilon\tau_Q(\tau_\varepsilon+\tau_Q),\label{stability_a}\\
&&(\tau_\varepsilon+\tau_Q)|B| D+\rho\tau_\varepsilon\tau_Q(\tau_\varepsilon+\tau_Q) E>\tau_\varepsilon\tau_Q D^2+\rho(\tau _\varepsilon+\tau_Q)^2 C,\label{stability_b}\\
&&c_s^2 D-E\ge \rho c_s^4(\tau_\varepsilon+\tau_Q),\label{stability_c}\\
&&(\tau_\varepsilon+\tau_Q)\left [|B|(c_s^2 D-2E)+2 c_s^2\rho\tau_\varepsilon\tau_Q E+CD\right ]>2c_s^2\rho(\tau_\varepsilon+\tau_Q)^2 C+ \tau _\varepsilon \tau _Q D(c_s^2D-E),\label{stability_d}\\
&&|B|D\left [C (\tau _\varepsilon+\tau _Q)+ E \tau _\varepsilon \tau_Q\right ]+2\rho\tau_\varepsilon \tau _Q(\tau_\varepsilon+ \tau _Q) C E >\rho C^2 (\tau_\varepsilon+ \tau _Q)^2+\tau_\varepsilon \tau_Q(C D^2 +\rho\tau_\varepsilon \tau_Q E^2)\nonumber\\
&&+B^2 E(\tau_\varepsilon+\tau _Q),\label{stability_e}
\eea
\eml
where $B$ and $C$ are given by
\bml
\bea
B&\equiv&-\tau_\varepsilon \left(\rho  c_s^2 \tau _Q+\zeta +\frac{4 \eta}{3} +\sigma \kappa _s\right)-\rho  \tau _P \tau _Q,\\
C&\equiv&\tau_P \left(\rho  c_s^2 \tau_Q+\sigma  \kappa_s\right)-\beta_\varepsilon \left ( \zeta +\frac{4\eta}{3} \right ),
\eea
\eml
as in Eq.\ \eqref{ABC}, with $|B|=-B> 0$ from \eqref{C_condition_d} in the strict form.
}

To prove the statement above, as before we may expand the perturbations $\delta\Psi= \left(\delta \varepsilon,\delta u^\mu, \delta n\right)$ in Fourier modes by means of the substitution $\delta\Psi(X)\to \exp[T(\Gamma t+k_i x^i)]\delta\Psi(K)$, where $K^\mu=(i\Gamma,k^i)$ is dimensionless due to the introduction of background temperature $T$ in the exponent. We begin by proving stability in the local rest frame, where the modes are the roots of the shear and sound polynomials
\bml
\label{modes}
\bea
\text{Shear channel:} &&\,\, \bar{\tau}_Q\Gamma^2+\bar{\eta}k^2+\Gamma=0,\label{shear}\\
\text{Sound channel:} &&\,\, a_0\Gamma^5 + a_1\Gamma^4+a_2\Gamma^3+a_3\Gamma^2+a_4\Gamma+a_5=0,\label{sound}
\eea    
\eml
where $k^2=k^ik_i$ and
\bml
\bea 
a_0&=&\bar{\tau}_\varepsilon\bar{\tau}_Q,\\
a_1&=&\bar{\tau}_\varepsilon+\bar{\tau}_Q,\\
a_2&=&1+k^2|\bar{B}|,\\
a_3&=&k^2 \bar{D},\\
a_4&=&c_s^2 k^2+k^4 \bar{C},\\
a_5&=& k^4 \bar{E}.
\eea
\eml
We defined the dimensionless quantities $\bar{\tau}_{Q}=T\tau_{Q} $, $\bar{\tau}_{\varepsilon}=T\tau_{\varepsilon} $, $\bar{\eta}=T\eta/\rho$, $\bar{B}=(T^2/\rho)B$, $\bar{C}=(T^2/\rho)C$, $\bar{D}=(T/\rho)D$, and $\bar{E}=(T/\rho)E$. From the second inequality in \eqref{C_condition_d} in its strict form one obtains that $\bar{B}< 0$ (see the definition of $a_2$). 
The analysis of stability in the LRF goes as follows:

\medskip

{\bf Shear stability conditions:} The second order polynomial \eqref{shear} has two roots with $\Gamma_R\le0$ only if $\tau_Q>0$ and $\eta\ge0$, which is in accordance with assumption (A1). One can see that $\tau_Q$ clearly acts as a relaxation time (the same role is played by the shear relaxation time coefficient $\tau_\pi$ present in MIS theory) for the shear channel, which ensures causality. In fact, the condition $\tau_Q>0$ is clear since the leading contribution to the non-hydrodynamic frequency in this channel goes as $1/\tau_Q $ at zero wavenumber. 

\medskip

{\bf Sound stability conditions:} As for the sound channel in the rest frame, by means of the Routh-Hurwitz criterion \cite{gradshteyn2007}, the necessary and sufficient conditions for $\Gamma_R< 0$ are (i) $a_0,a_1>0$, (ii) $a_1a_2-a_0a_3>0$, (iii) $a_3(a_1a_2-a_0a_3)-a_1(a_1a_4-a_0a_5)> 0$, (iv) $(a_1a_4-a_0a_5)[a_3(a_1a_2-a_0a_3)-a_1(a_1a_4-a_0a_5)]-a_5(a_1a_2-a_0a_3)^2> 0$, and (v) $a_5> 0$. Condition (i) is already satisfied from (A1). Condition (ii) corresponds to the first inequality in \eqref{stability_a}, while (iii) is the second inequality in \eqref{stability_a} and \eqref{stability_b}. Condition (iv) corresponds to \eqref{stability_c}--\eqref{stability_e}. Given that $E\ge 0$, thus, when $E=0$ and (i)--(iv) are observed, then $\Gamma_R\le 0$, which is in accordance with stability. Also, if $k=0$, then $\Gamma_R\le 0$ (three zero roots and two negative roots) because $a_0,a_1,a_2>0$ from (A1). Hence, the system is linearly stable in the local rest frame. 

We remark that our system displays three types of hydrodynamic modes and three non-hydrodynamic modes. In the small $k$ expansion that typically defines the linearized hydrodynamic regime, our shear channel gives a diffusive hydrodynamic mode with (real) frequency $\omega(k) = - i k^2 \eta/(\varepsilon+P) + \ldots$ while in the sound channel one finds proper sound waves with $\omega(k) = \pm c_s k - i k^2 \Gamma_{s}/2+\ldots$ and also a heat diffusion mode with $\omega(k)=-iD k^2 +\ldots$, where $D\sim \sigma$, and $\Gamma_{s} = \Gamma_{s}(\eta,\zeta,\sigma)$ just as in Eckart theory (see Ref.\ \cite{Hoult:2020eho} for their detailed expressions). Therefore, our theory has the same physical content of Eckart's theory in the hydrodynamic regime. On the other hand, the shear channel has a non-hydrodynamic mode with frequency given by $\omega(k) = -i/\tau_Q +\ldots$ while the sound channel has two non-hydrodynamic modes with frequency $\omega(k)=-i/\tau_\varepsilon+\ldots$ and $\omega(k)=-i/\tau_Q+\ldots$ in the low $k$ limit. These non-hydrodynamic modes parametrize the UV behavior of the system in a way that ensures causality and stability, making sure that the theory is well defined (though, of course, not accurate) even outside the typical domain of validity of hydrodynamics. 

The complete proof of linear stability demands an analysis of the linearized system around an equilibrium state at nonzero velocity. In this regard, we shall use the results presented in Sec.\ \ref{linearstability}. We first write the system in \eqref{EOMcurrent} as a first-order linear system of PDE's. Then, since we already have proven causality and also linear stability in the LRF, it remains to be shown that the first order counterpart of \eqref{EOMcurrent} is diagonalizable in the sense of \eqref{s_2-4}. This is done below.

\medskip

{\bf A first order system:} following Sec.\ \ref{firstorder}, we may define $\delta V=u^\alpha\partial_\alpha\delta\varepsilon$, $\delta \mathcal{V}^\mu=\Delta^{\mu\alpha}\partial_\alpha\delta\varepsilon$, $\delta W=u^\alpha\partial_\alpha \delta n$, $\delta \mathcal{W}^\mu=\Delta^{\mu\alpha}\partial_\alpha \delta W$, $\delta S^\mu=u^\alpha\partial_\alpha \delta u^\mu$, $\delta\mathcal{S}^{\nu}_{\lambda}=\Delta^{\alpha}_\lambda\partial_\alpha \delta u^\nu$. Since the current is ideal, i.e., $J^\mu=n u^\nu$, then the linearized conservation equation $\partial_\mu \delta J^\mu=\delta W+n\delta \mathcal{S}^\nu_\nu=0$ enables us to eliminate $\delta W$ from the new system of equations. Hence, the first order equations become
\bml
\label{first_order_EOM}
\bea
&&\tau_\varepsilon u^\alpha\partial_\alpha \delta V+\rho\tau_Q\partial_\alpha\delta S^\alpha+\beta_\varepsilon\partial_\alpha \delta\mathcal{V}^\alpha+\beta_n\partial_\alpha\delta\mathcal{W}^\alpha+\rho\tau_\varepsilon u^\alpha\partial_\alpha \delta\mathcal{S}^\nu_\nu+\delta V+\rho\delta \mathcal{S}^\nu_\nu=0,\\
&&\tau_P \Delta^{\mu\alpha}\partial_\alpha\delta V+\rho\tau_Q u^\alpha\partial_\alpha\delta S^\mu+\beta_\varepsilon u^\alpha\partial_\alpha\delta\mathcal{V}^\mu+\beta_n u^\alpha\partial_\alpha\delta\mathcal{W}^\mu+\Pi^{\mu\lambda\alpha}_\nu\partial_\nu \delta\mathcal{S}^\nu_\lambda+p'_\varepsilon\delta\mathcal{V}^\mu+p'_n \delta\mathcal{W}^\mu+\rho\delta S^\mu=0,\\
&&u^\alpha\partial_\alpha \delta \mathcal{V}^\mu-\Delta^{\mu\alpha}\partial_\alpha\delta V=0,\label{EOM_c}\\
&&u^\alpha\partial_\alpha \delta \mathcal{W}^\mu+n\Delta^{\mu\alpha}\partial_\alpha\delta \mathcal{S}^\nu_\nu=0,\label{EOM_d}\\
&&u^\alpha\partial_\alpha\delta\mathcal{S}^\mu_\lambda-\Delta^{\alpha}_\lambda\partial_\alpha\delta S^\mu=0,\label{EOM_e}
\eea
\eml
where $p'_n=(\partial P/\partial n)_\varepsilon$ and 
\be
\Pi^{\mu\lambda\alpha}_\nu=-\eta\left (\Delta^{\mu\lambda}\delta^\alpha_\nu+\Delta^{\lambda\alpha}\delta^\mu_\nu\right )+\left (\rho\tau_P-\zeta+\frac{2\eta}{3}\right )\Delta^{\mu\alpha}\delta^\lambda_\nu.
\ee
The supplemental equations \eqref{EOM_c}--\eqref{EOM_e} come from the identities $\partial_\alpha\partial_\beta\delta\varepsilon-\partial_\beta\partial_\alpha\delta\varepsilon=0$, $\partial_\alpha\partial_\beta\delta n-\partial_\beta\partial_\alpha\delta n=0$, and $\partial_\alpha\partial_\beta\delta u^\mu-\partial_\beta\partial_\alpha\delta u^\mu=0$, respectively, when contracted with $u^\alpha\Delta^{\beta\lambda}$. In particular, in Eq.\ \eqref{EOM_d} we have substituted $\delta W=-n\delta \mathcal{S}^\nu_\nu$ that comes from the conservation equation of $J^\mu$. Then, we may write \eqref{first_order_EOM} in matrix form $\mathbb{A}^\alpha\partial_\alpha \delta\Psi(X)+\mathbb{B}\Psi(X)=0$, were $\delta\Psi(X)$ is the $29\times 1$ column matrix with entries $\delta V,\delta S^\nu,\delta\mathcal{V}^\nu,\delta\mathcal{W}^\nu,\delta\mathcal{S}^\nu_0,\delta\mathcal{S}^\nu_1,\delta\mathcal{S}^\nu_2,\delta\mathcal{S}^\nu_3$, 
\be
\label{A}
\mathbb{A}^\alpha=\bbm
\tau_\varepsilon u^\alpha & \rho\tau_Q\delta^\alpha_\nu & \beta_\varepsilon\delta^\alpha_\nu & \beta_n \delta^\alpha_\nu & \rho\tau_\varepsilon u^\alpha \delta_\nu^0 & \rho\tau_\varepsilon u^\alpha \delta_\nu^1 & \rho\tau_\varepsilon u^\alpha \delta_\nu^2 & \rho\tau_\varepsilon u^\alpha \delta_\nu^3\\
\tau_P \Delta^{\mu\alpha} & \rho\tau_Q u^\alpha\delta^\mu_\nu & \beta_\varepsilon u^\alpha \delta^\mu_\nu & \beta_n u^\alpha \delta^\mu_\nu & \Pi^{\mu0\alpha}_\nu & \Pi^{\mu1\alpha}_\nu & \Pi^{\mu2\alpha}_\nu & \Pi^{\mu3\alpha}_\nu\\
-\Delta^{\mu\alpha} & 0_{4\times 4} & u^\alpha \delta^\mu_\nu & 0_{4\times 4} & 0_{4\times 4} & 0_{4\times 4} & 0_{4\times 4} & 0_{4\times 4}\\
0_{4\times 1} & 0_{4\times 4} & 0_{4\times 4} & u^\alpha \delta^\mu_\nu & n\Delta^{\mu\alpha}\delta_\nu^0 & n\Delta^{\mu\alpha}\delta_\nu^1 & n\Delta^{\mu\alpha}\delta_\nu^2 & n\Delta^{\mu\alpha}\delta_\nu^3\\
0_{4\times 1} & -\Delta^{\alpha}_0\delta^\mu_{\nu} & 0_{4\times 4} & 0_{4\times 4} & u^\alpha \delta^\mu_\nu & 0_{4\times 4} & 0_{4\times 4} & 0_{4\times 4}\\
0_{4\times 1} & -\Delta^{\alpha}_1\delta^\mu_\nu & 0_{4\times 4} & 0_{4\times 4} & 0_{4\times 4} & u^\alpha \delta^\mu_\nu & 0_{4\times 4} & 0_{4\times 4}\\
0_{4\times 1} & -\Delta^{\alpha}_2\delta^\mu_\nu & 0_{4\times 4} & 0_{4\times 4} & 0_{4\times 4} & 0_{4\times 4} & u^\alpha \delta^\mu_\nu & 0_{4\times 4}\\
0_{4\times 1} & -\Delta^{\alpha}_3\delta^\mu_\nu & 0_{4\times 4} & 0_{4\times 4} & 0_{4\times 4} & 0_{4\times 4} & 0_{4\times 4} & u^\alpha \delta^\mu_\nu\\
\ebm,
\ee
and
\be
\mathbb{B}=\bbm
1 & 0_{1\times 4} & 0_{1\times 4} & 0_{1\times 4} & \rho\delta_\nu^0 & \rho\delta_\nu^1 & \rho\delta_\nu^2 & \rho\delta_\nu^3\\
0_{4\times 1} & \rho\delta^\mu_\nu & p'_\varepsilon\delta^\mu_\nu & p'_n\delta^\mu_\nu & 0_{4\times 4} & 0_{4\times 4} & 0_{4\times 4} & 0_{4\times 4}\\
0_{4\times 1} & 0_{4\times 4} & 0_{4\times 4} & 0_{4\times 4} & 0_{4\times 4} & 0_{4\times 4} & 0_{4\times 4} & 0_{4\times 4}\\
0_{4\times 1} & 0_{4\times 4} & 0_{4\times 4} & 0_{4\times 4} & 0_{4\times 4} & 0_{4\times 4} & 0_{4\times 4} & 0_{4\times 4}\\
0_{4\times 1} & 0_{4\times 4} & 0_{4\times 4} & 0_{4\times 4} & 0_{4\times 4} & 0_{4\times 4} & 0_{4\times 4} & 0_{4\times 4}\\
0_{4\times 1} & 0_{4\times 4} & 0_{4\times 4} & 0_{4\times 4} & 0_{4\times 4} & 0_{4\times 4} & 0_{4\times 4} & 0_{4\times 4}\\
0_{4\times 1} & 0_{4\times 4} & 0_{4\times 4} & 0_{4\times 4} & 0_{4\times 4} & 0_{4\times 4} & 0_{4\times 4} & 0_{4\times 4}\\
0_{4\times 1} & 0_{4\times 4} & 0_{4\times 4} & 0_{4\times 4} & 0_{4\times 4} & 0_{4\times 4} & 0_{4\times 4} & 0_{4\times 4}
\ebm.
\ee
We must now obtain the eigenvectors of \eqref{s_2-1}. However, note that $\mathbb{A}^\alpha$ above is exactly the same as the matrix $\mathbb{A}^\alpha_m$ in \eqref{Sobolev_8} with the difference that now the coefficients of $\mathbb{A}^\alpha$ are constants. We have already proven in Sec.\ \eqref{sec:hyperbolicity} that the matrix $\mathbb{A}^\alpha_m$ in Eq.\ \eqref{s_2-1} has real eigenvalues and a complete set of eigenvectors in $\mathbb{R}^{29}$. The same solution is true for $\mathbb{A}^\alpha$ in \eqref{s_2-1} if we change $\xi_\alpha\to n_\alpha$ (and also $\mathbb{A}^\alpha_m\to \mathbb{A}^\alpha$) in the results for the matter sector in Sec.\ \eqref{sec:hyperbolicity}. Thus, the $29\times 29$ matrix $(-n_\alpha \mathbb{A}^\alpha)\zeta_\beta \mathbb{A}^\beta$ is diagonalizable, completing the requirements from Theorem III. This shows that the theory is linearly stable in any other reference frame $\mathcal{O}'$ connected via a Lorentz transformation. Therefore, one then obtains that our set of  linearized second order PDE's is stable in any equilibrium state.

\subsection{Fulfilling the causality, local well-posedness, and linear stability conditions}
\label{S:Non_empty} 

We now give a simple example that illustrates that the set of linear stability conditions (and consequently, causality and local well-posedness, since those are part of the linear stability conditions) is not empty. Let us analyze the case where $\tau_Q=\tau_\varepsilon$ and $\tau_P=c_s^2\tau_\varepsilon$, assuming an equation of state $P=P(\varepsilon)$, with  $c_s^2=p'_\varepsilon=1/2$. Also, assume that $\zeta+4\eta/3>0$ (their specific values are not relevant as far as they are positive and $\eta>0$ for the sake of the stability and well-posedness theorems). Then, one may easily verify that the causality conditions \eqref{C_conditions} hold in their strict form, as required, and that the remaining conditions \eqref{stability} are also observed when $\rho\tau_\varepsilon=8(\zeta+4\eta/3)$, $\kappa_\varepsilon=\kappa_s/2=1/4$, and in the three different situations, namely, $\sigma/(\zeta+4\eta/3)=0,\,1/4$, and 1.


\section{Conclusions and Outlook}
\label{sec:conclusions}

In this work, we presented the first generalization of relativistic Navier-Stokes theory that simultaneously satisfies the following properties: the system, with or without coupling to Einstein's equations, is causal, strongly hyperbolic, and 
locally well-posed,
see the content of Theorem I and II); equilibrium states in flat spacetime are stable (consequence of Theorem III); all dissipative contributions (shear viscosity, bulk viscosity, and heat conductivity) are included; and finally the effects from nonzero baryon number are also taken into account. All of the above holds without any simplifying symmetry assumptions and are mathematically rigorously established. 
In addition, entropy production is non-negative in the regime of validity of this effective theory.

This is accomplished in a natural way using a generalized Navier-Stokes theory containing only the original hydrodynamic variables, which is different than other approaches where the space of variables is extended (such as in M\"uller-Israel-Stewart theory). However, it is important to remark that the meaning of the hydrodynamic variables in our work is different than in standard approaches, such as \cite{LandauLifshitzFluids} and \cite{EckartViscous}. In fact, in the context of the formalism put forward by Bemfica, Disconzi, Noronha and Kovtun in Refs.\ \cite{Bemfica:2017wps,Kovtun:2019hdm,Bemfica:2019knx}, our formulation uses a definition for the hydrodynamic variables (i.e. our choice of hydrodynamic frame) that is not standard as there are nonzero out of equilibrium corrections to the energy density and there is energy/heat diffusion even at zero baryon density. Despite these necessary differences (imposed by causality and stability), the theory still provides the simplest causal and strongly hyperbolic generalization of Eckart's original theory \cite{EckartViscous}, sharing the same physical properties in the hydrodynamic regime (for instance, both theories have the same spectrum of hydrodynamic modes). However, differently than Eckart's approach, our formulation is fully compatible with the postulates of general relativity and its physical content in dynamical settings can be readily investigated using numerical relativity simulations. In fact, we hope that the framework presented here will provide the starting point for future systematic studies of viscous phenomena in the presence of strong gravitational fields, such as in neutron star mergers.   
 
Motivated by the task of establishing stability of general equilibrium states in flat spacetime, in this work we also proved a new general result (see Theorem III) concerning the stability of relativistic fluids. In fact, we found conditions that causal relativistic fluids should satisfy such that stability around the static equilibrium state directly implies stability in any other equilibrium state at nonzero background velocity. 
Theorem III is very general and its regime of applicability goes beyond BDNK theories and it could also be relevant when investigating the stability properties of other sets of linear equations of motion as well.  In this regard, see the discussion in Section \ref{S:Overview}, and
Sections \ref{S:Toy_model_example}
and \ref{S:MIS_stability_example} for further examples of the applicability of Theorem III.

Our generalized Navier-Stokes theory can be used to understand how matter in general relativity starts to deviate from equilibrium. An immediate application is in the modeling of viscous effects in neutron star mergers. Our approach can be useful in simulations that aim at determining the fate of the hypermassive remnant formed after the merger of neutron stars, hopefully leading to a better quantitative understanding of their evolution and eventual gravitational collapse towards a black hole. Differently than any other approach in the literature, the new features displayed by our formulation and its strongly hyperbolic character make it a suitable candidate to be used in such simulations. This will be especially relevant also when considering how viscous effects may modify the gravitational wave signals emitted soon after the merger \cite{Alford:2017rxf,Most:2021zvc}. In this regard, we remark that previous simulations performed in Ref.\ \cite{Shibata:2017xht} employed a formulation of relativistic viscous hydrodynamics where the key properties studied here (causality, strong hyperbolicity, and local well-posedness) are not known to hold in the nonlinear regime. 

Our work is applicable in the case of baryon-rich matter, such as that formed in neutron star mergers or in low energy heavy-ion collisions. The latter include the experimental efforts in the beam energy scan program at RHIC \cite{STARnote}, the STAR fixed-target program \cite{STARnote}, the HADES experiment at GSI \cite{Galatyuk:2014vha}, the future FAIR facility at GSI \cite{Durante:2019hzd}, and also NICA \cite{Kekelidze:2017tgp}. For a discussion of viscous effects in low energy heavy-ion collisions at nonzero density see \cite{Denicol:2018wdp,Fotakis:2019nbq,Dore:2020jye}. High energy heavy-ion collisions, such as those studied at the LHC, involve a different regime than the one considered here where the net baryon number can be very small and, thus, that case is better understood using a different formulation such as the one proposed in \cite{Hoult:2020eho}, also in the context of the BDNK formalism.  
    
In our approach, we only take into account first order derivative corrections to the dynamics. Therefore, the domain of validity of our theory is currently limited by the size of such deviations. Hence, further work is needed to extend our analysis, incorporating higher order derivative corrections, to get a better understanding of what happens as the system gets farther and farther from equilibrium. In this context, it would be interesting to extend our equations to include second order corrections and consider also, more generally, the large order behavior of the gradient expansion in an arbitrary hydrodynamic frame. The latter will be different than most approaches to the gradient expansion since in BDNK the constitutive relations contain time derivatives even in the local rest frame of the fluid. This essential difference has important consequences in a kinetic theory formulation, see the original references \cite{Bemfica:2017wps,Bemfica:2019knx}. The large order behavior of the relativistic gradient series has been recently the focus of several works  \cite{Heller:2013fn,Basar:2015ava,Buchel:2016cbj,Denicol:2016bjh,Heller:2016rtz,Strickland:2017kux, Casalderrey-Solana:2017zyh, Denicol:2017lxn,Denicol:2018pak,Withers:2018srf,Behtash:2019qtk, Denicol:2019lio,Grozdanov:2019kge,Grozdanov:2019uhi,Heller:2020uuy}, and it would be interesting to extend such analyses to include the type of theories investigated here. 

There are a number of ways in which our work could be extended or improved. First, it would be useful to obtain a better qualitative understanding why some hydrodynamic frames (such as the Landau-Lifshitz frame or the Eckart frame) are not compatible with causality and stability in the BDNK approach, given that the situation is different in other formulations. In fact, the Landau frame seems to display no significant issues in the case of
MIS-like theories in the nonlinear regime at least at zero chemical potential, as demonstrated in \cite{Bemfica:2020xym}. Perhaps a more in depth investigation of how BDNK emerges in kinetic theory, going beyond the original work done in \cite{Bemfica:2017wps,Bemfica:2019knx}, can be useful in this regard (see also the recent work \cite{Hoult:2021gnb}). Also, it would be interesting to use the BDNK approach to investigate causality and stability in more exotic cases, such as in relativistic superfluids. Furthermore, the inclusion of electromagnetic field effects in the dynamics of relativistic viscous fluids can also be of particular relevance, especially in the context of neutron star mergers \cite{Ciolfi:2020cpf} and high-energy heavy ion collisions \cite{Skokov:2009qp}. This problem has been recently investigated using other formulations of viscous fluid dynamics, see for instance Refs.\ \cite{Chandra:2015iza,Hernandez:2017mch,Grozdanov:2016tdf,Denicol:2018rbw,Most:2021rhr,Most:2021uck}, and also most recently in the BDNK approach in Ref.\ \cite{Armas:2022wvb}. Consistent modeling of relativistic viscous fluid dynamics coupled to electromagnetic fields can also be relevant to determine the importance of dissipative processes in the dynamics and radiative properties of slowly accreting black holes, as discussed in \cite{Chandra:2015iza}. 

Further work needs to be done to understand the global in-time features of solutions of relativistic viscous fluid dynamics. For instance, one may investigate the presence of shocks, which is a topic
widely investigated in the context of ideal fluids \cite{ChristodoulouShocks, ChristodoulouShockDevelopment, DisconziSpeckRelEulerNull, SpeckBook, ChoquetBruhatGRBook} and was done in \cite{Disconzi:2020ijk} for the 
MIS theory (see Section \ref{S:Overview} for further discussion on shocks). The importance of hydrodynamic shocks has been recognized both in an astrophysical setting
\cite{Chandra:2015iza} as well as in study of jets in the quark-gluon plasma \cite{CasalderreySolana:2004qm,Satarov:2005mv,Gubser:2007ga,Betz:2008wy,Betz:2008ka,Bouras:2009nn,Bouras:2010hm,Betz:2010qh,Bouras:2014rea,Tachibana:2014lja,Yan:2017rku,Tachibana:2017syd,Tachibana:2019hrn}. We also remark that one task that we have not done here was the construction of initial data for the full Einstein plus fluid system by solving the Einstein constraint equations. We believe that standard arguments to handle the constraints \cite{ChoquetBruhatGRBook} will be applicable in our case. This will be investigated in detail in a future work.

We believe our work will also be relevant to give insight into the physics of turbulent fluids embedded in general relativity. The fact that the equations of motion of the viscous fluid must be hyperbolic in relativity stands in sharp contrast to the parabolic nature of the non-relativistic Navier-Stokes equations, usually employed in studies of turbulence. Recent works in Refs.\ \cite{Eyink:2017zfz,Calzetta:2020wzr} tackled the problem of turbulence in the relativistic regime and our formulation may be very useful in this regard, as it provides a simple strongly hyperbolic generalization of Eckart's theory that is fully compatible with general relativity.    

In summary, in this paper we propose a new solution to the question initiated by Eckart in 1940 concerning the motion of viscous fluids in relativity. Our approach is rooted in well-known physical principles and solid mathematics, displays a number of desired properties, and extends the state-of-the-art of the field in a number of ways. Potential applications of the formalism presented here spread across a numbers of areas, including astrophysics, nuclear physics, cosmology, and mathematical physics. This work establishes for the first time a common unifying framework, from heavy-ion collisions to neutron stars, that can be used to discover the novel properties displayed by ultradense baryonic matter as it evolves in spacetime.

\section*{Acknowledgments} We thank P.~Kovtun, G.~S.~Denicol, and L.~Gavassino for insightful discussions. We also thank the anonymous referees
and the Editor for providing valuable feedback that helped improve the manuscript. 
MMD is partially supported by a Sloan Research Fellowship provided by the Alfred P. Sloan foundation, NSF grant DMS-2107701, and a Dean's Faculty Fellowship. JN is partially supported by the U.S. Department of Energy, Office of Science, Office for Nuclear Physics under Award No. DE-SC0021301. 

\appendix

\section{Proof of Theorem I}
\label{Theorem_I}

We only consider the 10 independent components of the metric and, thus, this system of equations can be written in terms of a $16\times 1$ column vector $\Psi=(\varepsilon,n,u^\nu,g_{\mu\nu})$ and its equation of motion in \eqref{EOM} can be expressed in matrix form as $\mathfrak{M}(\partial)\Psi=\mathfrak{N}$, where $\mathfrak{N}$ contains the $\mathcal{B}$ terms that do not enter in the principal part. The matrix $\mathfrak{M}(\partial)$ is given by
\be
\label{Matrix}
\mathfrak{M}(\partial)=\bbm
\mathbb{M}(\partial) & \mathfrak{b}(\partial)\\
0_{6\times 10} & I_{10}g^{\alpha\beta}\partial^2_{\alpha\beta}
\ebm
\ee
where the $6\times 10$ matrix $\mathfrak{b}(\partial)$ contains the $\tilde{\mathcal{B}}$ terms and
\be
\label{Matrix_A}
\mathbb{M}(\partial)=\bbm
0 & u^\alpha u^\beta & n\delta^{(\alpha}_\nu u^{\beta)} \\
(\tau_\varepsilon u^\alpha u^\beta+\beta_\varepsilon\Delta^{\alpha\beta})&\beta_n\Delta^{\alpha\beta}&\rho(\tau_\varepsilon+\tau_Q)u^{(\alpha}\delta^{\beta)}_\nu\\
(\beta_\varepsilon+\tau_P) u^{(\alpha}\Delta^{\beta)\mu} & \beta_n u^{(\alpha}\Delta^{\beta)\mu} & 
C^{\mu\alpha\beta}_\nu
\ebm\partial^2_{\alpha\beta}.
\ee
The system's characteristics are obtained by replacing $\partial_\alpha\to\xi_\alpha$ and 
determining the roots of $\det[\mathfrak{M}(\xi)]=0$. The system is causal when the solutions for $\xi_\alpha=(\xi_0(\xi_i),\xi_i)$ are such that (C1) $\xi_\alpha$ is real and (C2) $\xi_\mu \xi^\mu \geq 0$
\cite{ChoquetBruhatGRBook}. It is easy to see that $\det[\mathfrak{M}(\xi)]=  (\xi_\alpha\xi^\alpha)^{10}\det[\mathbb{M}(\xi)]$. The roots associated with the vanishing of the overall factor $(\xi_\alpha\xi^\alpha)^{10}=0$ coming from the gravitational sector are clearly causal. The remaining roots come from
$\det[\mathbb{M}(\xi)]=0$, which we will investigate next.

We first define $b\equiv u^\alpha\xi_\alpha$ and $v^\alpha\equiv \Delta^{\alpha\beta}\xi_\beta$, which gives $\xi_\alpha=-bu_\alpha+v_\alpha$ and $\xi_\alpha\xi^\alpha=-b^2+v\cdot v$, where $v\cdot v=\Delta^{\alpha\beta}\xi_\alpha\xi_\beta$. We proceed by also defining the tensor
\be
\label{Matrix_D}
\mathfrak{D}^\mu_\nu=\mathcal{C}^{\mu\alpha\beta}_\nu\xi_\alpha\xi_\beta=\left(\tau_P \rho -\zeta -\frac{\eta}{3}\right ) v^\mu\xi_\nu+[\rho\tau_Q b^2-\eta(v\cdot v)]\delta^\mu_\nu,
\ee
which gives 
\bml
\label{det}
\bea
&&\det[\mathbb{A}(\xi)]=\det\bbm
0 & b^2 & nb\xi_\nu\\
\tau_\varepsilon b^2+\beta_\varepsilon(v\cdot v) & \beta_n(v\cdot v) & \rho(\tau_\varepsilon+\tau_Q)b v_\nu\\
(\beta_\varepsilon+\tau_P) bv^\mu & \beta_n b v^\mu & \mathfrak{D}^{\mu}_\nu 
\ebm\nonumber\\
&&=-b^2[\rho\tau_Q b^2-\eta(v\cdot v)]^3\left [Ab^4+Bb^2(v\cdot v)+C(v\cdot v)^2\right ]\label{det_a}\\
&&=-\rho^4\tau_Q^4\tau_\varepsilon\,(u^\alpha\xi_\alpha)^2 \prod_{a=1,\pm}\left [(u^\alpha\xi_\alpha)^2-c_a\Delta^{\alpha\beta}\xi_\alpha\xi_\beta\right ]^{n_a},\label{det_b}
\eea
\eml
where, to shorten notation in \eqref{det_a} we defined
\bml
\label{ABC}
\bea
A&\equiv&\rho\tau_\varepsilon\tau_Q,\\
B&\equiv&-\tau_\varepsilon \left(\rho  c_s^2 \tau _Q+\zeta +\frac{4 \eta}{3} +\sigma \kappa _s\right)-\rho  \tau _P \tau _Q,\\
C&\equiv&\tau_P \left(\rho  c_s^2 \tau_Q+\sigma  \kappa_s\right)-\beta_\varepsilon \left ( \zeta +\frac{4\eta}{3} \right ),
\eea
\eml
and used the fact that $\beta_\varepsilon+n\beta_n/\rho=\tau_Q c_s^2+\sigma\kappa_s/\rho$. In Eq.\ \eqref{det_a} it becomes evident that assumption (A1) guarantees that $v^\mu \ne 0$, eliminating one of the possible acausal roots. From \eqref{det_a} to \eqref{det_b} we defined $n_1=3$, $n_\pm=1$, $c_1=\frac{\eta}{\rho\tau_s}$, and $c_\pm=\frac{-B\pm\sqrt{B^2-4AC}}{2A}$. Note that since $\xi^\alpha\xi_\alpha=-b^2+(v\cdot v)$, the roots in \eqref{det_b} can be cast as $b^2=c_a (v\cdot v)$. Then, (C1) demands that $c_a\in\mathbb{R}$ together with $c_a\ge 0$ and (C2) that $c_a< 1$ for causality  \footnote{Mathematically,
$c_a=1$, which corresponds to $\xi_\mu \xi^\mu=0$, is allowed for causality. But this would correspond to information propagating at the
speed of light in the matter sector, which physically is not expected. Thus, our results are in fact
a bit stronger than the statement of the theorem in that we obtain causality with the characteristic
speeds of the matter sector strictly less than the speed of light.
In particular, it is because we demand the strict inequality $c_a<1$
that we have strict inequalities in \eqref{C_condition_a}, \eqref{C_condition_d}, and \eqref{C_condition_e}.}, what comes from the fact that the root $b^2=c_a v\cdot v$ must obey $\xi_\mu \xi^\mu=-b^2+v\cdot v=(1-c_a)v\cdot v> 0$. Thus, causality is ensured if $0\le c_a< 1$ in the matter sector. Clearly, the root $b=u^\alpha\xi_\alpha=0$ is causal. Also, the 6 roots related to $c_1$ are causal when \eqref{C_condition_a} is observed. As for the roots $c_{\pm}$, they are real if $B^2-4AC\ge0$, i.e., if the first inequality in \eqref{C_condition_b} holds. On the other hand, $c_\pm\ge 0$ is obtained whenever  $c_-\ge 0$, which is guaranteed if $-B\ge0$ [second inequality in condition \eqref{C_condition_d}] together with $C\ge0$ [second inequality of \eqref{C_condition_a}], while $c_\pm<1$ is ensured if $c_+<1$, which demands that $2A+B> 0$ [first inequality in condition \eqref{C_condition_d}] and $A+B+C>0$ [condition \eqref{C_condition_e}]. 

\hfill $\Box$

\bigskip

We observe that, although we employed the harmonic gauge to calculate
the system's characteristics, the causality established in Theorem I does not depend on any
gauge choices. This follows from well-known properties of Einstein's equations
\cite{WaldBookGR1984}  and the
geometric invariance of the characteristics \cite{Courant_and_Hilbert_book_2}.
See the end of Section  \ref{localwellposedness_section} for further comments in this direction.

The analysis above and the conditions we obtained for causality are valid in the full nonlinear regime of the theory. However, we remark in passing that the principal part concerning only the fluid equations would have exactly the same structure if one were to linearize the fluid dynamic equations about 
equilibrium with nonzero flow in Minkowski spacetime. This is a generic feature of the BDNK approach (at least, when truncated at first order), i.e, 
the analysis of the system's characteristics, and thus of its causality properties,
is formally the same in the nonlinear regime and in the linearization
about a generic equilibrium state. This is not, however, a general feature of hydrodynamic models as it does not hold in MIS-like theories. In fact, as discussed at length in \cite{BemficaDisconziNoronha_IS_bulk,Bemfica:2020xym}, in MIS the thermodynamic fluxes explicitly enter in the calculation of the characteristics, but they are not present
in the linear analysis.

\section{Proof of Proposition I}
\label{Proposition_I}

To prove (i) we may compute the determinant $\det(\xi_\alpha\mathfrak{A}^\alpha)=\det(\xi_\alpha \mathbb{A}_m^\alpha)\det(\xi_\alpha \mathbb{A}^\alpha_g)(u^\alpha\xi_\alpha)^{16}$. Note that $u^ \alpha\xi_\alpha\ne 0$ if $\xi$ is timelike. We must then look into the matter and gravity sector in what follows. We again define $b=u^\alpha\xi_\alpha$ and $v^\mu=\Delta^{\mu\alpha}\xi_\alpha$, $v\cdot v=\Delta^{\mu\nu}\xi_\mu\xi_\nu$, and  introduce 
\be
\Xi^\mu_\nu=v_\lambda\Pi^{\mu\lambda\alpha}_\nu\xi_\alpha
=-\eta(v\cdot v)\delta^\mu_\nu-\eta v^{\mu}\xi_\nu+\left (\rho\tau_P-\zeta+\frac{2\eta}{3}\right )v^{\mu}v_\nu
\ee
to obtain
\bea
\label{Sobolev_140}
&&\det(\xi_\alpha \mathbb{A}^\alpha_m)=
\det\bbm
\tau_\varepsilon b & \rho\tau_Q\xi_\nu & \beta_\varepsilon\xi_\nu & \beta_n \xi_\nu & \rho\tau_\varepsilon b \delta_\nu^0 & \rho\tau_\varepsilon b \delta_\nu^1 & \rho\tau_\varepsilon b \delta_\nu^2 & \rho\tau_\varepsilon b \delta_\nu^3\\
\tau_P v^{\mu} & \rho\tau_Q b\delta^\mu_\nu & \beta_\varepsilon b \delta^\mu_\nu & \beta_n b \delta^\mu_\nu & \Pi^{\mu0\alpha}_\nu\xi_\alpha & \Pi^{\mu1\alpha}_\nu\xi_\alpha & \Pi^{\mu2\alpha}_\nu\xi_\alpha & \Pi^{\mu3\alpha}_\nu\xi_\alpha\\
-v^{\mu} & 0_{4\times 4} & b \delta^\mu_\nu & 0_{4\times 4} & 0_{4\times 4} & 0_{4\times 4} & 0_{4\times 4} & 0_{4\times 4}\\
0_{4\times 1} & 0_{4\times 4} & 0_{4\times 4} & b \delta^\mu_\nu & n v^{\mu}\delta_\nu^0 & n v^{\mu}\delta_\nu^1 & n v^{\mu}\delta_\mu^2 & n v^{\mu}\delta_\nu^3\\
0_{4\times 1} & -v_0\delta^\mu_{\nu} & 0_{4\times 4} & 0_{4\times 4} & b \delta^\mu_\nu & 0_{4\times 4} & 0_{4\times 4} & 0_{4\times 4}\\
0_{4\times 1} & -v_1\delta^\mu_\nu & 0_{4\times 4} & 0_{4\times 4} & 0_{4\times 4} & b \delta^\mu_\nu & 0_{4\times 4} & 0_{4\times 4}\\
0_{4\times 1} & -v_2\delta^\mu_\nu & 0_{4\times 4} & 0_{4\times 4} & 0_{4\times 4} & 0_{4\times 4} & b \delta^\mu_\nu & 0_{4\times 4}\\
0_{4\times 1} & -v_3\delta^\mu_\nu & 0_{4\times 4} & 0_{4\times 4} & 0_{4\times 4} & 0_{4\times 4} & 0_{4\times 4} & b \delta^\mu_\nu
\ebm\nonumber\\
&&=b^{19}
\det\bbm
\tau_\varepsilon b^2+\beta_\varepsilon(v\cdot v) & b^2(\rho\tau_Q\xi_\nu+\rho\tau_\varepsilon v_\nu)-n\beta_n (v\cdot v)v_\nu \\
(\tau_P+\beta_\varepsilon) v^{\mu} & \rho\tau_Q b^2\delta^\mu_\nu+\Xi^\mu_\nu-n\beta_n v^\mu v_\nu
\ebm\nonumber\\
&&=b^{19}\left [\rho\tau_Q b^2-\eta(v\cdot v)\right ]^3
\left [Ab^4+B b^2(v\cdot v)+C(v\cdot v)^2\right ]\nonumber\\
&&=\rho^4\tau_Q^4\tau_\varepsilon b^{19} \prod_{a=1,\pm}\left [b^2-c_a(v\cdot v)\right ]^{n_a},
\eea
where, as we have obtained in \eqref{det}, \eqref{ABC}, and in the text below it,
\bml
\bea
A&\equiv&\rho\tau_\varepsilon\tau_Q,\\
B&\equiv&-\tau_\varepsilon \left(\rho  c_s^2 \tau _Q+\zeta +\frac{4 \eta}{3} +\sigma \kappa _s\right)-\rho  \tau _P \tau _Q,\\
C&\equiv&\tau_P \left(\rho  c_s^2 \tau_Q+\sigma  \kappa_s\right)-\beta_\varepsilon \left ( \zeta +\frac{4\eta}{3} \right ),
\eea
\eml
$n_1=3$, $n_\pm=1$, $c_1=\frac{\eta}{\rho\tau_s}$, and $c_\pm=\frac{-B\pm\sqrt{B^2-4AC}}{2A}$. It is worth mentioning that the assumptions of Proposition I guarantee that $0<c_1,c_\pm<1$. 
Under assumptions (A1), $\eta>0$, and conditions \eqref{C_conditions} in the strict form, then one obtains that $\det(\xi_\alpha A^ \alpha_m)=0$ only if $0\le c_a< 1$ (with the equality holding only in the case $a=0$), i.e., the equation $b_a^2-c_a (v_a\cdot v_a)=0$ gives $\xi_{a,\alpha}$ such that $\xi_{a,\alpha}\xi_a^\alpha=-b^2_a+v_a\cdot v_a=(1-c_a)v_a\cdot v_a> 0$. Thus, if $\xi$ is timelike, then (i) is guaranteed for the matter sector as well. As for the gravity sector one obtains that
\bea
\label{Sobolev_14}
\det(\xi_\alpha \mathbb{A}^\alpha_g)&=&\det\bbm 
b I_{10} & -v_0I_{10} & -v_1I_{10} & -v_2I_{10} & -v_3I_{10}\\
-v^0 I_{10} & b I_{10} & 0_{10\times 10} & 0_{10\times 10} & 0_{10\times 10}\\
-v^1 I_{10} & 0_{10\times 10} & b I_{10} & 0_{10\times 10} & 0_{10\times 10}\\
-v^2 I_{10} & 0_{10\times 10} & 0_{10\times 10} & b I_{10} & 0_{10\times 10}\\
-v^3 I_{10} & 0_{10\times 10} & 0_{10\times 10} & 0_{10\times 10}  & b I_{10}
\ebm\nonumber\\
&=&\frac{1}{b^{10}}\det\bbm 
(b^2-v^\nu v_\nu )I_{10} & 0_{10\times 10} & 0_{10\times 10} & 0_{10\times 10} & 0_{10\times 10}\\
-v^0 I_{10} & b I_{10} & 0_{10\times 10} & 0_{10\times 10} & 0_{10\times 10}\\
-v^1 I_{10} & 0_{10\times 10} & b I_{10} & 0_{10\times 10} & 0_{10\times 10}\\
-v^2 I_{10} & 0_{10\times 10} & 0_{10\times 10} & b I_{10} & 0_{10\times 10}\\
-v^3 I_{10} & 0_{10\times 10} & 0_{10\times 10} & 0_{10\times 10}  & b I_{10}
\ebm\nonumber\\
&=&(u^\alpha\xi_\alpha)^{30}(\xi_\alpha\xi^\alpha)^{10}.
\eea
Again, note that if $\xi$ is timelike, then $\det(\xi_\alpha \mathbb{A}^ \alpha_g)\ne 0$. This completes the proof of (i).

As for (ii), let us define $\phi_\alpha=\zeta_\alpha+\Lambda\xi_\alpha$ and make the changes $\xi\to\phi$ in the determinant calculations above. Then, the eigenvalues $\Lambda$ are obtained from the roots of $\det(\phi_\alpha\mathfrak{A}^\alpha)=\det(\phi_\alpha \mathbb{A}_m^\alpha)\det(\phi_\alpha \mathbb{A}^\alpha_g)(u^\alpha\phi_\alpha)^{16}=0$. Note that the general form of the equations implies that the roots $\phi_\alpha=-u_\alpha u^\beta\phi_\beta+\Delta_\alpha^\beta\phi_\beta$ obey  
\be
\label{Sobolev_roots}
(u^\alpha\phi_\alpha)^2-\beta \Delta^{\alpha\beta}\phi_\alpha\phi_\beta=0,
\ee 
where, from causality, in any of the above cases we have that $0\le \beta\le 1$. Then, for each $\beta$, the eigenvalues $\Lambda$ are 
\be
\label{Sobolev_Eigenvalues}
\Lambda=\frac{\beta(\Delta^{\alpha\beta}\xi_\alpha\zeta_\beta)-(u^\alpha\xi_\alpha)(u^\alpha\zeta_\alpha)\pm \sqrt{\mathcal{Z}}}{(u^\alpha\xi_\alpha)^2-\beta\Delta^{\alpha\beta}\xi_\alpha\xi_\beta},
\ee
where, since $\xi_\alpha\xi^\alpha<0$, then $(u^\alpha\xi_\alpha)^2-\beta\Delta^{\alpha\beta}\xi_\alpha\xi_\beta>0$ because $0\le\beta\le 1$ and
\bea
\label{Sobolev_Discriminant}
\mathcal{Z}&=&\beta\big \{\Delta^{\alpha\beta}\zeta_\alpha\zeta_\beta (u^\mu\xi_\mu)^2+\Delta^{\alpha\beta}\xi_\alpha\xi_\beta (u^\mu\zeta_\mu)^2 
-2(u^\alpha\xi_\alpha)(u^\beta\zeta_\beta)\Delta^{\mu\nu}\xi_\mu\zeta_\nu\nonumber\\
&&-\beta\left [(\Delta^{\alpha\beta}\zeta_\alpha\zeta_\beta)(\Delta^{\mu\nu}\xi_\mu\xi_\nu)-(\Delta^{\alpha\beta}\xi_\alpha\zeta_\beta)^2\right ]\big\}\nonumber\\
&>&\beta\big [\Delta^{\alpha\beta}\zeta_\alpha\zeta_\beta (u^\mu\xi_\mu)^2+\Delta^{\alpha\beta}\xi_\alpha\xi_\beta (u^\mu\zeta_\mu)^2 
-2(u^\alpha\xi_\alpha)(u^\beta\zeta_\beta)\Delta^{\mu\nu}\xi_\mu\zeta_\nu\nonumber\\
&&-(\Delta^{\alpha\beta}\zeta_\alpha\zeta_\beta)(\Delta^{\mu\nu}\xi_\mu\xi_\nu)+(\Delta^{\alpha\beta}\xi_\alpha\zeta_\beta)^2\big]\nonumber\\
&=&\beta\big\{(-\xi^\alpha\xi_\alpha)(\zeta^\beta\zeta_\beta)+\left [(u^\alpha\xi_\alpha)(u^\beta\zeta_\beta)-\Delta^{\alpha\beta}\xi_\alpha\zeta_\beta\right ]^2\big\}>0.
\eea
In the operations above we used the fact that $0\le\beta\le 1$, $(\Delta^{\alpha\beta}\xi_\alpha\zeta_\beta)^2\le (\Delta^{\alpha\beta}\xi_\alpha\xi_\beta)(\Delta^{\mu\nu}\zeta_\mu\zeta_\nu)$ from the Cauchy-Schwarz inequality and that $\xi$ is timelike and $\zeta$ spacelike. Thus, causality guarantees reality of the eigenvalues.

Now we turn to the problem of completeness of the set of eigenvectors. We begin by counting the linearly independent eigenvectors of $\phi_{a,\alpha}^{(m)} \mathbb{A}^\alpha_m$, where $\phi^{(m)}_{a,\alpha}=\zeta_\alpha+\Lambda^{(m)}_a\xi_\alpha$ and $\Lambda^{(m)}_a$ are the eigenvalues of the matter sector and are obtained by means of \eqref{Sobolev_Eigenvalues} in the cases $\beta=c_0=0$ when $a=0$ and $\beta=c_a$ when $a=1,\pm$. Let us define an arbitrary vector
\be
\label{s_r}
r^{(m)}=\bbm F\\G^\nu\\H^\mu\\I^\mu\\J^{\nu}_0\\J^{\nu}_1\\J^{\nu}_2\\J^{\nu}_3\ebm.
\ee
Then, for each of the eigenvalues $\Lambda_a^{(m)}$, $a=0,1,\pm$, we must verify how many of the 29 variables in the vector \eqref{s_r} are free parameters under the equation $\phi^{(m)}_{a,\alpha} \mathbb{A}^\alpha_m r^{(m)}_a=0$. In fact, this is the dimension of the null space of the matrix $\phi^{(m)}_{a,\alpha} \mathbb{A}^\alpha_m$ and corresponds to the number of linearly independent (LI) eigenvectors of $\Lambda_a^{(m)}$. The eigenvectors are the following:
\begin{itemize}
\item $\Lambda_0^{(m)}$: this root has multiplicity 19. The eigenvector that obey $\phi^{(m)}_{0,\alpha}\mathbb{A}^\alpha r^{(m)}_0=0$ is
\be
\label{s_I14}
r^{(m)}_0=\bbm 0\\0_{4\times 1}\\H^\mu\\I^\mu\\J^{\nu}_0\\J^{\nu}_1\\J^{\nu}_2\\J^{\nu}_3\ebm,
\ee
where only $19$ out of the $24$ components $H^\mu,I^\mu,J^{\nu}_\lambda$ are free variables because of the $1+1+3$ constraints $\beta_\varepsilon \phi^{(m)}_{0,\nu}H^\nu+\beta_n \phi^{(m)}_{0,\nu} I^\nu=0$, $J^\lambda_\lambda=0$, and $\Delta^{\mu\lambda}\phi^{(m)}_{0,\nu}J^\nu_\lambda+\Delta^{\lambda\beta}\phi^{(m)}_{0,\beta}J^\mu_\lambda=0$ (note that the last 4 equations are not all independent since the contraction with $u_\mu$ is identically zero, resulting in 3 independent constraints). Thus, the multiplicity of $\Lambda_0$ equals the number of LI eigenvectors, i.e., 19. 
\item $\Lambda_1^{(m)\pm}$: in this case each of the two eigenvalues have multiplicity 3 since $n_1=3$ in \eqref{Sobolev_140} (note that since we assumed here that $\eta>0$, than $c_1\ne0$ and, thus, $c_1\ne c_0$ and the eigenvalues are different from the case $c_0=0$). We may perform some elementary row operations over the linear system $\phi^{(m)}_{1,\alpha} \mathbb{A}^\alpha r^{(m)}_1=0$ to obtain, by imposing $b^2-c_1(v\cdot v)=0$ (remember that $b=u^ \alpha\phi_\alpha$ and $v^\alpha=\Delta^{\alpha\beta}\phi_\beta$ after the change $\xi\to\phi$),
\be
\label{s_I13-1}
\bbm
\tau_\varepsilon b^2+\beta_\varepsilon(v\cdot v) & b\rho\tau_Q\phi_\nu+b\rho\tau_\varepsilon v_\nu-\frac{n\beta_n (v\cdot v)}{b}v_\nu  & 0_{1\times 4} & 0_{1\times 4} & 0_{1\times 4} & 0_{1\times 4} & 0_{1\times 4} & 0_{1\times 4}\\
0_{4\times 1} & \mathcal{K}_\nu v^{\mu} & 0_{4\times 4} & 0_{4\times 4} & 0_{4\times 4} & 0_{4\times 4} & 0_{4\times 4} & 0_{4\times 4}\\
-v^{\mu} & 0_{4\times 4} & b \delta^\mu_\nu & 0_{4\times 4} & 0_{4\times 4} & 0_{4\times 4} & 0_{4\times 4} & 0_{4\times 4}\\
0_{4\times 1} & \frac{n v^\mu v_\nu}{b} & 0_{4\times 4} & b \delta^\mu_\nu & 0_{4\times 4} & 0_{4\times 4} & 0_{4\times 4} & 0_{4\times 4}\\
0_{4\times 1} & -v_0\delta^\mu_{\nu} & 0_{4\times 4} & 0_{4\times 4} & b \delta^\mu_\nu & 0_{4\times 4} & 0_{4\times 4} & 0_{4\times 4}\\
0_{4\times 1} & -v_1\delta^\mu_\nu & 0_{4\times 4} & 0_{4\times 4} & 0_{4\times 4} & b \delta^\mu_\nu & 0_{4\times 4} & 0_{4\times 4}\\
0_{4\times 1} & -v_2\delta^\mu_\nu & 0_{4\times 4} & 0_{4\times 4} & 0_{4\times 4} & 0_{4\times 4} & b \delta^\mu_\nu & 0_{4\times 4}\\
0_{4\times 1} & -v_3\delta^\mu_\nu & 0_{4\times 4} & 0_{4\times 4} & 0_{4\times 4} & 0_{4\times 4} & 0_{4\times 4} & b \delta^\mu_\nu
\ebm r^{(m)}_1=0,
\ee
where 
\bea
\mathcal{K}_\nu&=&\left [-\eta \xi_\nu+\left (\rho\tau_P-\zeta+\frac{2\eta}{3}-n\beta_n \right )v_\nu\right ]\left [\tau_\varepsilon b^2+\beta_\varepsilon(v\cdot v)\right ]\nonumber\\
&&-(\tau_P+\beta_\varepsilon)\left [b^2\rho\tau_Q\xi_\nu+b^2\rho\tau_\varepsilon v_\nu-n\beta_n (v\cdot v)v_\nu\right ].
\eea
This enables us to find the eigenvectors
\be
\label{s_I15}
{}^{\pm}r_1^{(m)}=\bbm
F_\pm \\ G^\nu_\pm \\ H^\nu_\pm \\ I^\nu_\pm \\ {}^{\pm}J^{\nu}_{0} \\ {}^{\pm}J^{\nu}_{1} \\ {}^{\pm}J^{\nu}_{2} \\ {}^{\pm}J^{\nu}_{3}
\ebm,
\ee
where, from the $29+29=58$ components of the above eigenvectors (29 for $\Lambda_1^{(m)+}$ and 29 $\Lambda_1^{(m)-}$ cases), they are subjected to the following $26+26$ constraints: $1+1=2$ constraints
\[[\tau_\varepsilon b^2_\pm+\beta_\varepsilon(v_\pm\cdot v_\pm)]F_\pm + b_\pm\rho\tau_Q\,{}^{\pm}\phi_{1,\nu}^{(m)} G^\nu+b_\pm\rho\tau_\varepsilon v_\nu^\pm G^\nu-\frac{n\beta_n (v_\pm\cdot v_\pm)}{b_\pm}v^\pm_\nu G^\nu_\pm=0,\]
$1+1=2$ constraints $\mathcal{K}_\nu^\pm G_\pm^\nu=0$, $4+4=8$ constraints $b_\pm H^\mu_\pm=v^\mu_\pm F_\pm$, $4+4=8$ constraints $n v^\mu_\pm v_\nu^\pm G^\nu+b_\pm^2 I_\pm^\mu=0$, and the $16+16=32$ constraints $b_\pm \,{}^{\pm}J^\mu_{\pm\lambda}=v_\lambda^\pm G^\mu_\pm$, where ${}^\pm\phi^{(m)}_{1,\alpha}={}^\pm\Lambda_1^{(m)} \xi_\alpha+\zeta_\alpha$ and $b^{\pm}$ and $v_\pm^\alpha$ are defined in terms of ${}^{\pm}\phi_{1,\nu}^{(m)}$. Hence, there is a total of $3+3=6$ free parameters. Once again, the degeneracy equals the number of LI eigenvectors.
\item $(\Lambda_\pm)^\pm$: since there is no degeneracy in these four last eigenvalues and they are distinct from the others because $c_\pm\ne0$ in the strict form of the inequalities in \eqref{C_conditions} and different among them, then one has 4 LI eigenvectors. 
\end{itemize}
Thus, the system has $19+6+4=29$ LI eigenvectors. Therefore, there is a complete set in $\mathbb{R}^{29}$, namely, $\{r^{(m)}_b\}_{b=1}^{29}$ such that $\phi^{(m)}_a \mathbb{A}^\alpha_m r^{(m)}_b=0$. Hence, we can use the 29 linearly independent set $\mathcal{S}^{(m)}=\{R^{(m)}_b\}_{b=1}^{29}$ to verify that
\be
\label{11}
R^{(m)}_b=\bbm r^{(m)}_b \\ 0_{66\times 1}\ebm
\ee
obeys $(\zeta_\alpha+\Lambda^{(m)}_a\xi_\alpha) \mathfrak{A}^\alpha R^{(m)}_b=0$.

Now, before we discuss the gravity sector $\{F_A,\mathcal{F}^\delta_A\}$, let us look at the sector containing the original fields $\varepsilon$, $n$, $u^\nu$, and $g_{\mu\nu}$. In this case, let us define
\be
\label{Sobolev_12}
R^{(d)}=\bbm 0_{79\times 1}\\ r^{(d)}\ebm,
\ee 
where $r^{(d)}$ is a $16\times 1$ column vector.  Then, $(\zeta_\alpha+\Lambda^{(d)}_a \xi_\alpha)\mathfrak{A}^\alpha R^{(d)}_a=0$ reduces to the eigenvalue problem $u^\alpha\phi^{(d)}_\alpha I_{16}r^{(d)}=0$ whose eigenvalues are $u^\alpha\phi^{(d)}_\alpha=0$, i.e., $\Lambda^{(d)}=\zeta_\alpha u^\alpha/\xi_\alpha u^\alpha$. Thus, the eigenvectors may be any basis of $\mathbb{R}^{16}$. Let $\{r^{(d)}_a\}_{a=1}^{16}$ be a basis of $\mathbb{R}^{16}$. Then, the set $\mathcal{S}^{(d)}=\{R^{(d)}_a\}_{a=1}^{16}$ is a linearly independent set of 16 eigenvectors of $\phi_\alpha^{(d)}\mathfrak{A}^\alpha$.

To finalize the eigenvector counting we have to analyze the sector containing $F_A$ and $\mathcal{F}^\delta_A$. In this case, let us define
\be
\label{Sobolev_13}
R^{(g)}=\bbm w\\ r^{(g)}\\0_{16\times 1}\ebm,
\ee
where $w$ is some $29\times 1$ columns vector while $r^{(g)}$ is a $50\times 1$ columns vector. The eigenvalues of this sector are in \eqref{Sobolev_14} and are given by $\Lambda^{(g)}_0=u^\alpha\zeta_\alpha/u^\beta\xi_\beta$, coming from $u^\alpha\phi^{(g)}_{0,\alpha}=0$ (here $\phi^{(g)}_{a,\alpha}=\zeta_\alpha+\Lambda^{(g)}_a\xi_\alpha$) with multiplicity 30 and corresponding to $\beta=0$, and the two roots ${}^{\pm}\Lambda^{(g)}_1$ with multiplicity 10 each coming from ${}^{\pm}\phi^{(g)}_{1,\alpha}\,{}^{\pm}\phi^{(g)\alpha}_1=-[u^\alpha\,{}^{\pm}\phi^{(g)}_{1,\alpha}]^2+\Delta^{\alpha\beta}\,{}^{\pm}\phi^{(g)}_{1,\alpha}\,{}^{\pm}\phi^{(g)}_{1,\beta}=0$, which corresponds to $\beta=1$, i.e., gravitational waves moving at the speed of light. 
Then, the eigenvalue problem $\phi^{(g)}_{a,\alpha}\mathfrak{A}^\alpha R^{(g)}_a=0$ reduces to the two equations
\bml
\label{Sobolev_15}
\bea
\phi^{(g)}_{a,\alpha} \mathbb{A}^\alpha_m w_a=L^\alpha r^{(g)}_a,\label{Sobolev_15a}\\
\phi^{(g)}_{a,\alpha}\mathbb{A}^\alpha_g r^{(g)}_a=0.\label{Sobolev_15b}
\eea
\eml
For the eigenvalues ${}^{\pm}\Lambda^{(g)}_1$, one obtains that $\det[{}^{\pm}\phi^{(g)}_{1,\alpha} \mathbb{A}^\alpha_m]\ne0$ because the root $\beta=1$ has been eliminated from the matter sector (remember that $c_a<1$). Thus, there exists a solution of \eqref{Sobolev_15a} for each $r^{(g)}_a$ in \eqref{Sobolev_15b}. One needs to count the number of linearly independent $r^{(g)}_1$ for $\Lambda^{(g)}_1$, i.e., the number of vectors in the basis of the kernel of $\phi^{(g)}_{1,\alpha}\mathbb{A}^\alpha_g$.  In this case, after some elementary row operations [look at the second equality in \eqref{Sobolev_14} after setting $b^ 2=v\cdot v$] one obtains that
\be
{}^{\pm}\phi^{(g)}_{1,\alpha} \mathbb{A}^\alpha_g\sim 
\bbm 
0_{10\times 10} & 0_{10\times 10} & 0_{10\times 10} & 0_{10\times 10} & 0_{10\times 10}\\
-\Delta^{0\alpha}\,{}^{\pm}\phi^{(g)}_{1,\alpha} I_{10} & (u^\alpha\,{}^{\pm}\phi^{(g)}_{1,\alpha}) I_{10} & 0_{10\times 10} & 0_{10\times 10} & 0_{10\times 10}\\
-\Delta^{1\alpha}\,{}^{\pm}\phi^{(g)}_{1,\alpha} I_{10} & 0_{10\times 10} & (u^\alpha\,{}^{\pm}\phi^{(g)}_{1,\alpha}) I_{10} & 0_{10\times 10} & 0_{10\times 10}\\
-\Delta^{2\alpha}\,{}^{\pm}\phi^{(g)}_{1,\alpha} I_{10} & 0_{10\times 10} & 0_{10\times 10} & (u^\alpha\,{}^{\pm}\phi^{(g)}_{1,\alpha}) I_{10} & 0_{10\times 10}\\
-\Delta^{3\alpha}\,{}^{\pm}\phi^{(g)}_{1,\alpha} I_{10} & 0_{10\times 10} & 0_{10\times 10} & 0_{10\times 10}  & (u^\alpha\,{}^{\pm}\phi^{(g)}_{1,\alpha}) I_{10}
\ebm,
\ee
which has 40 pivots and 10 independent variables (corresponding to the variables associated to the first 10 columns). Thus, there are 10 linearly independent vectors for each eigenvalue ${}^{\pm}\Lambda^{(g)}_1$, i.e., there is a set $\{{}^{-}r^{(g)}_{1,b},{}^{+}r^{(g)}_{1,b}\}_{b=1}^{10}$ of 20 linearly independent vectors with corresponding $w^\pm_{1,b}=[{}^{\pm}\phi^{(g)}_{1,\alpha}\mathbb{A}^\alpha_m]^{-1} L^a {}^{\pm}r^{(g)}_{1,b}$ coming from \eqref{Sobolev_15a} such that
$\mathcal{S}^{(g)}_1=\{{}^{+}R^{(g)}_{1,b},{}^{-}R^{(g)}_{1,b}\}_{b=1}^{10}$, where
\[{}^{\pm}R^{(g)}_{1,b}=\bbm w^{\pm}_{1,b}\\ {}^{\pm}r^{(g)}_{1,b}\\0_{16\times 1}\ebm,\]
is a linearly independent set of 20 eigenvectors of $\phi^{(g)}_{1,\alpha}\mathfrak{A}^\alpha$.

As for the eigenvalue $\Lambda^{(g)}_0$, note that in this case $\det[\phi^{(g)}_{0,\alpha}\mathbb{A}^\alpha_m]=0$ because $\beta=c_0=0$ is also a root of this equation. Thus, for every solution $r^{(g)}_{a}$ in \eqref{Sobolev_15b}, \eqref{Sobolev_15a} can be either undetermined or have infinite solutions. However, for any two different solutions, say, $w_a^1$ and $w_a^2$ for one $r^{(g)}_{a}$, the difference between $R^{(g)1}_a-R^{(g)2}_a$ corresponds to a vector in the space spanned by $\mathcal{S}^{(m)}$, that lies in the Kernel of $\phi^{(g)}_{0,\alpha}\mathbb{A}^\alpha_m$. Therefore, since we are counting the number of linearly independent eigenvectors, we must choose one particular solution $w_a$, if it exists, for each $r^{(g)}_{a}$. We begin by solving Eq.\ \eqref{Sobolev_15b}. Let $\{l^\mu_1=u^\mu,l_2^\mu,l_3^\mu\}$ be a set of linearly independent vectors that are orthogonal to $\phi^{(g)}_{0,\alpha}=\zeta_\alpha+\Lambda^{(g)}_0\xi_\alpha$, to wit, $l_c^\alpha \phi^{(g)}_{0,\alpha}=0$ and $\{e_a\}_{a=1}^{10}$ be any basis of $\mathbb{R}^{10}$. Then, one may verify that the 30 linearly independent vectors 
\be
r^{(g)}_{0,ac}=\bbm 0_{10\times 1}\\l_c^0 e_a\\l_c^1 e_a\\l_c^2 e_a\\l_c^3 e_a \ebm 
\ee 
satisfy $\phi^{(g)}_{0,\alpha}\mathbb{A}^\alpha_g r^{(g)}_{0,ac}=0$. Now we must solve \eqref{Sobolev_15a}, where
\bea
\label{Sobolev_sol1}
\phi^{(g)}_{0,\alpha}L^\alpha r^{(g)}_{0,ac}&=&\bbm 0_{13\times 1}\\ 
\phi^{(g)}_{0,\alpha} \mathcal{Y}^{\mu A\alpha}_{0\delta}l_c^\delta (e_a)_A\\
\phi^{(g)}_{0,\alpha} \mathcal{Y}^{\mu A\alpha}_{1\delta}l_c^\delta (e_a)_A\\
\phi^{(g)}_{0,\alpha} \mathcal{Y}^{\mu A\alpha}_{2\delta}l_c^\delta (e_a)_A\\
\phi^{(g)}_{0,\alpha} \mathcal{Y}^{\mu A\alpha}_{3\delta}l_c^\delta (e_a)_A\ebm=K_a
\bbm
0_{13\times 1}\\
\phi^{(g)}_{0,0} l_c^\mu\\
\phi^{(g)}_{0,1} l_c^\mu\\
\phi^{(g)}_{0,2} l_c^\mu\\
\phi^{(g)}_{0,3} l_c^\mu
\ebm,
\eea
where we defined
\[K_a\equiv \frac{1}{2}\left [\sum_{\underset{\sigma \le \beta}{\sigma,\beta}}
 (2-\delta_{\sigma \beta})u^{(\sigma}u^{\beta)} (e_a)_{\sigma\beta}\right ].\]
Let us look for the particular solution 
\be
w_{ac}=\bbm 
0\\
-\beta_\varepsilon y^\nu_{ac}\\
\rho\tau_Q y^\nu_{ac}\\
0_{20\times 1} 
\ebm.
\ee
Note that
\be
\label{Sobolev_sol2}
\phi^{(g)}_{0,\alpha}\mathbb{A}^\alpha_m w_{ac}=\bbm
0_{13\times 1}\\
\beta_\varepsilon\phi^{(g)}_{0,0}y^\mu_{ac}\\
\beta_\varepsilon\phi^{(g)}_{0,1}y^\mu_{ac}\\
\beta_\varepsilon\phi^{(g)}_{0,2}y^\mu_{ac}\\
\beta_\varepsilon\phi^{(g)}_{0,3}y^\mu_{ac}
\ebm
\ee
and then, by inserting \eqref{Sobolev_sol1} and \eqref{Sobolev_sol2}  into Eq.\ \eqref{Sobolev_15a}, one finds that
\be
\beta_\varepsilon\phi^{(g)}_{0,\nu}y^\mu_{ac}=K_a \phi^{(g)}_{0,\nu}l_c^\mu.
\ee
This leads to the solution $y^\mu _{ac}=K_a l^\mu_c/\beta_\varepsilon$ and, thus,
\be
w_{ac}=\bbm 
0\\
-K_a l^\nu_{c}\\
\frac{\rho\tau_Q}{\beta_\varepsilon} K_a l^\nu_{c}\\
0_{20\times 1} 
\ebm.
\ee
As a consequence, the set $\mathcal{S}^{(g)}_0=\{R^{(g)}_{1,1},R^{(g)}_{1,2},R^{(g)}_{1,3},\cdots R^{(g)}_{10,1},R^{(g)}_{10,2},R^{(g)}_{10,3}\}$ with 
\[R^{(g)}_{ac}=\bbm w_{ac}\\ r^{(g)}_{0,ac}\\0_{16\times 1}\ebm\]
is a linearly independent set of 30 eigenvectors of $\phi^{(g)}_{0,\alpha}\mathfrak{A}^\alpha$. Thus, $\mathfrak{S}=\mathcal{S}^{(m)}\cup\mathcal{S}^{(d)}\cup\mathcal{S}^{(g)}_1\cup\mathcal{S}^{(g)}_0$ contains a complete set of eigenvectors $R$ of $\phi_\alpha\mathfrak{A}^\alpha R=0$ in $\mathbb{R}^{95}$. This completes the proof.
\hfill $\Box$

\medskip

We remark that the assumption that the inequalities hold in strict form is technical. If equality is allowed,
then the multiplicity of the eigenvalues might change. This is because with equality one can have
$c_a=0$ for $a=1$ or $\pm$ and thus the characteristics defined
by  $b^2 - c_a (v\cdot v) = 0$ can degenerate into the characteristics $b=0$. Since the latter
is already present in the system, the multiplicity of the characteristics would change.
This does not mean that the system would not be diagonalizable. Nor does it imply that 
local well-posedness, established in the next section, would fail  \footnote{We recall that we are
interested in the diagonalization because it allows us to invoke known techniques to prove
local well-posedness. If the system is not diagonalizable, it remains possible that 
different techniques would lead to local well-posedness.}. However,
 a different proof would be needed to show diagonalization in 
the case $c_{a} =0$ in the cases $a=1$ or $\pm$. We believe that treating this very special 
case here would be a distraction from the main points of the paper. We also recall that already
in the case of an ideal fluid, a different approach to local well-posedness has to be employed
when the characteristics degenerate \cite{MR3335528}.

\section{Proof of Theorem II}
\label{Theorem_II}

As usual in studies of the initial-value problem for Einstein's equations \cite{WaldBookGR1984}, 
we embed $\Sigma$ into $\mathbb{R}\times \Sigma$ and work in harmonic coordinates in the neighborhood
of a point. Observe that we already know the system to be causal under our assumptions thus 
localization arguments are allowed.

The equations to be studied read
\bml
\label{Sobolev_20}
\bea
&& u^\alpha u^\beta \partial^2_{\alpha \beta} n
+n u^\alpha \delta^\beta_\nu\partial^2_{\alpha\beta} u^\nu+\tilde{\mathcal{B}}_1( n, u, g)\partial^2 g=\mathcal{B}_1(\partial n,\partial u,\partial g)\\
&& u_\nu u^\alpha u^\beta\partial_\alpha\partial_\beta u^\nu+\tilde{\mathcal{B}}_2(n,\varepsilon,u,g)\partial^2 g=\mathcal{B}_2(\partial n,\partial\varepsilon,\partial u,\partial g),\\
&&\beta_n\left (u^\mu \Delta^{\alpha\beta}+\Delta^{\mu(\alpha}u^{\beta)}\right )\partial_\alpha\partial_\beta n
+\mathfrak{E}^{\mu\alpha\beta}\partial_\alpha\partial_\beta \varepsilon+\bar{\mathcal{C}}^{\mu\alpha\beta}_\nu \partial_\alpha\partial_\beta u^ \nu\nonumber\\
&&+\tilde{\mathcal{B}}_3^\mu(n,\varepsilon,u,g)\partial^2g=\mathcal{B}_3^\mu(\partial n,\partial\varepsilon,\partial u,\partial g),\\
&&g^{\alpha\beta}\partial_\alpha\partial_\beta g_{\mu\nu}=\mathcal{B}_{4\,\mu\nu}(\partial n,\partial\varepsilon,\partial u,\partial g),
\eea
\eml
where
\bml
\label{Sobolev_21}
\bea
\bar{\mathcal{C}}^{\mu\alpha\beta}_\nu&=&\left (\tau_P\rho-\zeta-\frac{\eta}{3}\right )\Delta^{\mu(\alpha}\delta^{\beta)}_\nu-\eta\Delta^{\alpha\beta}\delta^\mu_\nu+\rho(\tau_\varepsilon+\tau_Q) u^\mu \Delta^{(\alpha}_\nu u^{\beta)}+\tau_Q\rho u^\alpha u^\beta \delta^\mu_\nu,\\
\mathfrak{E}^{\mu\alpha\beta}&=&u^\mu(\beta_\varepsilon \Delta^{\alpha\beta}+\tau_\varepsilon u^\alpha u^\beta )+(\beta_\varepsilon+\tau_P)\Delta^{\mu(\alpha}u^{\beta)},
\eea
\eml
and the notation for the $\tilde{\mathcal{B}}$'s and $\mathcal{B}$'s follow the same construction as in Section \ref{sec:causality}.

We can write \eqref{Sobolev_20} in matrix form as
\be
\label{Sobolev_22}
\mathfrak{M}(\partial)\Psi=\mathfrak{N}(\partial\Psi),
\ee
where $\Psi=(\varepsilon,n,u^\nu,g_{\mu\nu})^T$ is a $16\times 1$ column vector (we count only the 10 independent $g_{\mu\nu}$), $\mathfrak{B}(\partial\Psi)$ is also a $16\times 1$ column vector containing the $\mathfrak{N}$'s, i.e., the lower order terms in derivatives of each equation, and
\be
\label{Sobolev_23}
\mathfrak{M}(\partial)=\bbm \mathbb{M}(\partial) & \mathfrak{b}(\partial)\\ 0_{10\times 6} & g^{\alpha\beta}\partial_\alpha\partial_\beta I_{10}\ebm.
\ee
The $6\times 10$ matrix $\mathfrak{b}(\partial)$ contains the terms $\tilde{\mathcal{B}}\partial^2g$ while
\be
\label{Sobolev_24}
\mathbb{M}(\partial)=\bbm
0 & u^\alpha u^\beta & n\delta^{(\alpha}_\nu u^{\beta)}\\
0 & 0 & u_\nu u^\alpha u^\beta\\
\mathfrak{E}^{\mu\alpha\beta}  & \beta_n\left (u^\mu \Delta^{\alpha\beta}+\Delta^{\mu(\alpha}u^{\beta)}\right ) 
& \bar{\mathcal{C}}^{\mu\alpha\beta}_\nu  
\ebm\partial^2_{\alpha\beta}.
\ee

Let us compute the characteristic determinant of the system and its roots, i.e., 
$\det[\mathfrak{M}(\xi)]=\det[\mathbb{M}(\xi)](\xi^\alpha\xi_\alpha)^{10}=0$, where the substitution $\partial\to\xi$ takes place. 
The pure gravity sector has the roots $\xi^\alpha\xi_\alpha=0$.
As for the matter sector, by again defining $b=u^\alpha\xi_\alpha$, $v^\mu=\Delta^{\mu\nu}\xi_\nu$, $v\cdot v=v^\mu v_\mu$, and 
\bml
\bea
\tilde{\mathcal{C}}^\mu_\nu&=&\bar{\mathcal{C}}^{\mu\alpha\beta}_\nu\xi_\alpha\xi_\beta=[\tau_Q\rho b^2-\eta(v\cdot v)]\delta^\mu_\nu+ \left (\tau_P\rho-\zeta-\frac{\eta}{3}\right )v^\mu \xi_\nu+\rho(\tau_\varepsilon+\tau_Q)b u^\mu v_\nu ,\\
\mathfrak{D}^\mu_\nu&=&\left (\tau_P\rho-\zeta-\frac{\eta}{3}-n\beta_n\right )v^\mu\xi_\nu+[\tau_Q\rho b^2-\eta(v\cdot v)]\delta^\mu_\nu,\\
\tilde{\mathfrak{E}}^\mu&=&\mathfrak{E}^{\mu\alpha\beta}\xi_\alpha\xi_\beta=[\beta_\varepsilon (v\cdot v)+\tau_\varepsilon b^2 ]u^\mu+(\beta_\varepsilon+\tau_P) b v^\mu,
\eea
\eml
where $\mathfrak{D}^\mu_\nu$ is the same as the one defined in \eqref{Matrix_D}, we obtain that (by carrying out some elementary row operations)
\bea
\label{Sobolev_25}
\det[\mathbb{M}(\xi)]&=&\det\bbm
0 & b^2 & nb\xi_\nu\\
0 & 0 &  b^2 u_\nu\\
\tilde{\mathfrak{E}}^{\mu}  & \beta_n\left [u^\mu (v\cdot v)+b v^\mu\right ] & \tilde{\mathcal{C}}^{\mu}_\nu 
\ebm\nonumber\\
&=&\frac{b^3}{\tau_Q\rho b^2-\eta(v\cdot v)}\det\bbm
0 & b & n\xi_\nu\\
\tau_\varepsilon b^2+\beta_\varepsilon(v\cdot v) & \beta_n (v\cdot v) &  \rho(\tau_\varepsilon+\tau_Q)b v_\nu\\
(\beta_\varepsilon+\tau+P)b v^\mu  & \beta_n b v^\mu  & \mathfrak{D}^{\mu}_\nu 
\ebm.
\eea
The last determinant is the same as the one obtained in \eqref{det} and the result turns out to be
\bea
\label{Sobolev_26}
\det[\mathbb{M}(\xi)]&&=-b^4[\rho\tau_Q b^2-\eta(v\cdot v)]^2\left [Ab^4+Bb^2(v\cdot v)+C(v\cdot v)^2\right ]\nonumber\\
&=&-\rho^4\tau_Q^4\tau_\varepsilon\,(u^\alpha\xi_\alpha)^4 \prod_{a=1,\pm}\left [(u^\alpha\xi_\alpha)^2-c_a\Delta^{\alpha\beta}\xi_\alpha\xi_\beta\right ]^{\tilde{n}_a},
\eea
where, as in Eqs.\ \eqref{ABC}, 
\bml
\bea
A&\equiv&\rho\tau_\varepsilon\tau_Q,\\
B&\equiv&-\tau_\varepsilon \left(\rho  c_s^2 \tau _Q+\zeta +\frac{4 \eta}{3} +\sigma \kappa _s\right)-\rho  \tau _P \tau _Q,\\
C&\equiv&\tau_P \left(\rho  c_s^2 \tau_Q+\sigma  \kappa_s\right)-\beta_\varepsilon \left ( \zeta +\frac{4\eta}{3} \right ),
\eea
\eml
while $c_1=\frac{\eta}{\rho\tau_s}$ and $c_\pm=\frac{-B\pm\sqrt{B^2-4AC}}{2A}$, while $\tilde{n}_1=2$ and $\tilde{n}_\pm=1$. Note that the characteristics are still the same as in section \ref{sec:causality}, as expected, although the multiplicity of the roots changed
(and there was no reason for the multiplicities to be the same). We conclude that the characteristic determinant of the system is a product of strictly hyperbolic polynomials.
We verify at once that the system is a Leray-Ohya system \cite{LerayOhyaNonlinearReprint,ChoquetBruhatGRBook} for which the results
of \cite{CB_diagonal} (see also \cite{DisconziFollowupBemficaNoronha}) apply. Thus, if the initial data is 
quasi-analytic  (see  \cite{F:Quasi-analytic}) we obtain quasi-analytic solutions.

Denote the initial-data set in the theorem by $\mathcal{D}$ and let $\mathcal{D}_\ell$ be a sequence
of quasi-analytic initial-data converging to $\mathcal{D}$ in $H^N$ (see \cite{F:Sobolev} 
for the definition of $H^N$). Let $\mathbf{\Psi}_\ell$
solutions corresponding to $\mathcal{D}_\ell$ (which exist by the foregoing). In order
to finish the proof of the theorem, it suffices to show that $\mathbf{\Psi}_\ell$ has a limit in $H^N$.
The limit will then be a solution with the desired properties because we can pass to the limit in the equations
since $N\geq 5$. 

According to the arguments given in section 16.2 of \cite{TaylorPDE3} or 
in \cite{DisconziBemficaRodriguezShaoSobolevConformal,DisconziBemficaGraber}, the diagonalization 
obtained in section \ref{S:diagonalization} implies
that $\Psi$ defined in \eqref{E:vector_U} admits a uniform bound in $H^{N-1}$,
and uniform difference bounds in $H^{N-2}$ also holds.
We apply these bounds to 
the vector $\Psi_\ell$ corresponding to $\mathbf{\Psi}_\ell$.
We see at once that the uniform $H^{N-1}$ bounds for 
$\Psi_\ell$ imply uniform $H^N$ bounds for $\mathbf{\Psi}_\ell$, and the difference
bounds imply that $\mathbf{\Psi}_\ell$ is a Cauchy sequence in $H^{N-3}$, thus 
converging in this space. But low-norm convergence combined with high-norm 
boundedness implies that the limit is in fact in  $H^N$
\cite{PalaisSeminar}. 
\hfill $\Box$

\bigskip

We observe that a similar local well-posedness result holds for the fluid equations in a fixed background.

We recall that a standard tensorial argument \cite{WaldBookGR1984} guarantees that the solution established
in Theorem II is intrinsically defined, i.e., given the data, which is defined independently
of coordinates or gauge choices, there exists a spacetime where Einstein's equations are satisfied,
and this spacetime is defined without any reference to coordinates or gauge choices -- even if 
in the process of proving that this spacetime exists one has to work in a specific gauge and
coordinate system. Therefore, even though we used the harmonic gauge in the proof, the existence of the solution is guaranteed for other choices as well. This logic is similar to showing that a map from a finite-dimensional vector
space into itself is invertible: one can choose a basis, write the matrix of the linear transformation
with respect to that basis, and compute its determinant. The map is invertible if and only if 
the determinant is non-zero, and this conclusion (the invertibility or not of the linear map) is 
independent of any basis choice -- even if to show that the map is invertible we picked a basis
and computed the determinant with respect to that basis.

We note, however, the following subtlety which is very relevant for numerical simulations.
The fact that a unique solution is guaranteed to exist
for given initial data, and that this solution is well-defined regardless of gauge choices,
does not imply that such a solution can always be \emph{reconstructed} from an arbitrary gauge.
In other words, suppose we write the equations in a different gauge. \emph{If}
we can numerically integrate them, we will obtain the solution found in Theorem II written on that gauge
(modulo numerical accuracy). However, it is possible that the gauge we chose is not adequate to solve
the equations numerically, so that our numerical simulation will not produce a solution. This does not
mean, of course, that solutions do not exist; it simply means that the guaranteed-to-exist
solution given by Theorem II cannot be accessed from that specific gauge. 
To use again our analogy with determinants: suppose we computed the determinant on a basis $b_1$
and found it to be non-zero, but now we are interested in computing the determinant numerically
using another basis $b_2$. Depending on the basis $b_2$ and the numerical algorithm we use,
this might not be possible, which, of course, does not mean that the determinant is zero or ill-defined.

Thus, the practical matter of solving the equations numerically is not settled by an abstract
existence and uniqueness result as Theorem II. Such theorems are naturally important as
they provide the foundations on which numerical investigations can be built, i.e., it makes
sense to look for solutions numerically because solutions do exist. But these theorems do not, in general,
point to how to recover solutions numerically. That is why there is a great deal of work dedicated
to writing Einstein's equations in different forms and special gauges, even if basic existence
results for Einstein's equations coupled to most matter models are known, as reviewed in 
\cite{RezzollaZanottiBookRelHydro,baumgarte_shapiro_2010}.

\section{Proof of Theorem III}
\label{Theorem_III}

From causality one obtains that $\det(n_\alpha \mathbb{A}^\alpha)\ne 0$ as far as $n$ is timelike. Thus, we can rewrite \eqref{s_3} as 
\be
\label{s_2-2}
i\Omega \delta\Psi(K)=-i \kappa (-n_\alpha \mathbb{A}^\alpha)^{-1}\zeta_\beta \mathbb{A}^\beta\delta\Psi(K)-(-n_\alpha \mathbb{A}^\alpha)^{-1}\mathbb{B}\delta\Psi(K).
\ee
Since the eigenvalue problem \eqref{s_2-1} contains $N$ linearly independent vectors $\mathfrak{r}_a$, one may write \eqref{s_2-1} as
\be
\label{s_2-4}
(-n_\alpha \mathbb{A}^\alpha)^{-1}\zeta_\beta \mathbb{A}^\beta \mathfrak{r}_a=\Lambda_a \mathfrak{r}_a
\ee
and define the $N\times N$ invertible matrix $R=[\mathfrak{r}_1\,\cdots\,\mathfrak{r}_N]$ whose columns are the eigenvectors $\mathfrak{r}_1,\cdots,\mathfrak{r}_n$ and the $N\times N$ matrix \[L\equiv R^{-1}=\bbm \mathfrak{l}_1\\ \vdots \\ \mathfrak{l}_N\ebm,\] where the rows $\mathfrak{l}_a$ are the left eigenvectors of $(-n_\alpha \mathbb{A}^\alpha)\zeta_\beta \mathbb{A}^\beta$ which, consequently, obey $\mathfrak{l}_a \mathfrak{r}_b=\delta_{ab}$ (because $RL=I_N$). Then, we can write
\be
\label{s_2-41}
\delta\Psi(K)=RL\delta\Psi(K)=\sum_a c_a(K)\mathfrak{r}_a=R\mathfrak{c},
\ee
where $c_a(K)=\mathfrak{l}_a\delta\Psi(K)$ is a $c-$number and $\mathfrak{c}$ is the $N\times 1$ matrix
\[\mathfrak{c}=L\delta\Psi(K)=\bbm c_1(K)\\ \vdots \\ c_N(K)\ebm.\]
Therefore, \eqref{s_2-2} becomes
\be
\label{s_2-42}
i\Omega R\mathfrak{c}=-i \kappa R\mathsf{D}\, \mathfrak{c}-(-n_\alpha \mathbb{A}^\alpha)^{-1}\mathbb{B}R\mathfrak{c},
\ee
where $\mathsf{D}$ is the $N\times N$ real diagonal matrix $\mathsf{D}=diag(\Lambda_1,\cdots,\Lambda_N)$ and, thus, $(-n_\alpha \mathbb{A}^\alpha)^{-1}\zeta_\beta \mathbb{A}^\beta R = R\mathsf{D}$. By multiplying \eqref{s_2-42} by $\mathfrak{c}^\dagger R^{-1}$ from the left one obtains that
\be
\label{s_2-43}
i\Omega |\mathfrak{c}|^2=-i\kappa \mathfrak{c}^\dagger \mathsf{D} \mathfrak{c}-\mathfrak{c}^\dagger R^{-1}(-n_\alpha \mathbb{A}^\alpha)^{-1}\mathbb{B}R\mathfrak{c}.
\ee
Since $\mathsf{D}$ is real and diagonal (which gives $\mathfrak{c}^\dagger \mathsf{D}\mathfrak{c}\in\mathbb{R}$), $\Omega=\gamma_n(-i\Gamma+c^i k_i)$, and $\kappa=\gamma_\zeta(-i \hat{d}^jc_i\Gamma+\hat{d}^jk_j)$, then 
\be
\label{s_2-44}
\Gamma_R \mathfrak{c}^\dagger (\gamma_n I_N+\gamma_\zeta \hat{d}^jc_j \mathsf{D})\mathfrak{c}=-\Re[\mathfrak{c}^\dagger R^{-1}(n_\alpha \mathbb{A}^\alpha)^{-1}\mathbb{B}R\mathfrak{c}].
\ee
On the other hand, note that $\gamma_n I_N+\gamma_\zeta \hat{d}^jc_j \mathsf{D}$ is diagonal with elements
\be
\label{s_2-45}
(\gamma_n I_N+\gamma_\zeta \hat{d}^jc_j \mathsf{D})_{aa}=\gamma_n+\gamma_\zeta\hat{d}^jc_j\Lambda_a>0
\ee
because $|\hat{d}^jc_j|\le |c^i|<1$, $\Lambda\in[-1,1]$, and $\gamma_n\ge\gamma_\zeta$ from \eqref{s_3-1}. Hence, $\gamma_n I_N+\gamma_\zeta \hat{d}^jc_j \mathsf{D}$ is a  positive Hermitian matrix and $\mathfrak{c}^\dagger (\gamma_n I_N+\gamma_\zeta \hat{d}^jc_j \mathsf{D})\mathfrak{c}>0$. The consequence is that $\Gamma_R\le 0$ if and only if 
\be
\label{s_2-46}
\Re[\mathfrak{c}^\dagger R^{-1}(n_\alpha \mathbb{A}^\alpha)^{-1}\mathbb{B}R\mathfrak{c}]\ge 0.
\ee

Now, let $\mathcal{O}$ be the LRF and $\mathcal{O}'$ some other boosted frame. The connection between the two frames is given by the Lorentz transform $t'=\gamma(t-v^ix_i)$, ${x'}^i_\parallel=\gamma(x^i_\parallel-v^i t)$, and ${x'}^i_\perp=x^i_\perp$, where $\parallel$ and $\perp$ stand for the components parallel and perpendicular to $v^i$, respectively. This can be compactly written as ${X'}^\mu=\Lambda^\mu_\nu X^\nu$. Thus, one obtains that ${K'}^\mu=\Lambda^\mu_\nu K^\nu$ and $\delta\Psi'(K')=M\delta\Psi(K)$ from the structure of \eqref{s_2} (where $M$ is an $N\times N$ invertible matrix), leading to ${\mathbb{A}'}^\mu=\Lambda^\mu_\nu M\mathbb{A}^\mu M^{-1}$ and $\mathbb{B}'=M\mathbb{B}M^{-1}$. In particular, $\zeta_\alpha \mathbb{A}^\alpha=M^{-1}\zeta'_\alpha {\mathbb{A}'}^\alpha M$ and $n_\alpha \mathbb{A}^\alpha=M^{-1}(n'_\alpha {\mathbb{A}'}^\alpha)M$. From \eqref{s_2-1}, these relations give $R'=MR$, with the same eigenvalue $\Lambda$ in both frames. Then, since $\delta\Psi(K)=R\mathfrak{c}$ and $\delta\Psi'(K')=R'\mathfrak{c}'=MR\mathfrak{c}$, one concludes that $\mathfrak{c}=\mathfrak{c}'$, i.e., $c_a'(K')=c_a(K)$. Therefore, one arrives at the following identity:   
\be
\label{s_6}
{\mathfrak{c}'}^\dagger {R'}^{-1}(-n'_\alpha {\mathbb{A}'}^\alpha)^{-1}\mathbb{B}'R'\mathfrak{c}'=\mathfrak{c}^\dagger R^{-1}(-n_\alpha \mathbb{A}^\alpha)^{-1}\mathbb{B}R\mathfrak{c}.
\ee
However, if the system is stable in the LRF, then \eqref{s_2-46} holds and, from \eqref{s_6}, one automatically obtains that $\Gamma'_R\le 0$, proving that the system is also stable in any other frame $\mathcal{O}'$ obtained via a Lorentz transformation.

\hfill $\Box$

\bibliography{References.bib}

\begin{thebibliography}{230}
\bibitem [{F:S()}]{F:Sobolev}%
  \BibitemOpen
  \href@noop {} {}\bibinfo {note} {For readers interested in technical
  mathematical aspects of our results, we recall the following point. Local
  well-posedness always refers to local well-posedness in a certain function
  space $\mathsf{X}$. This means that given initial conditions in $\mathsf{X}$,
  the solution exists in the same function space, in a sense that can be made
  precise. The relevance of $\mathsf{X}$ is that it determines the regularity
  of the solution, e.g., analytic, infinitely many times differentiable, etc.
  Here, by local well-posedness, we always mean local well-posedness in Sobolev
  spaces, which is the space of functions with square integrable derivatives up
  to order $N$ ($N$ is a fixed integer depending on the structure of the
  equations), with norm given by $\|f \|^2_N = \sum_{k=0}^N \int |\partial^k
  f|^2 \, dx$, where $\partial^k$ represent all derivatives of order $k$ and
  the derivatives are understood in a distributional sense. It is known that
  Sobolev spaces are the appropriate function spaces for most physical
  applications, including the investigation of numerical algorithms
  \cite{RezzollaZanottiBookRelHydro, AnileBook}.}\BibitemShut {Stop}%
\bibitem [{F:Q()}]{F:Quasi-analytic}%
  \BibitemOpen
  \href@noop {} {}\bibinfo {note} {In addition, a weak form of local
  well-posedness was also obtained under these assumptions. Recalling that
  local well-posedness always refers to a certain function space, existence and
  uniqueness in the space of quasi-analytic functions has been obtained. This
  is the space of functions satisfying $\partial^k f(x) \sim (k!)^s |x|^k$,
  $s>1$; $s=1$ corresponds to analytic functions. Unfortunately, this space of
  functions is too restrictive for physical applications. That is why
  throughout this work by local well-posedness we always mean local
  well-posedness in Sobolev spaces; see \cite{F:Sobolev}.}\BibitemShut {Stop}%
\end{thebibliography}

\end{document}